\documentclass[titlepage,a4paper,12pt]{report}
\usepackage{amsmath}
\usepackage{amsfonts}
\usepackage{amssymb}

\usepackage{latexsym}
\usepackage{color}
\usepackage{graphicx}
\usepackage{epsfig}
\usepackage{times}
\usepackage{pifont}
\usepackage{fancybox}
\usepackage{setspace}
\usepackage{comment}
\usepackage{epic}

\newcommand{\mb}{\mathbf}
\newcommand{\g}{\mathfrak{g}}
\newcommand{\kf}{\mathfrak{k}}
\newcommand{\h}{\mathfrak{h}}
\newcommand{\s}{\mathfrak{s}}
\newcommand{\lf}{\mathfrak{l}}
\newcommand{\of}{\mathfrak{o}}
\newcommand{\uf}{\mathfrak{u}}
\newcommand{\p}{\mathfrak{p}}
\newcommand{\vv}{\vee}
\newcommand{\pf}{\mathrm{pf}}
\newcommand{\gal}{\mathrm{Gal}}
\newcommand{\com}{\mathrm{com}}
\newcommand{\diag}{\mathrm{diag}}
\newcommand{\sgn}{\mathrm{sgn}}
\newcommand{\eff}{\mathrm{eff}}
\newcommand{\ch}{\mathrm{ch}}
\newcommand{\ad}{\mathrm{ad}}
\newcommand{\Tr}{\mathrm{Tr}}
\newcommand{\End}{\mathrm{End}}
\newcommand{\Sym}{\mathrm{Sym}}
\newcommand{\Li}{\mathrm{Li}}
\newcommand{\id}{\mathrm{id}}

\newtheorem{thm}{Theorem}
\newtheorem{lemma}{Lemma}

\begin{document}

\begin{titlepage}
\begin{center}
{\large ALGEBRAIC K-THEORY AND PARTITION FUNCTIONS IN CONFORMAL FIELD THEORY}\\

\vspace{10mm}

{\bf Sin\'ead Keegan}\\

\vspace{20mm}

{\scshape The thesis is submitted to\\
University College Dublin\\
for the degree of PhD\\
in the College of\\
Engineering, Mathematical and Physical Sciences}\\

\vspace{10mm}

{\it September 2007}

\vspace{15mm}

Based on research conducted in the\\
Dublin Institute for Advanced Studies\\
and the School of Physics, UCD\\
{\footnotesize {\it (Head of School: Prof. Gerry O'Sullivan)}}

\vspace{10mm}

under the supervision of\\
{\bf Prof. Werner Nahm}

\end{center}

\end{titlepage}

\pagenumbering{roman}

\pagenumbering{roman}

\tableofcontents

\chapter*{Acknowledgements} I would like to thank Prof. Werner Nahm for giving me the
opportunity to work with him. I particularly appreciate his
willingness to answer so many questions; needless to say I have
learnt a great deal from him over the last three years. Many thanks
to my family whose support was essential to me while completing my
thesis. Thanks also to my friends, in particular Anne-Marie, for
plenty of kindness and encouragement. Their company made my time in
UCD much more enjoyable. I would like to acknowledge financial
support from IRCSET, and to thank Ronan Byrne for administering the
funds. I am grateful to Prof. Peter Hogan for the times he helped me
during my PhD, and I also wish to acknowledge the involvement of
other members of the School of Physics including Prof. Gerry
O'Sullivan, Prof. Martin Gr\"{u}newald, and Dr. Emma Sokell. Many
thanks also to my aunt \'{A}ine Hyland for plenty of good advice
over the last few months. Finally a special word of thanks to Prof.
Don Zagier for the time and effort he spent reading my thesis, and
for raising very interesting questions and providing many new
insights.

\chapter*{Notation} \begin{tabular}{lcl}
$\g$ &$\quad-\quad$& Lie algebra of rank $r$\\
$\alpha_1,\ldots,\alpha_r$ &$\quad-\quad$& Simple roots of $\g$\\
$\omega_1,\ldots,\omega_r$ &$\quad-\quad$& Fundamental weights of $\g$\\
$\Delta^+$ &$\quad-\quad$& Set of positive roots of $\g$\\
$\rho$ &$\quad-\quad$& Weyl vector\\
$C(\g)$ &$\quad-\quad$& Cartan matrix of $\g$\\
$W(\g)$ &$\quad-\quad$& Weyl group of $\g$\\
$g$ &$\quad-\quad$& Coxeter number of $\g$\\
$h$ &$\quad-\quad$& Dual Coxeter number of $\g$\\
$V(\lambda)$ &$\quad-\quad$& Irreducible representation of $\g$, of highest
weight $\lambda$\\
$\chi_{\lambda}=\ch(\lambda)$ &$\quad-\quad$& Character of the
representation
$V(\lambda)$\\
$Y(\g)$ &$\quad-\quad$& Yangian associated with $\g$\\
$W_i^j$ &$\quad-\quad$& Irreducible representation of $Y(\g)$\\
$Q_j^i$ &$\quad-\quad$& Character of the representation $W_i^j|_\g$\\
$Q_0$ &$\quad-\quad$& Weyl denominator\\
\end{tabular}

\newpage

\setcounter{page}{1}
\pagenumbering{arabic}
\chapter{Introduction} Quantum field theory (QFT) plays a major role in modern theoretical
physics. From its beginnings in the late 1920s, it has grown to be
the most successful physical theory in existence today. It provides
the best working description of the fundamental laws of physics and
is an extremely useful tool for investigating the behaviour of
complex systems. Its unrivalled ability to accurately calculate
physical quantities only adds to its reputation. Therefore it is
hardly surprising that quantum field theory plays such a central
role in our description of nature.

In addition to being a significant theory in its own right, QFT
provides essential tools to many other branches of physics, for
example to condensed matter physics. The influence of QFT is
far-reaching, with its very mathematical nature helping to build new
bridges between physics and mathematics. Despite its many great
successes, very little is known about the deep mathematical
structure underlying this theory. Progress in this area would be
beneficial to both mathematics and physics.

With any quantum field theory, the main aim is to find an exact
solution, by no means an easy task. In fact the solution of any
non-trivial QFT, whether or not it is physically relevant, would be
a major step forward in the search for a better understanding of the
subject. By an exact solution we mean the explicit calculation of
the n-point functions (or correlation functions) of the theory. This
is sufficient since Wightman's reconstruction theorem \cite{Wi}
states that with this knowledge the entire field content and
physical state space of the theory can be computed.

The attempt to solve any QFT exactly is a very ambitious
undertaking. The most likely candidates for success are those QFTs
whose symmetries give rise to a large number of conservation laws.
These laws might impose enough restrictions on the theory to allow
it to be solved exactly.

Unfortunately this approach is useless in 3+1 dimensions because of
the Coleman-Mandula theorem \cite{CM}. This states that besides
Poincar\'e invariance and an internal gauge group describing the
degeneracy of the particle spectrum, any additional symmetries cause
the scattering matrix of each massive QFT to be trivial. There is a
similar result due to Lochlainn O'Raifeartaigh~\cite{LR}, which
states that it is impossible to combine internal and relativistic
symmetries other than in a trivial way.

{\bf 1+1 dimensional integrable models}

This motivates us to consider the 1+1 dimensional case, in which the
Coleman-Mandula theorem no longer holds. Here there is nothing to
rule out the existence of a theory with an infinite number of
conservation laws, meaning that the hope of finding exact solutions
is much more realistic. Such theories are called integrable.
Integrable models are hugely relevant in physics, having found
applications in off-critical descriptions of statistical mechanics
and condensed matter systems reduced to two dimensions.

Restrictions to this low-dimensional case may seem unrealistic.
However, by facilitating the construction of exact solutions, 1+1
dimensional theories can be used to shed new light on the structure
of more general QFTs. This easily justifies the study of QFT in 1+1
dimensions.

{\bf Links to conformal field theory}

The recent wave of interest in integrable models lies in their
interpretation as deformed conformal field theories (CFTs)
\cite{Z1}. CFTs form a particular class of integrable field
theories. They are characterised by scale invariance and describe
massless relativistic particles or statistical mechanical systems at
a critical point. As with other theories, they become extremely
powerful in two dimensions, being characterised by an infinite
number of conserved currents that ensure their solvability. Here the
infinite set of conservation laws corresponds to conformal
space-time symmetry. Interest in this topic was rekindled when
Belavin, Polyakov and Zamolodchikov \cite{BPZ} showed that the
so-called minimal models are particular examples of solvable
massless QFTs.

Given a CFT, what happens to the infinite set of conservation laws
arising from conformal invariance when the system moves away from
the critical point and scale invariance is lost? When the action of
the critical point theory is perturbed by particular fields, the
conformally invariant structure is in general destroyed. However,
Zamolodchikov showed that for particular deformations of the CFT, an
infinite set of the conserved charges may survive the breaking of
conformal symmetry. This results in the corresponding massive field
theory being integrable.

{\bf Scattering matrices}

The scattering matrix (or S-matrix) is central to the study of
quantum field theory. It determines the on-shell structure of the
model and describes the collision of quantum particles. It must obey
certain constraints inspired by the physics of the system. These
include crossing symmetry, Lorentz invariance, analyticity in the
energy variables, and the conservation of probability. In 1+1
dimensions these constraints are often restrictive enough that they
enable one to conjecture the S-matrix and hence determine the theory
completely. An essential feature of any 1+1 dimensional integrable
field theory is an infinite set of conserved charges. This strongly
restricts the dynamics of the system, imposing the following
conditions on the scattering process: conservation of the total
number of particles, conservation of individual particle momenta,
and factorisation of the S-matrix into 2-particle scattering
amplitudes.

Consistency of different ways of decomposing an amplitude into
two-particle amplitudes leads to cubic relations between the
two-particle amplitudes. These cubic relations are essentially the
Yang-Baxter (or star-triangle) relations \cite{Y,Bx}.

In principle, exact construction of the scattering matrix serves as
a first step in the calculation of the n-point or correlation
functions of a system, for example via the form factor
programme~\cite{We, KW, Smi}. Although the relevant calculations are
highly non-trivial and have been carried out in only the simplest
cases, this nevertheless demonstrates the importance of the S-matrix
in the search for a complete solution of any quantum field theory.

\pagebreak
{\bf The thermodynamic Bethe ansatz}

On the other hand it is possible to start with a massive integrable
field theory and proceed in the opposite direction to recover
conformal invariance. In the high-energy limit the masses of
particles become negligible, with the result that scale and
therefore conformal invariance are approximately restored. The high
energy limit of an integrable QFT is studied by means of the
thermodynamic Bethe ansatz (TBA).

The TBA was developed over twenty years ago by Yang and Yang
\cite{YY1,YY2}, as a technique to calculate thermodynamic quantities
for a system of bosons interacting dynamically through factorisable
scattering. The method was later generalised \cite{Z2,KM1,KM2} to a
system of relativistic particles interacting dynamically through the
scattering matrix of an integrable QFT. It has become one of the
most effective techniques for exploring the close relationship
between conformal and integrable field theories.

Using the TBA approach, information can be extracted from a massive
integrable quantum field theory once its scattering matrix is known.
In particular, the effective central charge of the corresponding UV
conformal field theory can be calculated. The information gained in
this way is often sufficient to determine the field content at the
critical point. Once the perturbing operator has been determined,
the integrable field theory can then be formally described in terms
of a classical Lagrangian. The TBA also has applications in
scattering theory, where it can be used to test conjectured
S-matrices for consistency.

This brief account of quantum field theory sets the scene for the
work in this thesis.

{\bf Outline of PhD thesis}

By a mixture of analytical and numerical techniques, physicists have
formulated many beautiful and well-supported hypotheses for the
integrable massive perturbations of conformally invariant theories,
though mathematical proofs are not yet available. In particular it
has been found that conformal dimensions are given in terms of the
dilogarithm formulas~\cite{GT}, by finite order elements of the
Bloch group, a basic tool in algebraic K-theory. This shows a deep
connection between physics and a very active domain of number
theory.

Among the integrable models for which this relationship holds are
those described by pairs $(X,Y)$ of ADET Dynkin diagrams~\cite{N}.
Such models have equations of the form $AU=V$, where $A=C(X)^{-1}
\otimes C(Y)$, $C$ denotes a Cartan matrix, and $X$ and $Y$ are the
Dynkin diagrams of simple Lie algebras of ranks $m$ and $n$
respectively. Moreover $U=\log(x)$ and $V=\log(1-x)$ satisfy
$e^U+e^V=1$, the vector $x$ is given by $x=(x_{11}, \ldots,
x_{mn})$, and $f(x)$ is to be interpreted as
$\left(f(x_{11}),\ldots,f(x_{mn})\right)$. The relationship between
the matrix $A$ and the scattering matrix of the integrable quantum
field theory is described in chapter 3.

It seems that the resulting algebraic equations for $x$ are solvable
in terms of roots of unity. The case $(X,Y)=(A_m,A_n)$ has already
been studied \cite{N}. In this thesis we consider mainly the case
$(X,Y)=(D_m,A_n)$. We study the algebraic equations of the model and
find all solutions in explicit form using the representation theory
of Lie algebras and related quantum groups. The solutions seem to be
torsion elements of the Bloch group, allowing the effective central
charge of the corresponding CFT to be calculated using the dilog
formula mentioned above.

On a more mathematical note, we are interested in a closely related
problem involving certain q-hypergeometric series. We investigate
the overlap between series of this type and modular functions. We
study a particular class of $2$-fold q-hypergeometric series,
denoted $f_{A,B,C}$, depending on some $2\times 2$ matrix $A$. It
turns out that for certain choices of the matrix $A$, the function
$f$ can be made modular. We calculate the corresponding values of
$B$ and $C$. It is expected that functions $f$ arising in this way
are characters of some rational conformal field theory. We show that
this is true in at least one case. The content of the thesis is
arranged as follows:

{\bf Chapter 2} introduces the mathematical and physical concepts
important for subsequent chapters. The first section presents the
theory of simple Lie algebras, as well as a number of other useful
mathematical definitions. The second section is a short introduction
to conformal field theory, concentrating on the aspects most
relevant to this thesis.

{\bf Chapter 3} is the core of the thesis. We study the integrable
models described by pairs of Dynkin diagrams $(D_m,A_n)$. The
equations of these models are solved in the general case, and their
relation to Yangian representation theory is discussed. Results of
the effective central charge calculations for many different models
are summarised. The realisation of such models as coset models is
also discussed.

{\bf Chapter 4} places our results in a more general context. We
examine the relationship between certain q-hypergeometric series and
modular functions. This is very closely related to the conformal
field theory of the previous chapter.

{\bf Chapter 5} contains initial steps for the solutions of the
remaining cases. A number of integrable models described by pairs of
`non-classical' Dynkin diagrams is studied. In particular the pairs
$(E_6,T_1)$, $(E_7,T_1)$, and $(E_8,T_1)$ are considered. In each
case the equations of the models are solved, and corresponding
values of the effective central charge are calculated. All
calculations in this chapter are carried out using only elementary
algebra.

\chapter{Overview} \section{Mathematical Overview}
\subsection{Lie algebras} This section gives a brief introduction to
simple Lie algebras, focusing mainly on concepts that arise in the
study of conformal field theory. As far as possible the discussion
is self-contained. There are numerous good books on the subject, and
for a more detailed account of the material presented here see e.g.
\cite{diF,FH}.

\subsubsection{Simple Lie algebras, generators, and roots}
A vector space $\g$ is called a {\bf Lie algebra} if it is equipped
with an anti-symmetric bilinear map $[\cdot,\cdot]: \g\times \g \rightarrow \g$,
called the {\bf Lie bracket}, satisfying the Jacobi identity
$$\left[x,[y,z]\right]+[y,[z,x]]+[z,[x,y]]=0\ ,$$
for all $x,y,z\in \g$.

We consider only finite-dimensional Lie algebras, meaning that $\g$
is finite-dimensional when viewed as a vector space. Moreover we
restrict ourselves to {\bf simple Lie algebras}. These are
non-abelian, i.e. $[\g,\g]\neq0$, and contain no proper ideal, i.e.
there is no subalgebra $\kf$ of $\g$ such that $[\kf,\g]\subset\kf$.
This particular class of Lie algebras is special in that each of its
members is classified uniquely by its Dynkin diagrams.

Below are some examples of simple Lie algebras that can be realised in terms of
matrix algebras.
\begin{eqnarray*}
A_n &=& \s\uf(n+1)\ ,\\
B_n &=& \s\of(2n+1)\ ,\\
C_n &=& \s\p(2n)\ ,\\
D_n &=& \s\of(2n)\ .
\end{eqnarray*}
Here $A_n$ denotes the set of anti-hermitian complex
$(n+1)\times(n+1)$ matrices of trace zero. Elements of $B_n$ and
$D_n$ are anti-symmetric complex matrices of trace zero. Elements of
$C_n$ are $2n\times 2n$ matrices of form
$$\left(\begin{array}{cc}A&B\\C&D\\
\end{array}\right)$$
where $B$ and $C$ are symmetric and $D=-A^T$.

A Lie algebra can be specified by a set of generators $\{I^a\}$ and
their commutation relations $[I^a,I^b]=\sum_c f^{abc}I^c$, where
$f^{abc}$ denote the structure constants of $\g$. There are
dim($\g$) generators in total.

In the {\bf Cartan-Weyl basis} generators are constructed as follows:
\begin{itemize}
\item{Choose a maximal commuting subspace $\h$ of $\g$ (called a
{\bf Cartan subalgebra}). Choose the first r generators
$\{H^1,\ldots,H^r\}$ to be a basis of $\h$. The dimension of $\h$,
denoted $r$, is called the {\bf rank} of $\g$.}
\item{One can simultaneously diagonalise the actions
$x\rightarrow [H^i,x]$ of the $H^i$ on $x$. The remaining generators
are chosen to be eigenvectors $X_{\beta}$ of these maps, so that
$[H^i,X^{\beta}]=\beta^iX^{\beta}$.}
\end{itemize}

A {\bf root} $\beta=\left(\beta^1,\ldots,\beta^r\right)$ is an
element of $\h^{\star}$ which is a non-zero eigenvector of the
action of $X^{\beta}$ on $\h$. The set of all roots is denoted by
$\Delta$. $\Delta$ spans $\h^{\star}$ but is not a linearly
independent set, therefore it is natural to choose a subset $\Pi$ of
$\Delta$ that forms a basis of $\h^{\star}$. There exists a subset
$\Pi=\{\alpha_1,\ldots,\alpha_r\}$ of $\Delta$, called the set of
{\bf simple roots}, such that $\Pi$ is a linearly independent set
and each root $\beta$ can be written as an integral linear
combination of simple roots
$$\beta=\sum_{i=1}^r n_i\alpha_i\ ,$$
where $n_i\in\mathbb{Z}$, and either all $n_i\geq0$ or all
$n_i\leq0$. A root $\beta$ is called a {\bf positive root} if all
$n_i\geq0$ and a {\bf negative root} if all $n_i\leq0$. The sets of
positive and negative roots are denoted by $\Delta^+$ and $\Delta^-$
respectively. Clearly $\Delta^-= - \Delta^+$ and $\Delta=\Delta^+
\cup\Delta^-$.

The {\bf length} of a root $\beta$ is defined by
$$l(\beta)=\sum_{i=1}^r n_i\ .$$

The length functional $l:\Delta\rightarrow\mathbb{Z}$ allows a
partial ordering, $<$, to be defined on the set of roots as follows.
For $\alpha, \beta \in \Delta$ we write
\begin{equation}
\alpha>\beta \Leftrightarrow l(\alpha)>l(\beta)\ . \label{eq:order}
\end{equation}

\subsubsection{The adjoint representation and the Killing form}
The most natural representation of a Lie algebra is the {\bf adjoint
representation}, in which $\g$ is represented as an operator algebra
acting on itself. Every element $x\in\g$ can be viewed as an
operator by means of the adjoint action, $x\mapsto \ad(x)$, which
acts on $\g$ as
$$\ad(x)(y)=[x,y]\ .$$

This leads in a natural way to the introduction of a symmetric bilinear form on
$\g$. This is the so-called {\bf Killing form} $K:\g \times \g \rightarrow
\mathbb{C}\,$, defined by
\begin{equation}
(x,y)\mapsto K(x,y)=\frac{1}{2h}\Tr\left(\ad(x)\circ \ad(y)\right)\
, \label{eq:killing}
\end{equation}
for all $x,y\in\g$. The factor $h$ in the normalisation constant is
the dual Coxeter number of $\g$, see~(\ref{eq:dual}).

The Killing form establishes an isomorphism between $\h$ and $\h^{\star}$, and
can be used to induce a positive definite scalar product on $\h^{\star}$ through
$(\alpha,\beta)=K(H^{\alpha},H^{\beta})$. Here the element $H^{\alpha}\in\h$ is
uniquely defined by the relation $\alpha(H)=K(H^{\alpha},H)$. Since all roots of
$\g$ sit in $\h^{\star}$, this defines a scalar product on the root space.
From now on this is the scalar product we use between roots.

The partial ordering introduced in~(\ref{eq:order}) identifies a unique {\bf
highest root} $\theta$ for which
$$l(\theta)>l(\alpha)\ ,$$
for all other roots $\alpha$. The particular
normalisation~(\ref{eq:killing}) of the Killing form was chosen to
ensure that $$(\theta,\theta)=2\ .$$ This follows the standard
convention in which the square length of a long root is $2$.

The highest root can be written as
$$\theta = \sum_{i=1}^r a_i\alpha_i=\sum_{i=1}^ra_i^\vv\alpha_i^\vv\ ,$$
where the $a_i$ and $a_i^\vv$ are natural numbers, called the {\bf Kac labels}
and {\bf dual Kac labels} respectively. Here $\alpha_i^\vv$ is the {\bf simple
coroot} corresponding to the simple root $\alpha_i$. It is defined as
$$\alpha_i^\vv:=\frac{2\alpha_i}{|\alpha_i^2|}\ .$$

The sums of the Kac labels and dual Kac labels define two important
constants in the theory of simple Lie algebras. These are the {\bf
Coxeter number}
$$g=\sum_{i=1}^r a_i +1\ ,$$
and the {\bf dual Coxeter number}
\begin{equation}
h=\sum_{i=1}^r a_i^\vv +1\ . \label{eq:dual}
\end{equation}

\subsubsection{Cartan matrices and Dynkin diagrams}
To each simple Lie algebra $\g$ we can associate a unique {\bf
Cartan matrix} $C(\g)$. This is an positive definite $r\times r$
matrix, whose elements $C_{ij}$ are defined in terms of the simple
roots by
\begin{equation}
C_{ij}=(\alpha_i,\alpha_j^\vv)\ . \label{eq:cartan}
\end{equation}
The Cartan matrix has the following properties:
\begin{itemize}
\item{$C_{ij}\in\mathbb{Z}$\ ,}
\item{The diagonal entries satisfy $C_{ii}=2$\ ,}
\item{The off-diagonal entries $C_{ij}$ can take values 0, -1,
-2, or -3\ ,}
\item{$C_{ij}=0 \, \Leftrightarrow \, C_{ji}=0$\ ,}
\item{$\det(C)>0$\ .}
\end{itemize}

The {\bf Dynkin diagram} of a simple Lie algebra is a connected
planar diagram encoding the Cartan matrix in the following way. To
every simple root $\alpha_i$ we associate a vertex, and vertices $i$
and $j$ are joined by $C_{ij}C_{ji}$ lines. Let $\theta_{ij}$ be the
angle between the simple roots $\alpha_i$ and $\alpha_j$. It can
easily be shown that $C_{ij}C_{ji}=4\cos^2\theta_{ij}$, and hence
that any two vertices can be joined by either 0, 1, 2, or 3 links.
Where there is only one link joining two vertices, the corresponding
simple roots are both of the same length. Where there is more than
one link joining two vertices, an arrow is drawn pointing from the
longer to the shorter root. The Dynkin diagrams of simple Lie
algebras are separated into two classes, those in which only single
links occur (the ADE series), and those where multiple links are
allowed (the BCFG series). For obvious reasons these are called {\bf
simply-laced} and {\bf non simply-laced} respectively. Clearly in
the former case all roots must have the same length, while in the
latter they can be different. In the set of all roots of a given Lie
algebra, at most two different lengths are possible.

A simple Lie algebra uniquely determines its Dynkin diagram, and
hence the classification of simple Lie algebras corresponds exactly
to the classification of Dynkin diagrams. The Dynkin diagram
contains all of the information necessary to reconstruct the entire
root system of the algebra. For simple Lie algebras the set of all
allowed Dynkin diagrams is as follows.

{\it Simply-laced:}
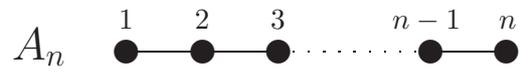
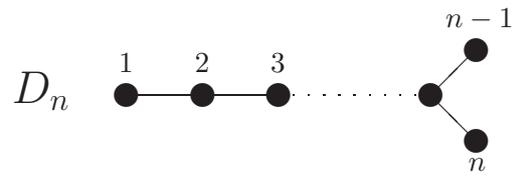
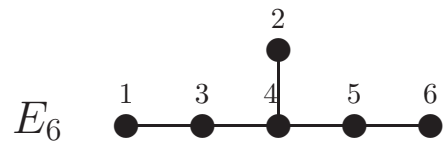
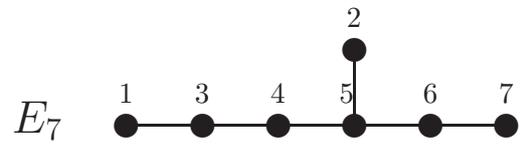
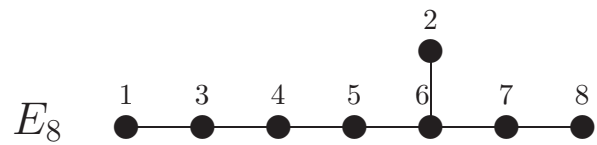
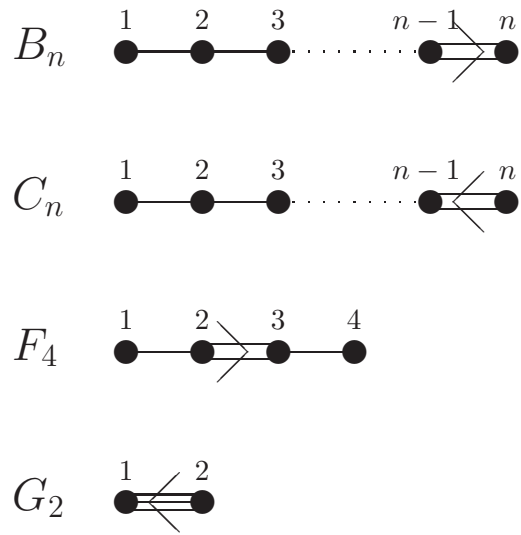
\begin{center}
\begin{picture}(50,10)
\put(-15,-1){\Large{$A_n$}}
\put(0,0){\circle*{3}}
\put(-1,3){\small{$1$}}
\put(0,0){\line(1,0){10}}
\put(10,0){\circle*{3}}
\put(9,3){\small{$2$}}
\put(10,0){\line(1,0){10}}
\put(20,0){\circle*{3}}
\put(19,3){\small{$3$}}
\dottedline{2}(20,0)(40,0)
\put(40,0){\circle*{3}}
\put(35,3){\small{$n-1$}}
\put(40,0){\line(1,0){10}}
\put(50,0){\circle*{3}}
\put(49,3){\small{$n$}}
\end{picture}
\end{center}

\begin{center}
\begin{picture}(50,20)
\put(-15,-1){\Large{$D_n$}}
\put(0,0){\circle*{3}}
\put(-1,3){\small{$1$}}
\put(0,0){\line(1,0){10}}
\put(10,0){\circle*{3}}
\put(9,3){\small{$2$}}
\put(10,0){\line(1,0){10}}
\put(20,0){\circle*{3}}
\put(19,3){\small{$3$}}
\dottedline{2}(20,0)(40,0)
\put(40,0){\circle*{3}}
\put(40,0){\line(1,1){6}}
\put(46,6){\circle*{3}}
\put(42,9){\small{$n-1$}}
\put(40,0){\line(1,-1){6}}
\put(46,-6){\circle*{3}}
\put(45,-10){\small{$n$}}
\end{picture}
\end{center}

\begin{center}
\begin{picture}(50,25)
\put(-15,-1){\Large{$E_6$}}
\put(0,0){\circle*{3}}
\put(-1,3){\small{$1$}}
\put(0,0){\line(1,0){10}}
\put(10,0){\circle*{3}}
\put(9,3){\small{$3$}}
\put(10,0){\line(1,0){10}}
\put(20,0){\circle*{3}}
\put(19,13){\small{$2$}}
\put(20,0){\line(1,0){10}}
\put(30,0){\circle*{3}}
\put(30,0){\line(1,0){10}}
\put(20,0){\line(0,1){10}}
\put(40,0){\circle*{3}}
\put(39,3){\small{$6$}}
\put(29,3){\small{$5$}}
\put(18,3){\small{$4$}}
\put(20,10){\circle*{3}}
\end{picture}
\end{center}

\begin{center}
\begin{picture}(50,20)
\put(-15,-1){\Large{$E_7$}}
\put(0,0){\circle*{3}}
\put(-1,3){\small{$1$}}
\put(0,0){\line(1,0){10}}
\put(10,0){\circle*{3}}
\put(9,3){\small{$3$}}
\put(10,0){\line(1,0){10}}
\put(20,0){\circle*{3}}
\put(29,13){\small{$2$}}
\put(20,0){\line(1,0){10}}
\put(30,0){\circle*{3}}
\put(30,0){\line(1,0){10}}
\put(30,0){\line(0,1){10}}
\put(40,0){\circle*{3}}
\put(39,3){\small{$6$}}
\put(28,3){\small{$5$}}
\put(19,3){\small{$4$}}
\put(30,10){\circle*{3}}
\put(50,0){\circle*{3}}
\put(49,3){\small{$7$}}
\put(40,0){\line(1,0){10}}
\end{picture}
\end{center}

\begin{center}
\begin{picture}(50,20)
\put(-15,-1){\Large{$E_8$}}
\put(0,0){\circle*{3}}
\put(-1,3){\small{$1$}}
\put(0,0){\line(1,0){10}}
\put(10,0){\circle*{3}}
\put(9,3){\small{$3$}}
\put(10,0){\line(1,0){10}}
\put(20,0){\circle*{3}}
\put(39,13){\small{$2$}}
\put(20,0){\line(1,0){10}}
\put(30,0){\circle*{3}}
\put(30,0){\line(1,0){10}}
\put(40,0){\line(0,1){10}}
\put(40,0){\circle*{3}}
\put(38,3){\small{$6$}}
\put(29,3){\small{$5$}}
\put(19,3){\small{$4$}}
\put(40,10){\circle*{3}}
\put(50,0){\circle*{3}}
\put(49,3){\small{$7$}}
\put(40,0){\line(1,0){10}}
\put(60,0){\circle*{3}}
\put(59,3){\small{$8$}}
\put(50,0){\line(1,0){10}}
\end{picture}
\end{center}

\pagebreak
{\it Non simply-laced:}
\begin{center}
\begin{picture}(50,10)
\put(-15,-1){\Large{$B_n$}}
\put(0,0){\circle*{3}}
\put(-1,3){\small{$1$}}
\put(0,0){\line(1,0){10}}
\put(10,0){\circle*{3}}
\put(9,3){\small{$2$}}
\put(10,0){\line(1,0){10}}
\put(20,0){\circle*{3}}
\put(19,3){\small{$3$}}
\dottedline{2}(20,0)(40,0)
\put(40,0){\circle*{3}}
\put(35,3){\small{$n-1$}}
\put(40,1){\line(1,0){10}}
\put(40,-1){\line(1,0){10}}
\put(50,0){\circle*{3}}
\put(49,3){\small{$n$}}
\put(47,0){\line(-1,1){4}}
\put(47,0){\line(-1,-1){4}}
\end{picture}
\end{center}

\begin{center}
\begin{picture}(50,10)
\put(-15,-1){\Large{$C_n$}}
\put(0,0){\circle*{3}}
\put(-1,3){\small{$1$}}
\put(0,0){\line(1,0){10}}
\put(10,0){\circle*{3}}
\put(9,3){\small{$2$}}
\put(10,0){\line(1,0){10}}
\put(20,0){\circle*{3}}
\put(19,3){\small{$3$}}
\dottedline{2}(20,0)(40,0)
\put(40,0){\circle*{3}}
\put(35,3){\small{$n-1$}}
\put(40,1){\line(1,0){10}}
\put(40,-1){\line(1,0){10}}
\put(50,0){\circle*{3}}
\put(49,3){\small{$n$}}
\put(43,0){\line(1,1){4}}
\put(43,0){\line(1,-1){4}}
\end{picture}
\end{center}

\begin{center}
\begin{picture}(50,10)
\put(-15,-1){\Large{$F_4$}}
\put(0,0){\circle*{3}}
\put(-1,3){\small{$1$}}
\put(0,0){\line(1,0){10}}
\put(10,0){\circle*{3}}
\put(9,3){\small{$2$}}
\put(10,1){\line(1,0){10}}
\put(10,-1){\line(1,0){10}}
\put(16,0){\line(-1,1){4}}
\put(16,0){\line(-1,-1){4}}
\put(20,0){\circle*{3}}
\put(19,3){\small{$3$}}
\put(20,0){\line(1,0){10}}
\put(30,0){\circle*{3}}
\put(29,3){\small{$4$}}
\end{picture}
\end{center}

\begin{center}
\begin{picture}(50,10)
\put(-15,-1){\Large{$G_2$}}
\put(0,0){\circle*{3}}
\put(-1,3){\small{$1$}}
\put(0,0){\line(1,0){10}}
\put(10,0){\circle*{3}}
\put(9,3){\small{$2$}}
\put(0,1){\line(1,0){10}}
\put(0,0){\line(1,0){10}}
\put(0,-1){\line(1,0){10}}
\put(3,0){\line(1,1){4}}
\put(3,0){\line(1,-1){4}}
\end{picture}
\end{center}

The algebras of types $A$, $B$, $C$, and $D$ are known as {\bf classical Lie
algebras}, while $E_6$, $E_7$, $E_8$, $F_4$, and $G_2$ are the {\bf
exceptional Lie algebras}.

\subsubsection{Weights and the Weyl vector}
Let $\{\omega_1,\ldots,\omega_r\}$ be the basis dual to the basis of
simple coroots, i.e. $$(\omega_i,\alpha_j^\vv)=\delta_{ij}\ .$$ Then
$\omega_1,\ldots,\omega_r$ are called the {\bf fundamental weights}
of $\g$.

For an arbitrary representation of $\g$, a basis $\{|\lambda\rangle\}$ can
always be found such that
$$H^i|\lambda\rangle=\lambda^i|\lambda\rangle\ .$$
The eigenvalues $\lambda^i$ form a vector $\lambda=(\lambda^1,\ldots,
\lambda^r)$ called a {\bf weight}. As in the case of roots, weights live in the
space $\h^{\star}$; $\,\lambda(H^i)=\lambda^i$. Therefore, the scalar
product between weights is also fixed by the Killing form. That roots and
weights occupy the same space makes perfect sense, since roots are just a
special name given to the weights of the adjoint representation.

Every weight can be written as an integral linear combination of the fundamental
weights
$$\lambda=\sum_{i=1}^r\lambda_i\omega_i\ ,$$
where $\lambda_i\in\mathbb{Z}$. The expansion coefficients $\lambda_i$ of a
weight $\lambda$ in the fundamental weight basis are called {\bf Dynkin labels}.
From now on when a weight is written in component form
$$\lambda=(\lambda_1,\ldots,\lambda_r)\ ,$$
it is assumed that the components are the Dynkin labels.

One weight of particular importance is the {\bf Weyl vector}, $\rho$. This is
written in terms of fundamental weights as
$$\rho=\sum_{i=1}^r \omega_i =(1,1,\ldots,1)\ .$$
Alternatively it can be expressed in terms of positive roots as
$$\rho=\frac{1}{2}\sum_{\alpha\in\Delta^+}\alpha\ .$$

The Weyl vector is uniquely determined by the property that
$(\rho,\alpha_i^\vv)=1$. This follows immediately from the definition of $\rho$
in terms of fundamental weights.

\subsubsection{Weyl reflections and the Weyl group}
Given roots $\alpha$ and $\beta$ of $\g$, it can be shown that the quantity
$\beta-(\alpha^\vv,\beta)\,\alpha$ is also a root. This leads to the
introduction of a new operation $s_{\alpha}:\Delta\rightarrow\Delta$ defined by
$$s_{\alpha}\beta:=\beta-(\alpha^\vv,\beta)\alpha\ .$$
Such a mapping is called a {\bf Weyl reflection}. It is a reflection
with respect to the hyperplane perpendicular to $\alpha$. The set of
all such reflections with respect to roots forms the {\bf Weyl
group} of $\g$, denoted $W(\g)$. The $r$ {\bf simple Weyl
reflections}
$$s_i\equiv s_{\alpha_i}\ ,$$
generate the whole of the Weyl group via composition. Since only a
limited number of composite simple Weyl reflections can differ from
the identity, $W(\g)$ is always a finite group. The action of the
Weyl group on the simple roots yields the set of all roots of the
algebra
$$\Delta=W(\g)\{\alpha_1,\ldots,\alpha_r\}\ .$$

\subsubsection{The Weyl character formula}
A character is a useful way of encoding all of the information about
a representation. Suppose $\pi:\g\rightarrow \End(V)$ is a
representation of $\g$. Then the {\bf character} of $\pi$ is the
function
\begin{eqnarray*}
\chi:\g&\rightarrow& \mathbb{C}\\
x&\mapsto&\Tr(\pi(x))\ .
\end{eqnarray*}
The character is independent of the choice of basis for $V$.

The {\bf Weyl character formula} allows us to calculate the
character of an irreducible representation given its highest weight.
The character of the irreducible representation of highest weight
$\lambda$ is given by
$$\chi_{\lambda}=\frac{\sum_{w\in W(\g)}\det(w)e^{w(\lambda+\rho)}}{\sum_{w\in
W(\g)}\det(w)e^{w\rho}}\ .$$

The Weyl denominator can be written as
$$\sum_{w\in
W(\g)}\det(w)e^{w\rho}=\prod_{\alpha\in\Delta^+}\left(e^{\alpha/2}-e^{-\alpha/2}
\right)\ .$$ In many calculations it is more convenient to use the
multiplicative form of this expression.

\subsubsection{Chevalley generators}
The Lie algebra $\g$ can also be described as the Lie algebra with
generators $H_i$, $X_i^{\pm}$ for $i=1,\ldots,r$, and defining
relations
\begin{eqnarray*}
\left[H_i,H_j\right] &=& 0\ ,\\
\left[H_i,X_j^{\pm}\right] &=& C_{ij}X_j^{\pm} \ ,\\
\left[X_i^{+},X_j^{-}\right] &=& \delta_{ij}H_i\ ,\\
\left(ad_{X_i^{\pm}}\right)^{1-C_{ij}}\left(X_j^{\pm}\right) &=&
0\,\,\, \rm{for}\,\, i\neq j\ .
\end{eqnarray*}
The $H_i,\ X_i^{\pm}$ are called the {\bf Chevalley generators} of
$\g$.

\subsubsection{Universal enveloping algebra}
For a finite-dimensional complex Lie algebra $\g$, its tensor
algebra $T(\g)$ is given by
$$T(\g)=\mathbb{C} \oplus \g \oplus (\g\otimes\g) \oplus
(\g\otimes\g\otimes\g)\oplus\ldots\ .$$ Let $I$ be the two-sided
ideal of $T(\g)$ generated by all elements of the form
$$x\otimes y - y\otimes x -[x,y]\ ,$$ where $x,\,y\in\g$. Then the
universal enveloping algebra of $\g$, denoted $U\g$, is defined to
be the quotient
$$U\g = T(\g)/I\ .$$

\subsubsection{Affine Kac-Moody algebras} To every finite-dimensional
Lie algebra $\g$, we can associate an affine extension $\hat{\g}$ by
adding an extra node to the Dynkin diagram of $\g$. For example if
$\g=E_8$ the new node is added to the first node on the left hand
side of the original $E_8$ Dynkin diagram. The resulting algebra is
called an {\bf affine Kac-Moody algebra}. Affine Kac-Moody algebras
have important applications in conformal field theory, in particular
in the study of WZW models.

The fundamental concepts of roots, weights, Cartan matrices and the
Weyl group extend easily from the finite to the affine case. However
the addition of the extra simple root in the affine case results in
both the root system and the Weyl group of $\hat{\g}$ becoming
infinite. Consequently all highest-weight representations are
infinite-dimensional. For simplicity these representations are
arranged in terms of a new parameter, $k$, called the {\bf level}.
The level of a weight (now described by $r+1$ Dynkin labels) is just
the sum of all of its Dynkin labels, each multiplied by the
corresponding comark.

Affine Kac-Moody algebras form a large subclass of the more general
Kac-Moody algebras. In particular they consist of those Kac-Moody
algebras whose Cartan matrix is positive semi-definite. For a good
overview of affine Kac-Moody algebras see e.g. \cite{diF,JF}.

\subsection{Hopf algebras and Yangians}
Let $A$ be an associative unitary algebra over a commutative ring
$k$, and let ${\mathbf 1}\in A$ be the unit element of $A$. Let
$\id:A\rightarrow A$ denote the identity transformation, given by
$\id(a)=a$ for all $a\in A$. $A$ is a {\bf Hopf algebra} if it
contains linear operators
\begin{itemize}
\item{multiplication $m:A\otimes A\rightarrow A$, given by
$m(a\otimes b)=ab$ for all $a,b\in A$, and satisfying
$m(m\otimes\id)=m(\id\otimes m)$\ ,}
\item{co-multiplication $\Delta:A\rightarrow A\otimes A$\ ,}
\item{antipode $s:A\rightarrow A$\ ,}
\item{unit $\eta:k\rightarrow A$, given by $\eta(c)=c{\mathbf 1}$ for all $c\in k$\ ,}
\item{co-unit $\epsilon:A\rightarrow k$\ ,}
\end{itemize}
subject to the following axioms for all $a,b\in A$
\begin{itemize}
\item{Associativity of $\Delta: \quad
(\id\otimes\Delta)\Delta(a)=(\Delta\otimes\id)\Delta(a)$\ ,}
\item{Definition of $s: \quad
m(\id\otimes s)\Delta(a)=m(s\otimes\id)\Delta(a)=\eta\epsilon(a)$\
,}
\item{Definition of $\epsilon:\quad
(\epsilon\otimes\id)\Delta(a)=(\id\otimes\epsilon)\Delta(a)=a$,\\
(Note that we identify the spaces $k$, $k\otimes A$, and $A\otimes
k$, which are naturally isomorphic)\ .}
\end{itemize}
$\Delta$ and $\epsilon$ are homomorphisms
\begin{eqnarray*}
\Delta(ab)&=&\Delta(a)\Delta(b)\ ,\\
\epsilon(ab) &=& \epsilon(a)\epsilon(b)\ ,
\end{eqnarray*}
and $s$ is an anti-homomorphism of $A$
$$s(ab)=s(b)s(a)\ .$$
The map $s^2$ is an automorphism of $A$, and if $A$ is commutative
or co-commutative it can be shown that $s^2=\id$.

For example, given a Lie algebra $\g$, its universal enveloping
algebra $U\g$ is a Hopf algebra. Take $m$ to be the usual formal
multiplication on $U\g$, and define a co-multiplication $\Delta$, a
co-unit $\epsilon$, and an antipode $s$ for all $x_1,\ldots,
x_r\in\g$ as follows
\begin{eqnarray*}
\epsilon(x_1\ldots x_r) &=& \delta_{r\,0}1\ ,\\
s(x_1\ldots x_r) &=& (-1)^r x_r\ldots x_1\ ,\\
\Delta(x_1\ldots x_r) &=& \prod_{i=1}^r \left( x_i\otimes 1 +
1\otimes x_i \right)\\
&=& \sum_{s=0}^r \sum_{\mathbf{i},\mathbf{j}} x_{i_{1}}\ldots
x_{i_s} \otimes x_{j_1}\ldots x_{j_{r-s}}\ ,
\end{eqnarray*}
where the second summation is over $i_1<\ldots<i_s$,
$j_1<\ldots<j_{r-s}$, and $\{i_1,\ldots,
i_s\}\cup\{j_1,\ldots,j_{r-s}\}=\{1,\ldots,r\}$.

\subsubsection{Yangians}
Throughout this discussion $\g$ will denote a finite-dimensional
complex simple Lie algebra of dimension $d$, whose generators
$I^1,\ldots,I^d$ are orthonormal with respect to the Killing form,
and satisfy the commutation relations $[I^a,I^b]=f^{abc}I^c$. To
every such $\g$ we can associate an (infinite-dimensional) Hopf
algebra, $Y(\g)$, called a Yangian. One has $U\g\subset Y(\g)$.

Let $Y(\g)$ be the algebra generated by $\{I^a,J^a\}$, with the
additional constraints that
\begin{eqnarray}
\left[I^a,J^b\right]&=&f^{abc}J^c\ , \nonumber\\
\left[J^a,\left[J^b,I^c\right]\right]-\left[I^a,\left[J^b,J^c\right]\right]
&=& a_{abcdeg}\left\{I^d,I^e,I^g\right\}\ , \label{eq:terrific}\\
\left[\left[J^a,J^b\right],\left[I^l,J^m\right]\right]
+\left[\left[J^l,J^m\right],\left[I^a,J^b\right]\right] &=&
\left(a_{abcdeg}f^{lmc}+a_{lmcdeg}f^{abc}\right)\left\{I^d,I^e,I^g\right\}\
, \nonumber
\end{eqnarray} where
$$a_{abcdeg}=\frac{1}{24}f^{adi}f^{bej}f^{cgk}f^{ijk}\ ,$$
and
$$\left\{x_1,x_2,x_3\right\}=\sum_{\{i,\ j,\ k\}=\{1,\ 2,\ 3\}}x_ix_jx_k\ .$$

The final two constraints in~(\ref{eq:terrific}) are chosen such
that the coproduct $\Delta:Y(\g)\rightarrow Y(\g)\otimes Y(\g)$,
given by
$$\Delta(J^a)=J^a\otimes 1+1\otimes J^a+\frac{1}{2}f^{abc}I^c\otimes I^b\
,$$ becomes a homomorphism.

The first of equations~(\ref{eq:terrific}) says that the $J^a$'s
form a basis of a representation isomorphic to the adjoint
representation of $\g$ (on a new vector space that has no Lie
algebra properties).

To see that the Yangian is in fact a Hopf algebra, define a co-unit
$\epsilon:Y(\g)\rightarrow \mathbb{C}$ by
\begin{eqnarray*}
\epsilon(I^a) &=& 0\ ,\\
\epsilon(J^a) &=& 0\ ,
\end{eqnarray*}
and an antipode $s:Y(\g)\rightarrow Y(\g)$ by
\begin{eqnarray*}
s(I^a) &=& -I^a\ ,\\
s(J^a) &=& -J^a+\frac{f^{abc}}{2}I^c\otimes I^b\ .
\end{eqnarray*}

Yangians were introduced in this way by Drinfeld \cite{D1} as part
of his work on solutions of the Yang-Baxter equation. He later gave
a second description \cite{D2} of Yangians, in terms of generators
and relations, which we now show.

Let $C=(C_{ij})$, $i,j=1,\ldots,r$, denote the Cartan matrix of
$\g$, where $r$ is the rank of $\g$. Let $d_1,\ldots,d_r$ be a set
of coprime positive integers such that the matrix $d_iC_{ij}$ is
symmetric. These $d_i$ are uniquely determined. Then the Yangian
$Y(\g)$ is isomorphic to the associative algebra with generators
$X_{ik}^{\pm}$ and $H_{ik}$, $i=1,\ldots r$ and $k\in\mathbb{N}$,
and defining relations
\begin{itemize}
\item{$[H_{ik},H_{jl}]=0$\ ,}
\item{$[H_{i0},X_{jl}^{\pm}]=\pm d_iC_{ij}X_{jl}^{\pm}$\ ,}
\item{$[X_{ik}^{+},X_{jl}^{-}]=\delta_{ij}H_{i,k+l}$\ ,}
\item{$[H_{i,k+1},X_{jl}^{\pm}]-[H_{ik},X_{j,l+1}^{\pm}]=\pm
\frac{1}{2}d_iC_{ij}\left(H_{ik}X_{jl}^{\pm}+X_{jl}^{\pm}H_{ik}\right)$\
,}
\item{$[X_{i,k+1}^{\pm},X_{jl}^{\pm}]-[X_{ik}^{\pm},X_{j,l+1}^{\pm}]=\pm\frac{1}{2}
d_iC_{ij}\left(X_{ik}^{\pm}X_{jl}^{\pm}+X_{jl}^{\pm}X_{ik}^{\pm}\right)$\
,}
\item{$i\neq j \textrm{ and } n=1-C_{ij} \Rightarrow
\Sym\left[X_{i,k_1}^{\pm}\left[X_{i,k_2}^{\pm} \ldots \left[
X_{i,k_n}^{\pm},X_{jl}^{\pm}\right]\ldots\right]\right]=0$\ ,}
\end{itemize}
where $\Sym$ is the sum over all permutations of $k_1,\ldots,k_n$.
There are similarities between this realisation of the Yangian, and
the Chevalley description of a Lie algebra $\g$.

The isomorphism $\phi$ between the two different realisations of
$Y(\g)$ is given by
\begin{eqnarray*}
\phi(H_i) &=& d_i^{-1}H_{i0}\ ,\\
\phi(J(H_i)) &=& d_i^{-1}H_{i1}+\phi(v_i)\ ,\\
\phi(X_i^{\pm}) &=& X_{i0}^{\pm}\ ,\\
\phi(J(X_i^{\pm})) &=& X_{i1}^{\pm} + \phi(w_i^{\pm})\ ,
\end{eqnarray*}
where
\begin{eqnarray*}
v_i &=& \frac{1}{4}\sum_{\beta\in\Delta^+} \frac{d_{\beta}}{d_i}\,
K(\beta,\ \alpha_i) \left(X_{\beta}^+X_{\beta}^- + X_{\beta}^-
X_{\beta}^+\right) - \frac{d_i}{2}(H_i)^2\ ,\\
w_i^{\pm} &=& \pm\frac{1}{4}\sum_{\beta\in\Delta^+} d_{\beta}
\left(\left[X_i^{\pm},\ X_{\beta}^{\pm}\right] X_{\beta}^{\mp} +
X_{\beta}^{\mp} \left[X_i^{\pm},\ X_{\beta}^{\pm}\right] -
\frac{1}{4}d_i \left(X_i^{\pm}H_i + H_i X_i^{\pm}\right)\right)\ .
\end{eqnarray*}

Here $H_i$ and $X_i^{\pm}$ are Lie algebra generators (i.e.
$I^a$'s), and $J(H_i)$ and $J(X_i^{\pm})$ are the corresponding
Yangian generators $J^a=J(I^a)$.

While the coalgebra structure of $Y(\g)$ in the second realisation
can, in principle, be determined by the isomorphism $\phi$ and the
first Yangian definition, no explicit formula for the action of the
co-multiplication on the generators $X_{ik}^{\pm},\ H_{ik}$ is
known.

For a more complete introduction to Yangians see \cite{CP1,MK}.

\subsection{The dilogarithm and related functions}
The {\bf dilogarithm} is the function defined by the
power series
$$\Li_2(z)=\sum_{n=1}^{\infty}\frac{z^n}{n^2}\ ,\qquad\quad
z\in\mathbb{C}\ ,\,\,|z|<1\ .$$ It has an analytic continuation to
$z \in \mathbb{C}-(1,\infty)$, given by
$$\Li_2(z)=-\int_0^z{\frac{\log(1-u)}{u}\,du}\ .$$ There are only eight
special values for which this function can be computed exactly.
These are
$$0,\,\pm1,\,\frac{1}{2},\,\frac{3-\sqrt{5}}{2},\,
\frac{-1+\sqrt{5}}{2},\,\frac{1-\sqrt{5}}{2},\,\frac{-1-\sqrt{5}}{2}\
.$$ Nevertheless it satisfies many functional equations, for example
\begin{eqnarray*}
\Li_2\left(\frac{1}{z}\right) &=& -\Li_2(z)-\frac{\pi^2}{6}-\frac{1}{2}\log^2(-z)\ ,\\
\Li_2(1-z) &=& -\Li_2(z)+\frac{\pi^2}{6}-\log(z)\log(1-z)\ .
\end{eqnarray*}
A more detailed discussion of the dilogarithm is found in \cite{L,Z}.

The {\bf Bloch-Wigner function} is closely related to the
dilogarithm, and is defined for all $z\in\mathbb{C}$ by
$$D(z)=\mathrm{Im}\left(\Li_2(z)\right)+\mathrm{arg}(1-z)\log|z|\ .$$
$D(z)$ is a continuous function, and is real analytic on
$\mathbb{C}/\{0,1\}$. All of the functional equations satisfied by
$\Li_2(z)$ lose the elementary correction terms when expressed in
terms of $D(z)$. This function is particularly useful when studying
torsion in the Bloch group.

The {\bf Rogers dilogarithm} is defined by
$$L(x)=-\frac{1}{2}\int_0^x{\left[\frac{\log|1-u|}{u}+\frac{\log|u|}{1-u}\right
]du}\ , \,\,\,\, x\in\mathbb{R}\ .$$ On the interval $(0,1)$ it is
related to the ordinary dilogarithm $\Li_2$ by
$$L(x)=\Li_2(x)+\frac{1}{2}\log(x)\log(1-x)\ ,$$
and outside this interval one can set $L(0)=0$, $L(1)=\pi^2/6$, and
\begin{equation}
L(x)=\begin{cases}
2L(1)-L(\frac{1}{x}) & \text{if $x > 1$\ ,}\\
-L\left(\frac{x}{x-1}\right) &\text{if $x <0$}\ .
\end{cases}
\end{equation}

This function has many intriguing properties. It appears in various branches
of mathematics, including number theory, algebraic K-theory, and the geometry of
hyperbolic 3-manifolds.

As well as the functional equations
\begin{equation}
L(x) + L(1-x) = L(1)\ , \quad \text{for all $x\in\mathbb{R}$}\ ,
\label{eq:rl1}
\end{equation}
and
\begin{equation}
L(x)+L\left(\frac{1}{x}\right)=\begin{cases} 2L(1) & \text{if $x>0$} \ ,\\
-L(1) & \text{if $x<0$} \ ,
\end{cases} \label{eq:rl2}
\end{equation}
$L(x)$ satisfies the 5-term relation
\begin{equation*}
L(x) + L(y) + L\left(\frac{1-x}{1-xy}\right) + L(1-xy) + L\left(
\frac{1-y}{1-xy}\right)
\end{equation*}
\begin{equation}=
\begin{cases}
-3L(1) & \text{if $x,\ y<0$,\ $xy>1$\ ,}\\
+3L(1) & \text{otherwise}\ .
\end{cases} \label{eq:rl3}
\end{equation}
(Notice that this equation is cyclically symmetric in its five
arguments, and that the right hand side is $-3L(1)$ when all five
arguments are negative). The function $L(x)$ is not continuous at
infinity, but is continuous if we consider it modulo $\pi^2/2$.

Define a Riemann surface $\hat{\mathbb{C}}$ by
$$\hat{\mathbb{C}}=\left\{(u,v)\in\mathbb{C}^2\,|\ e^u+e^v=1\right\}
\cup(\infty,0)\cup(0,\infty)\ .$$ The Rogers dilogarithm can be
extended to a holomorphic function
\begin{equation}
\hat{L}:\hat{\mathbb{C}}\rightarrow \mathbb{C}/(2\pi
i)^{2}\mathbb{Z}\ ,\label{eq:ed}
\end{equation}
given by $\hat{L}(u,v)=F(v)+\frac{uv}{2}\ ,$ where
$$F(v)=\Li_2(1-e^v)\in\mathbb{C}/(2\pi i)^2\mathbb{Z}\ ,$$
and $v\in\mathbb{C}-2\pi i\mathbb{Z}$. $F$ is well-defined modulo
$(2\pi i)^2\mathbb{Z}$ since $\Li_2(1-e^v)=\int_0^v
\frac{t\,dt}{e^t-1}$, where the integrand $\frac{t}{e^t-1}$ has a
pole at every $t\in 2\pi in$, with residue $2\pi in$.

The function $\hat{L}$ has the properties
\begin{eqnarray*}
\hat{L}(u+2\pi i,v) &=& \hat{L}(u,v)+\pi i v\ ,\\
\hat{L}(u,v+2\pi i) &=& \hat{L}(u,v)-\pi i u\ ,
\end{eqnarray*}
of which the first characterises $\hat{L}$ up to some additive
constant. Here $(u,v)\in\hat{\mathbb{C}}$. There is an embedding
$(0,1)\rightarrow\hat{\mathbb{C}}$ given by
$$x\mapsto\left(\log(x),\log(1-x)\right)=(u,v)\ .$$ Clearly $\hat{L}(u,v)=L(x)$
for $x\in(0,1)$.

\subsection{The Bloch group and related structures}
Let $\mathbb{F}$ be a field. Consider the free abelian group with
basis $[z],\ z\in\mathbb{F}^{\star}$. Let
$\mathcal{A}=\mathcal{A}(\mathbb{F})$ be the subgroup of elements
$\sum_{i=1}^n n_i[z_i]$, ($z_i\in\mathbb{F}^{\star},\
n_i\in\mathbb{Z}$), satisfying
$$\sum_{i=1}^n{n_i\ (z_i) \wedge (1-z_i)=0}\ ,$$ where $1\wedge 0$
is to be interpreted as $0$.

Here the sum is taken in the abelian group
$\Lambda^2\mathbb{F}^{\star}$ (the set of all formal linear
combinations of symbols $x \wedge y$, for $x,y \in
\mathbb{F}^{\star}$, subject to the relations $x \wedge x=0$ (and
hence $x\wedge y=-y\wedge x$) and $(x_1x_2)\wedge y=x_1 \wedge y +
x_2 \wedge y$).

For example $6.[2/3] -[8/9] \in \mathcal{A}(\mathbb{Q})$ since
\begin{eqnarray*}
&& 6.\left(2/3\right)\wedge\left(1-2/3\right) -
\left(8/9\right)\wedge \left(1-8/9\right)\\
&=& 6.\left(2/3\right)\wedge \left(1/3\right) -
\left(8/9\right)\wedge \left(1/9\right)\\
&=& 6.(2)\wedge \left(1/3\right) -8\wedge \left(1/9\right)\\
&=& -6.(2) \wedge (3) + (8) \wedge (9)\\
&=& -6.(2)\wedge (3) + 6.(2)\wedge (3)\\
&=& 0\ .
\end{eqnarray*}

For all $x,y \in \mathbb{F}^{\star}-\{1\}$ with $xy\neq1$,
$\mathcal{A}$ contains the elements
$$2\left(\left[x\right] + \left[\frac{1}{x}\right]+\left[1\right]\right)\ ,$$
$$\left[x\right] + \left[1-x\right]-\left[1\right]\ ,$$
$$\left[x\right] + \left[y\right] + \left[\frac{1-x}{1-xy}\right] +
\left[1-xy\right] + \left[\frac{1-y}{1-xy}\right]\ .$$ Let
$\mathcal{C}=\mathcal{C}(\mathbb{F})$ be the subgroup of
$\mathcal{A}$ generated by all such elements. Then the {\bf Bloch
group} of $\mathbb{F}$ is defined as
$$\mathcal{B}({\mathbb{F}})=\mathcal{A}/\mathcal{C}\ .$$

The Bloch group is closely related to the Rogers dilogarithm
described in the previous section. Namely the
equations~(\ref{eq:rl1}),~(\ref{eq:rl2}), and~(\ref{eq:rl3}) imply
that $L:\mathbb{R}\rightarrow \mathbb{R}$ can be extended to a
function $$L:\mathcal{B}(\mathbb{R})\rightarrow
\frac{\mathbb{R}}{3L(1)\mathbb{Z}}=\mathbb{R}\mod\frac{\pi^2}{2}\mathbb{Z}\
,$$ by setting $L\left(\sum n_i[x_i]\right)=\sum n_iL(x_i) \mod
L(1)\mathbb{Z}$.

The {\bf torsion subgroup} of the Bloch group is the subgroup
consisting of all elements of finite order. An element
$z\in\mathcal{B}[\bar{\mathbb{Q}}]$ is torsion if and only if its
Bloch-Wigner dilogarithm, $D(z)$, is zero in all complex embeddings
(in which case its Rogers dilogarithm, $L(z)$, is a rational
multiple of $\pi^2$ in all real embeddings.)

For example, consider the inverse golden ratio
$\alpha=\frac{\sqrt{5}-1}{2}$. Setting $x=y=\alpha$ in the 5-term
relation $\left[x\right] + \left[y\right] +
\left[\frac{1-x}{1-xy}\right] + \left[1-xy\right] +
\left[\frac{1-y}{1-xy}\right]$, it follows that the element
$\left[\alpha\right]\in\mathbb{Z}\left[\mathbb{Q}\left(\sqrt{5}\right)\right]$
is killed by 5 in the Bloch group. That the corresponding value of
the Rogers dilogarithm is
$L(\alpha)=\frac{\pi^2}{10}\notin\frac{\pi^2}{2}\mathbb{Z}$ proves
that $\alpha$ is indeed 5-torsion and not trivial in the Bloch group
of $\mathbb{Q}(\sqrt{5})$ (or even in the Bloch group of
$\mathbb{R}$).

It is interesting to note that the same element $\alpha$ becomes
zero in $\mathcal{B}(\mathbb{C})$, and similarly in
$\mathcal{B}(\mathbb{F})$ for the field
$\mathbb{F}=\mathbb{Q}(\zeta)$, where $\zeta$ is a $5^{th}$ root of
unity. In both of these cases $[\alpha]$ itself can be written as a
5-term relation.

One weakness of the Bloch group for $F=\mathbb{C}$, is that it does
not take into account the multi-valued nature of the dilog function.
To deal with this problem we introduce an extension
$\hat{\mathcal{B}}(\mathbb{C})$ of $\mathcal{B}(\mathbb{C})$, called
the {\bf extended Bloch group}.

Let $[\hat{\mathbb{C}}]$ be the free abelian group with basis
$[(u,v)]$ for $(u,v)\in\hat{\mathbb{C}}$. The extended Bloch group
$\hat{\mathcal{B}}(\mathbb{C})$ can be introduced as an extension of
$[\hat{\mathbb{C}}]$ as follows. There is a natural linear map
$\sigma: [\hat{\mathbb{C}}] \rightarrow \Lambda^2\mathbb{C}$ induced
by $\sigma(u,v)=u\wedge v$, with
$\sigma(0,\infty)=\sigma(\infty,0)=0$. Let $\mathcal{P}$ be the
kernel of this map. Define
$$\hat{\mathcal{B}}(\mathbb{C})=\mathcal{P}/\mathcal{P}_0\ ,$$
where $\mathcal{P}_0$ is the subgroup of $\mathcal{P}$ generated by
all elements of the form
\begin{eqnarray*}
&& (u,v) + (v,u) -(0,\infty)\ ,\\
&& (u-2\pi i, v) + 2(u-v-\pi i, -v) + (u,v)\ ,\\
&& \sum_{i=1}^5 (u_i,v_i) -2(0,\infty)\ ,
\end{eqnarray*}
where $u_i=v_{i-1}+v_{i+1}$ for $i=1,\ldots,5$, and $v_0=v_5$,
$v_1=v_6$ for cyclic symmetry.

There is a map $\hat{\mathcal{B}}(\mathbb{C})\rightarrow
\mathcal{B}(\mathbb{C})$ given by $\sum_i n_i(u_i,v_i)\mapsto\sum_i
n_ie^{u_i}$. For a more detailed explanation of the extended Bloch
group see~\cite{N,DZ}.

One significant difference between the extended Bloch group and the
ordinary Bloch group of the complex numbers is that the torsion
subgroup of the extended Bloch group is non-trivial, while that of
the ordinary Bloch group is trivial.

The torsion subgroup of $\hat{\mathcal{B}}(\mathbb{C})$ plays an
important role in quantum field theory. On elements of this
subgroup, the map $(2\pi i)^{2}L$ takes values in
$\mathbb{Q}/\mathbb{Z}$. These values yield the conformal dimensions
(more precisely the exponents $h_i-c/24$) of the fields of the
theory.

\section{Introduction to Conformal Field Theory}
Here we give a short account of some important aspects of conformal
field theory. The emphasis is on the ideas relevant to subsequent
discussions. For a more detailed introduction to the subject see
e.g. \cite{diF,G,MG,KV}.

A conformal field theory is a special type of quantum field theory.
Many quantum field theories are attained from quantisation of
classical field theories. Their key ingredients are a fixed
spacetime and an action, $S[\Phi]$, defined on a set of fields
$\Phi_i$, $i\in\Delta$, which are real or complex-valued functions
on the spacetime. When renormalisation problems are handled properly
one can calculate the vacuum expectation values of the corresponding
quantum field theory by
$$\langle \Phi_1(x_1)\ldots \Phi_M(x_M)\rangle
= \mathcal{N}^{-1}\int \prod[\mathcal{D}\Phi_{\Delta}]
\Phi_1(x_1)\ldots \Phi_M(x_M)e^{-S[\Phi_{\Delta}]}\ ,$$ where
$$\mathcal{N} =
\int\prod[\mathcal{D}\Phi_{\Delta}]e^{-S[\Phi_{\Delta}]}\ ,$$ is the
normalisation factor. Such a vacuum expectation value is called a
{\bf correlation function}. A theory is considered to be solved once
all of its correlation functions have been calculated. The vacuum
expectation values have certain properties, in particular
invariances and operator product expansions, which can be used for
an axiomatic description. There is no general method of doing this,
however in certain QFTs the existence of symmetries places
sufficient constraints on the correlation functions to allow them to
be calculated exactly. This approach is especially likely to be
successful in the case of conformal field theories, those particular
QFTs that are invariant under conformal transformations.

\subsection{Conformal invariance}
A {\bf conformal transformation} is a restricted general coordinate
transformation $\mb{x}\rightarrow \tilde{\mb{x}}$, for which the
metric $g_{\mu\nu}$ is invariant up to a scale factor
\begin{equation}
g_{\mu\nu}(\mb{x})\rightarrow
{\tilde{g}}_{\mu\nu}(\tilde{{\mb{x}}})=\Lambda(\mb{x})g_{\mu\nu}(\mb{x})\
;\quad\Lambda(x)\equiv e^{\omega(x)}\ , \label{eq:scale}
\end{equation}
where $\omega$ is some function of $x$. The set of all conformal
transformations forms the {\bf conformal group}. In two dimensions
the metric $g_{\mu\nu}$ is given by $dt^2+dx^2$ on the torus, or by
the standard metric on $S^2$.

Two-dimensional conformal field theories have an infinite number of
conserved quantities (corresponding to local conformal symmetry),
and are therefore completely solvable by symmetry considerations
alone. We now take a closer look at this important special case.

\subsubsection{Conformal group in two dimensions}
In two dimensions there is an infinite number of coordinate
transformations that, although not everywhere well-defined, are
locally conformal. These are the analytic maps from the complex
plane to itself. This set is known as the {\bf local conformal
group}, although strictly speaking it is not a group since the
mappings are not necessarily one-to-one and do not map the Riemann
sphere to itself. Hence the need to distinguish these local
transformations from global conformal transformations, which are
well-defined everywhere.

The set of all analytic maps contains the six-parameter {\bf global
conformal group}, which is the subset of mappings that are
invertible and defined everywhere. The group formed by these global
transformations is often called the {\bf special conformal group}.
It can be shown that the functions $f(z)=(az+b)/(cz+d)$, satisfying
$ad-bc=1$, are the only globally defined invertible analytic maps.
Hence we can write the special conformal group as the set
$$\left\{f(z)=\frac{az+b}{cz+d};\quad ad-bc=1,\quad
a,b,c,d\in\mathbb{C}\right\}\ .$$ It is easy to see that this group
can be parametrised by the set of complex $2\times 2$ matrices with
unit determinant, modulo the negative unit matrix, i.e.
$SL(2,\mathbb{C})/\mathbb{Z}_2$.

The distinction between local and global conformal groups is unique
to the two-dimensional case. In higher dimensions all local
conformal transformations are global. As already stated, local
properties of conformal invariance are of more immediate interest
than global ones, since it is the infinite-dimensionality of the
local conformal group that allows so much to be known about
conformally invariant field theories in two dimensions. With this in
mind we now proceed to find the algebra of generators of the local
conformal group. This is the {\bf Witt algebra}, also known as the
{\bf conformal algebra}.

\subsubsection{Conformal generators and the Witt algebra}
The infinitesimal analytic functions parametrising the conformal
transformations, can be defined by restricting the plane to a finite
region around the origin and assuming that all singularities of the
analytic functions are outside the region chosen.

A suitable basis for the infinitesimal coordinate transformations
$$z\rightarrow \tilde{z}=z+\epsilon(z),\quad
\bar{z}\rightarrow\tilde{\bar{z}}+\bar{\epsilon}(\bar{z})\ ,$$ is
generated by the operators
$$l_n=-z^{n+1}\frac{d}{dz},\ {\bar{l}}_n=-\bar{z}^{n+1}\frac{d}{d\bar{z}},\quad
n\in\mathbb{Z}\ .$$ The holomorphic generators $\{l_n\}$ form a Lie
algebra with commutation relations
$$\left[l_n,l_m\right] = (n-m)\,l_{n+m}\ .$$ This is the so-called
{\bf Witt algebra}. The corresponding anti-holomorphic generators
form an isomorphic Lie algebra with commutation relations
$$\left[\bar{l}_n,\bar{l}_m\right] = (n-m)\,\bar{l}_{n+m}\ .$$
These generators also satisfy the relation
$$\left[l_n,\bar{l}_m\right]=0\ .$$

Each of these infinite-dimensional algebras contains a finite
subalgebra generated by $\{l_{-1},\, l_0,\, l_1\}$. This is the
subalgebra associated with the global conformal group. In the
quantum case the Witt algebra will be corrected to include an extra
term proportional to the central charge (the so-called {\bf central
extension}). The unique central extension will be given by the {\bf
Virasoro algebra}.

\subsubsection{Primary and quasi-primary fields}
Representations of the global conformal algebra (after quantisation)
assign quantum numbers to physical states. We can assume the
existence of the {\bf vacuum state} $|\,0 \rangle$ among the
physical states; it has vanishing quantum number and is invariant
under the transformation $z\rightarrow\frac{az+b}{cz+d}$, where
$a,b,c,d\in\mathbb{C}$ and $ad-bc=1$. The eigenvalues $h$ and
$\bar{h}$ of $l_0$ and ${\bar{l}}_0$ respectively, are called the
{\bf conformal weights} of a state.

Given a state with conformal dimensions $h$ and $\bar{h}$, its
scaling dimension, $\Delta$, and planar spin, $s$, are defined by
$$\Delta=h+\bar{h}\quad\mathrm{and}\quad s=h-\bar{h}\ .$$

Any field $\Phi$ that satisfies the transformation property
\begin{equation}
\Phi(\tilde{z},\tilde{\bar{z}})=\Phi(z,\bar{z})\left(\frac{dz}{d\tilde{z}}\right)^
{h}\left(\frac{d\bar{z}}{d\tilde{\bar{z}}}\right)^{\bar{h}}\
,\label{eq:pf}
\end{equation} is called a {\bf primary field}. The remaining CFT fields are
called {\bf secondary fields}. As we will see later, the importance
of primary fields lies in their ability to generate the whole field
content of a theory.

The operator product expansion should be constant under the
equation~(\ref{eq:pf}). For global conformal transformations the
correlation functions are conserved.

\subsubsection{Operator product expansion, central charge, and conformal families}
The {\bf operator product expansion} (OPE) expresses a product of
two operator-valued fields, at different points $z$ and $w$, as an
infinite sum of single fields. In two-dimensional CFTs it is a
convergent expansion. Although written without brackets, it is
understood that the OPE is meaningful only within correlation
functions.

In general the OPE of a holomorphic field $A(z)$ with an arbitrary
field $B(w)$ can be written as
$$A(z)B(w)= \sum_i C_i(z-w)\mathcal{O}_i(w)\ ,$$
where $\left\{{\mathcal{O}}_i\right\}$ is a complete set of local
operators, and the $C_i$ are (singular) numerical coefficients, and
the OPE is understood to be meaningful only within an correlation
function. Here we have
$$\langle A(z_1)B(z_2)C(z_3)\ldots\rangle =
\sum_i f_i(z_1-z_2)\Phi_i(z_2) C(z_3)\ldots\ ,$$ with the functions
$f_i$ depending only on $A,B$, and $\Phi_i$ a basis of fields.

There exists a particular field, $T$, such that the expansion of $T$
with a primary field (of conformal dimensions $h,\bar{h}$) is given
by
$$
T(z)\Phi(w,\bar{w})=\frac{h}{(z-w)^2}\Phi(w,\bar{w})+\frac{1}{z-w}
\partial_w\Phi(w,\bar{w})+\Phi^{(-2)}(w,\bar{w})+(z-w)\Phi^{(-3)}(w,\bar{w})+\ldots\
,$$ where $\ldots$ represents an infinite set of regular terms
depending on the new local fields, called the {\bf descendant
fields}, of the primary field $\Phi$. $T$ is called the {\bf
energy-momentum tensor}.  The descendant fields are determined by
$$\Phi^{(-n)}(w,\bar{w})=L_{-n}\Phi(w,\bar{w})\equiv\oint_w
\frac{dz}{2\pi i}(z-w)^{-n+1}T(z)\Phi(w,\bar{w})\ .$$

The OPE of $T$ with non-primary fields contains correction terms.
For example, the general OPE of the energy-momentum tensor, $T$,
with itself is
$$T(z)T(w)\sim\frac{c/2}{(z-w)^4}+ \frac{2T(w)}{(z-w)^2} +
\frac{{\partial}_w T(w)}{z-w}+{\rm holomorphic\ terms}\ ,$$ where
${\partial}_w$ denotes differentiation with respect to $w$. $c$ is a
constant that depends on the specific model under consideration.
This constant is called the {\bf central charge}. Physically the
central charge describes how a specific system reacts to the
introduction of a macroscopic length scale.

We have introduced the operators $L_n$, which appear in the formal
expansion of the energy-momentum tensor $T(z)$ around a point $w$
$$T(z)=\sum_{n\in\,\mathbb{Z}}\frac{L_n}{(z-w)^{n+2}}\ .$$

In particular we have
\begin{eqnarray*}
L_0\Phi(z,\bar{z})&=&h\Phi(z,\bar{z})\ ,\\
L_{-1}\Phi(z,\bar{z}) &=& \partial_z \Phi(z,\bar{z})\ ,\\
L_n\Phi(z,\bar{z}) &=& 0,\quad n\geq1\ ,
\end{eqnarray*}
where $\Phi_{-n}=L_{-n}\Phi$ are the new descendant fields.

For each primary field $\Phi$, there exists an infinite {\bf
conformal family} $[\Phi]$ of descendant fields. These are generated
by the repeated use of the operators $L_{-n}$
$$[\Phi]:=\left\{L_{-k_1}\ldots L_{-k_n}\Phi:\
k_1\geq k_2\geq\ldots\geq k_n>0\right\}\ .$$ It can be shown that
every conformal family defines a highest-weight representation of
the Virasoro algebra.

\subsubsection{Virasoro algebra}
We have already seen that the classical generators of local conformal
transformations obey the Witt algebra. We now show that the corresponding
quantum generators obey a similar algebra with an added central extension term.
This is the well-known {\bf Virasoro algebra}.

We saw above that the energy-momentum tensor, $T$, has a Laurent
expansion in terms of modes $L_n$. These modes are themselves
operators, and their action on the operator $\Phi(w)$ can be written
as
$$
L_n\Phi(w)=\frac{1}{2\pi i}\oint_w dz \, (z-w)^{n+1}T(z)\Phi(w)\ .$$
The mode operators $L_n$, and their anti-holomorphic counterparts
$\bar{L}_n$, are the quantum generators of the local conformal
transformations on the Hilbert space. They obey the algebra
\begin{eqnarray*}
\left[L_n,L_m\right] &=& (n-m)L_{n+m}+\frac{c}{12}n(n^2-1)\delta_{n+m,0}\ ,\\
\left[L_n,\bar{L}_m\right] &=& 0\ ,\\
\left[\bar{L}_n,\bar{L}_m\right] &=&
(n-m)\bar{L}_{n+m}+\frac{c}{12}n(n^2-1)\delta_{n+m,0}\ .
\end{eqnarray*}
Each set of generators $\{L_n\}$ and $\{\bar{L}_n\}$ constitutes a copy of the
so-called {\bf Virasoro algebra}. It is worth noting that the central term is
absent for the subalgebra $\{L_{-1},L_0,L_1\}$ belonging to the global
conformal group.

{\it Proof of commutation relations:}
\begin{eqnarray*}
&& \left[L_n,L_m\right]\\ &=& \frac{1}{(2\pi i)^2} \oint_0
dw\,w^{m+1}\oint_w dz\,z^{n+1}\left\{ \frac{c/2}{(z-w)^4}
+\frac{2T(w)}{(z-w)^2} + \frac{\partial T(w)}{(z-w)} +
\mbox{ reg.} \right\}\\
&=&
\frac{1}{2\pi i}\oint_0 dw\,w^{m+1}\left\{ \frac{c}{12}(n+1)n(n-1)w^{n-2} +
2(n+1)w^nT(w) + w^{n+1}\partial T(w)\right\}\\
&=&
\frac{c}{12}n(n^2-1)\,\delta_{n+m,0}+2(n+1)L_{m+n}-\frac{1}{2\pi i}\oint_0
dw(n+m+2)w^{n+m+1}T(w)\\
&=& \frac{c}{12}n(n^2-1)\,\delta_{n+m,0}+(n-m)L_{m+n}\ .
\end{eqnarray*}

Every conformal field theory determines a representation of the Virasoro algebra
for some value of $c$. For $c=0$ there is no central extension term, and the
Virasoro algebra reduces to the classical Witt algebra.

\subsubsection{Verma modules}
To introduce the Hilbert space of states, we begin by defining the
{\bf vacuum state}, $|0\rangle$, of the theory by the condition
$$L_n|0\rangle=0\ , \quad \mbox{for all}\, n\geq0\ .$$
To each primary field $\Phi_{h,\bar{h}}$ we can associate a {\bf
highest-weight state} $|h,\bar{h}\rangle$ by
$$|\,h,\bar{h}\rangle={\mathrm{lim}}_{z,\bar{z}\rightarrow 0}
\Phi_{h,\bar{h}}(z,\bar{z})|\,0\rangle\ .$$ It follows that
\begin{eqnarray*}
L_n|\,h,\bar{h}\rangle &=& 0\quad n>0\ ,\\
L_0|\,h,\bar{h}\rangle &=& h |\,h,\bar{h}\rangle\ ,\\
\bar{L_0}|\,h,\bar{h}\rangle &=& \bar{h} |\,h,\bar{h}\rangle\ .
\end{eqnarray*}

The states of the associated {\bf Verma module} are created by
acting on the primary state $|\,h,\bar{h}\rangle$ with arbitraty
polynomials in
$$\left\{L_{-n},\,{\bar{L}}_{-m}:\ m,n\geq1\right\}\ ,$$ and no relations
between these states except those given by the Virasoro algebra. The
Verma module is an infinite-dimensional representation of the
Virasoro algebra, completely characterised by its central charge and
the dimension of the highest-weight state. Physical representations
arise from Verma modules by reducing them modulo maximal submodules.
Their descendant states can be viewed as the result of the action of
descendant field on the vacuum
$$L_{-n}|\,h\rangle = \Phi^{(-n)}(0)|\,0\rangle\ .$$

Note that the descendant state $L_{-k_1}\ldots L_{-k_n}|\,h\rangle$
is itself an eigenvector of $L_0$ through $L_0L_{-k_1}\ldots
L_{-k_n}|\,h\rangle=(h+l)L_{-k_1}\ldots L_{-k_n}|\,h\rangle$. The
integer $\ l=\sum_{i=1}^n k_i\ \ (k_i>0)\ $ is called the {\bf
level} of the state.

The CFT vacuum is a trivial highest-weight state which defines the
trivial module corresponding to the identity operator. The Virasoro
operators $L_n$ in a Verma module act like raising and lowering
operators. Since $[L_0,L_{-n}]=nL_{-n}$, $L_0$ can be viewed as a
grading operator measuring the conformal dimension of a state.

All of the above can equally be applied to the anti-holomorphic
counterparts.

\subsubsection{Null states and the Kac determinant}
A descendant state $|\,v\rangle$ satisfying the equations
$$L_0|\,v\rangle=(h+N)|\,v\rangle,\quad
L_n|\,v\rangle=0\quad \mathrm{for }\  n>0\ ,$$ is called a {\bf null
state}. It is simultaneously a primary and a descendant state, and
is also a highest-weight state. To get an irreducible representation
of the Virasoro algebra we must eliminate all null states and their
descendant states, and consider the reduced theory.

The scalar product of two states at level $l$ is given by
$$\langle h\,|(L_{r_k}\ldots L_{r_1}) (L_{-s_1}\ldots L_{-s_t})
|\,h\rangle \equiv M_{\{r\}\{s\}}^{(l)}\ ,$$ where $\sum r_i=\sum
s_i=l$. $M$ is a block diagonal matrix, with blocks $M^{(l)}$
corresponding to states of level $l$. $M$ is called the {\bf Gram
matrix}. The matrices $M$ related to the lowest levels of a generic
Verma module can easily be calculated. For example, the states of
level $2$ are $L_{-1}^2|h\rangle$ and $L_{-2}|h\rangle$. Therefore
\begin{eqnarray*}
M_{12}^{(2)} &=&
\langle h|L_1L_1L_{-2}|h\rangle\\
&=& \langle h|L_1(L_{-2}L_1+3L_{-1})|h\rangle\\
&=& 3\langle h|L_1L_{-1}|h\rangle\\
&=& 6h \langle h|h \rangle\ .
\end{eqnarray*}
Similar calculations give the other three entries, resulting in the
matrix
$$M^{(2)}= \left(\begin{array}{cc} 4h(2h+1)&6h\\6h&4h+c/2
\end{array}\right)\ .$$

The determinant of this matrix is known as the {\bf Kac
determinant}, and null states in the Verma module correspond to
zeros of the Kac determinant. In the above example we have
$$\det M^{(2)}=32h^3+(4c-20)h^2+2ch\ .$$ Writing
$$\det M=32(h-h_{11})(h-h_{12})(h-h_{21})\ ,$$ we can see that the
roots of the Kac determinant are given by
\begin{eqnarray*}
h_{1,1} &=& 0\ ,\\
h_{1,2} &=& \frac{1}{16}\left(5-c-\sqrt{(1-c)(25-c)}\right)\ ,\\
h_{2,1} &=& \frac{1}{16}\left(5-c+\sqrt{(1-c)(25-c)}\right)\ .
\end{eqnarray*}

There exists a general formula for calculating the Kac determinant.
It is given by
$$\det M^{(l)}(c,h)=\prod_{k=1}^l\prod_{rs=k}\left[ h-h_{r,s} \right] ^{p(l-k)}\ ,$$
where $r$ and $s$ are positive integers, and $p\,(l-k)$ denotes the
number of partitions of the integer $l-k$.

There are many ways to express the roots of the Kac determinant. One
way is to write
\begin{equation}
h_{r,s}(m) =\frac{\left[(m+1)r-ms\right]^2-1}{4m(m+1)}\ ,
\label{eq:hrs}
\end{equation} where
$$m=-\frac{1}{2}\pm\frac{1}{2}\sqrt{\frac{25-c}{1-c}}\ .$$

\subsection{Models in conformal field theory}
\subsubsection{Minimal models}
CFTs that have a finite number of primary fields are called rational
conformal field theories. The minimal models are particular rational
CFTs of central charge $c<1$. They are called minimal since they are
based on a finite number of scalar primary fields, they have no
multiplicities in their spectra of conformal dimensions, and they
contain no additional symmetries except for conformal symmetry.
These models are characterised by
\begin{eqnarray*}
c &=& 1-6\frac{(p-p')^2}{pp'}\ ,\\
&\vspace{5mm}\\
h_{r,s} &=& \frac{(pr-p's)^2-(p-p')^2}{4pp'}\ ,
\end{eqnarray*}
where $p$ and $p'$ are positive integers having no non-trivial
common divisors. Notice that setting $m=p/p'$ in~(\ref{eq:hrs}),
with $\mathrm{gcd}(p,p')=1$, is equivalent to the above expressions
for $c$ and $h_{r,s}$. We can restrict values of $r,s$ to the
rectangle $0<r<p'$,\ $0<s<p$. This rectangle in the $(r,s)$ plane is
called the {\bf Kac table}. The symmetry $h_{r,s}=h_{p'-r,p-s}$
makes half of this table redundant. Minimal models usually describe
discrete statistical models at their critical points, and their
simplicity allows in principle for a complete solution.

\subsubsection{WZW models} A {\bf Wess-Zumino-Witten model} is a
simple model of conformal field theory whose solutions are realised
by affine Kac-Moody algebras. It has holomorphic fields with $h=1$,
the so-called currents $J^a$. Using the $J^a_m$ instead of $L_m$ one
can repeat most of the preceeding discussion. Given a Kac-Moody
algebra $\hat{g}_k$, of level $k$, the central charge of the
corresponding WZW model is given by
\begin{equation}
c(\hat{\g}_k)=\frac{k\,{\rm dim}(\g)}{k+h(\g)}\ . \label{eq:wzwc}
\end{equation}
Every WZW model has central charge $c>1$.

\subsubsection{Coset models} We now introduce a third class of models
by means of the {\bf coset construction}. This greatly increases the
number of known solvable models. A {\bf coset model} is a quotient
of two WZW models, with the central charge of the coset being the
difference of the central charges of the two WZW components. It is
expected that the coset construction will provide a framework for
the complete classification of all rational conformal field
theories. Coset models incorporate the two classes of models that we
have already looked at
\begin{itemize}
\item{WZW models are represented by trivial cosets,}
\item{Models with $c<1$ can be represented by the coset
construction, since the central charge of a coset is the difference
of the central charges of the two WZW components. However all RCFTs
with $c<1$ are known to be minimal models. Hence any coset with
$c<1$ must provide a new representation of a minimal model.}
\end{itemize}




\subsubsection{Characters}
Let $c,h\in\mathbb{Q}$ denote central charge and conformal dimension
respectively. Let $V_{c,h}$ denote the Verma module, (before
excluding the null submodules), generated by the Virasoro generators
$L_{-n}\ (n>0)$ acting on the highest-weight state
$|\,h,\bar{h}\rangle$. To each such Verma module we can associate a
generating function $\chi_{c,h}(\tau)$, called the character of the
module. This is defined by
\begin{eqnarray*}
\chi_{c,h}(\tau)&=& \Tr\,q^{L_0-c/24}\qquad (q\equiv e^{2\pi i\tau})\\
&=&\sum_{n=0}^{\infty}{\rm d}(h+n)\,q^{n+h-c/24}\ .
\end{eqnarray*}
Here ${\rm d}(n+h)$ denotes the number of linearly independent
states at level $n$ in the module and $\tau$ is a complex variable.
Characters can be viewed as generating functions for the number of
states at any given level.

Now let $V_{r,s}$ denote the Verma module
$V(c(p,p'),h_{r,s}(p,p'))$, built on the highest weight $h_{r,s}$
appearing in the Kac table. This reducible Verma module, with
highest weight $h_{r,s}$, contains null states that must be
eliminated in order to get the corresponding irreducible Verma
module, denoted $M_{r,s}$.

Define a new function by
$$K_{r,s}^{(p,p')}(q)=\frac{q^{-1/24}}{\phi(q)} \sum_{n\in\mathbb{Z}}
q^{(2pp'n+pr-p's)^2/4pp'}\ .$$ Then the character of the irreducible
Verma module $M_{r,s}$ is given by
$$\chi_{r,s}(q)=K_{r,s}^{(p,p')}(q)-K_{r,-s}^{(p,p')}(q)\ .$$

\subsection{Partition functions and modular invariance} We have so
far assumed conformal field theories to be defined on the whole
complex plane. Physically this is not a very realistic situation as
the holomorphic and antiholomorphic parts of the theory decouple
completely. To impose more realistic physical constraints on a
conformal field theory, we look to couple the holomorphic and
antiholomorphic sectors through the geometry of the space on which
the theory is defined. For this purpose we consider conformal field
theory on a torus. The interaction of the holomorphic and
antiholomorphic sectors in this case is given by {\bf modular
transformations}.

\subsubsection{Conformal field theory on the torus}
Define a torus on the complex plane by specifying two linearly
independent lattice vectors. These vectors are represented by
complex numbers $\omega_1$ and $\omega_2$ respectively, called the
{\bf periods} of the lattice. Identify points that differ by an
integer combination of these vectors. The properties of the
conformal field theory defined on the torus do not depend on the
overall scale of the lattice or on the absolute orientation of the
lattice vectors. The relevant scale parameter is
$\tau=\omega_2/\omega_1$, called the {\bf modular parameter}.

\subsubsection{The partition function}
The {\bf partition function} $Z$ of a CFT can be expressed in terms
of the Virasoro characters (of the Verma modules forming the Hilbert
space of the theory) as
$$Z=\sum_{h,\bar{h}}n_{h,\bar{h}}\chi_h(\tau)\chi_{\bar{h}}(\bar{\tau})\ ,$$
where $h$ and $\bar{h}$ label a certain highest-weight state
$|h,\bar{h}\rangle$, and $n_{h,\bar{h}}$ is the multiplicity of such
a state. For rational conformal field theories this sum\footnote{For
simplicity we will usually write this sum as
$Z=\sum_{i,j}n_{ij}\chi_i(\tau)\chi_j(\bar{\tau})$\ .} is always
finite.

\subsubsection{Modular invariance}
The partition function must be independent of the particular periods
$\omega_{1,2}$ chosen for a given torus. This has the advantage of imposing
certain constraints on the conformal field theory defined on the torus.

Suppose $\omega'_{1,2}$ are two periods describing the same lattice
as $\omega_{1,2}$. Since they belong to the same lattice, the points
$\omega'_1$ and $\omega'_2$ must be integer combinations of
$\omega_1$ and $\omega_2$, and vice versa. Moreover, the unit cell
of the lattice should have the same area no matter what periods are
chosen. Writing
$$\left(\begin{array}{c}
\omega'_1\\
\omega'_2\\
\end{array}\right)=
\left(\begin{array}{cc}
a&b\\
c&d\\
\end{array}\right)
\left(\begin{array}{c}
\omega_1\\
\omega_2\\
\end{array}\right)\ ,$$
the conditions above restrict the choice of entries to
$a,b,c,d\in\mathbb{Z}$ with $ad-bc=1$. This directs us to consider
the group of matrices $SL(2,\mathbb{Z})$.

Under the change of period above, the modular parameter transforms as
$$\tau\rightarrow \frac{a\tau+b}{c\tau+d}\ ,\quad ad-bc=1\ .$$
Clearly the signs of all parameters $a,b,c,d$ can be simultaneously
changed without affecting the overall transformation. Hence the
symmetry of interest here is the {\bf modular group}
$SL(2,\mathbb{Z})/\mathbb{Z}_2$. It can be shown that the modular
transformations $T:\tau\rightarrow\tau+1$ and
$S:\tau\rightarrow-\frac{1}{\tau}$ generate the whole of the modular
group.

Conformal invariance requires the partition function $Z$ to be
invariant under the modular group. This places some restrictions on
the characters $\chi_i$; in particular, the space generated by the
characters must be invariant under the modular transformation
$\tau\rightarrow -1/\tau$.

\subsubsection{The connection to algebraic K-theory}
For certain conformal field theories with integrable perturbations,
their characters $\chi_i$ can be described combinatorially as
\begin{equation}
\chi_i(\tau)=\sum_m \frac{q^{Q_i(m)}}{(q)_{m_1}\ldots(q)_{m_r}}\ ,
\label{eq:akt}
\end{equation}
where $Q_i({\mathbf m})=\frac{{\mathbf m}A{\mathbf
m}}{2}+b_i{\mathbf m}+h_i-\frac{c}{24},\ \
(q)_n=\prod_{i=1}^n(1-q^i),\ \  {\mathbf m}=(m_1,\ldots,m_r)$, and
$r$ is the rank of the matrix $A$. Here the matrix $A$ is the same
for all characters $\chi_i$ of a given CFT.

It is expected~\cite{N} that a general sum of the
form~(\ref{eq:akt}), with rational coefficients, can only be modular
when all solutions of the equations $$\sum_j
A_{ij}\log(x_j)=\log(1-x_i)\ ,$$ give finite order elements
$\left(\log(x_i), \log(1-x_i)\right)$ of the extended Bloch group.
That the elements $\sum_i[x_i]$ belong to the Bloch group at all is
shown at the beginning of Chapter 3.

There are many matrices $A$ for which $\sum_j
A_{ij}\log(x_j)=\log(1-x_i)$ yield torsion elements of the extended
Bloch group. The best known examples are related to Dynkin diagrams.
Given a pair of Dynkin diagrams $(X,Y)$, the matrix $A$ is given by
$A(X,Y)=C(X)\otimes C(Y)^{-1}$. In chapter 3 we will consider the
case $(X,Y)=(D_m,A_n)$.

It is expected that whenever $\sum_j A_{ij}\log(x_j)=\log(1-x_i)$
yields finite order elements of the Bloch group, the resulting $x_i$
should be rational linear combinations of roots of unity, possibly
apart from some $\mathbb{Z}_2$ extension. This has turned out to be
true for all examples considered so far. Moreover there is a map
from finite order elements of the Bloch group to the central charges
and scaling dimensions of conformal field theories. This mapping is
given by the dilogarithm function and will be described in chapter
3.

This suggests a clear, and very interesting, relationship between
certain integrable quantum field theories in two dimensions and the
algebraic K-theory of the complex numbers.

\chapter{Yangians and the $(D_m,A_n)$ Models} It is clear from the previous chapter that a study of integrable
models described by pairs of Dynkin diagrams should yield
interesting results. In this chapter we consider the particular
models described by the pairs $(D_m,A_n)$. Using the representation
theory of Yangians we solve the equations of these models in the
general case. We demonstrate how to calculate the effective central
charge using the dilogarithm formula, and finally we relate these
models to the coset models described in chapter 2.

\section{Overview}
\subsubsection{Equations of the model $(X,Y)$}
Consider the integrable model described by the pair of Dynkin
diagrams $(X,Y)$. The equations of this model are of the form
$AU=V$, where the matrix $A$ is given by
\begin{equation}
A(X,Y)=C(X)^{-1}\otimes C(Y)\ ,\label{eq:A}
\end{equation}
and $X$ and $Y$ are Dynkin diagrams of ranks $m$ and $n$
respectively. Here $U=\log(x)$ and $V=\log(1-x)$, such that the
equation $e^U+e^V=1$ is satisfied. $x$ denotes the vector
$(x_{11},\ldots,x_{mn})$. The matrix $A$, defined in~(\ref{eq:A}),
is positive definite and symmetric.

Exponentiation of $AU=V$ leads to a set of purely algebraic
equations.
\begin{eqnarray}
AU=V &\equiv& \sum_j A_{ij}\log(x_j)=\log(1-x_i) \label{eq:log}\\
&\Rightarrow& \prod_j x_j^{A_{ij}}=1-x_i\ . \label{eq:nolog}
\end{eqnarray}
Notice that this exponentiation transforms a set of equations with
infinitely many solutions~(\ref{eq:log}) into a set with a finite
number of solutions~(\ref{eq:nolog}). For simplicity we choose to
solve the equations in the form~(\ref{eq:nolog}); however, care must
be taken to choose logarithms in such a way that the original
equations~(\ref{eq:log}) of the model are satisfied.

For any solution $(x_{11},\ldots,x_{mn})$ of~(\ref{eq:nolog}), the
element $[x_{11}]+\ldots+[x_{mn}]$ belongs to the Bloch group:
\begin{eqnarray*}
\sum_i x_i\wedge(1-x_i) &=& \sum_i x_i\wedge \left(\prod_j
x_j^{A_{ij}} \right)\ ,\\
&=& \sum_i\sum_j x_i\wedge A_{ij}x_j\ ,\\
&=& \sum_i \sum_j A_{ij}x_i\wedge x_j\ ,\\
&=& 0\ ,
\end{eqnarray*}
since $A_{ij}$ is symmetric and $x_i\wedge x_j$ is anti-symmetric in
$i$ and $j$.

In some cases the algebraic equations~(\ref{eq:nolog}) can be solved
using nothing more than elementary algebra. However as the matrix
$A$ grows in size this becomes more difficult. By a suitable change
of variables, the equations~(\ref{eq:nolog}) can be written in a
form that allows them to be solved relatively easily using the
representation theory of Lie algebras and related quantum groups.

For this purpose we introduce the new variable $z=(z_{11},\ldots,
z_{mn})$, where
\begin{equation}
x=z^{-C(X)\otimes\,I_Y}\ . \label{eq:xz}
\end{equation}
The algebraic equations~(\ref{eq:nolog}) can be rewritten in terms of
$z$.
\begin{eqnarray}
x^A=1-x &\Rightarrow& x^{C(X)^{-1}\otimes\,C(Y)}=1-x \nonumber\\
&\Rightarrow& z^{-I_X\otimes\,C(Y)}=1-z^{-C(X)\otimes\,I_Y} \nonumber\\
&\Rightarrow& z^{2-C(Y)}+z^{2-C(X)}=z^2\ . \label{eq:zeqn}
\end{eqnarray}
Equation ~(\ref{eq:zeqn}) has many solutions for which some
components of $z$ vanish. These are called non-admissable since they
do not yield solutions of~(\ref{eq:log}). We discard these solutions
immediately.

We impose the boundary condition $z_{i,n+1}=1$ on the
equations~(\ref{eq:zeqn}), because with these boundary conditions
the solutions $z_{ij}$ arise naturally in representation theory, see
\cite{N} and references therein. (After imposing $z_{i,n+1}=1$,
equations~(\ref{eq:zeqn}) are exactly the equations discussed by
Kirillov and Reshetikhin~\cite{KR}, for the Lie algebra $X$, whose
solutions arise as characters of the Yangian $Y(X)$). In particular,
for the model $(D_m,A_n)$, the components $z_{ij}$ of $z$ are the
characters $Q_j^i$ of the Yangian $Y(D_m)$, that satisfy
$Q_{n+1}^i=1$ for $i=1,2,\ldots,m$. Hence a solution
of~(\ref{eq:zeqn}) amounts to finding a matrix $g\in SO(2m)$, whose
Yangian characters satisfy $Q_{n+1}^i(g)=1$ for $i=1,2,\ldots,m$.
Once such a matrix has been found, the equations~(\ref{eq:zeqn}) can
be solved using the relation
$$z_{ij}=Q_j^i(g)\ .$$
These $z_{ij}$ can easily be transformed into
solutions $x_{ij}$ of~(\ref{eq:nolog}) using equation~(\ref{eq:xz}).

\subsubsection{Effective Central Charge Calculations}
Let $x^i=(x_{11}^i,\ldots,x_{mn}^i)$ denote a solution of the system
of equations~(\ref{eq:nolog}). Here $i$ is an index to distinguish
between solutions, so $0\leq i\leq I$, where $I$ is the number of
solutions.

If $A$ is any positive definite matrix then the system of
equations~(\ref{eq:nolog}) has a unique solution with all $x_{ij}$
real and between 0 and 1. This fact is proved in~\cite{DZ}. Denote
this solution by $x^0=(x_{11}^0,\ldots,x_{mn}^0)$. Then
$0<x_{jk}^0\in\mathbb{R}<1$ for all $j$ and $k$. For future
reference we refer to $x^0$ as the `minimal solution'.

We are interested in the values taken by the solutions $x^i$ under the
mapping
\begin{equation}
\frac{6}{\pi^2}\sum_{jk=1,\ldots,mn}L(u_{jk}^i,v_{jk}^i)=c-24h_i\mod 24\mathbb{Z}\ .\label{eq:chi}
\end{equation}
Here $u_{jk}^i=\log(x_{jk}^i)$ and $v_{jk}^i=\log(1-x_{jk}^i)$ form
solutions of the equations $AU=V$, provided logarithms of the
complex numbers are chosen appropriately. Here $L$ is the analytic
continuation~(\ref{eq:ed}) of the Rogers dilogarithm, so that for
the real solution $x^0$ one has simply
$$L(u_{jk}^0,v_{jk}^0)=L(x_{jk}^0)\ .$$

The {\bf effective central charge} is defined as the particular
value of $c-24h_i$ that arises from the minimal solution $x^0$. In this case we have
\begin{equation}
c_{\eff}=\frac{6}{\pi^2}\sum_{jk=1,\ldots,mn}L(u_{jk}^0,v_{jk}^0)\
.\label{eq:ch0}
\end{equation}

The factor of $24\mathbb{Z}$ in~(\ref{eq:chi}) essentially arises
because the dilogarithm function is multi-valued outside its region
of convergence $|z|<1$. The only value of $c-24h_i$ that can be
calculated exactly is $c_{\eff}$, since it corresponds to the
minimal solution (all of whose components are real and between $0$
and $1$). Although it will not always be mentioned, this mod
$24\mathbb{Z}$ term of course applies throughout the thesis.

\subsubsection{The matrix $A$}
The matrix $A$ defined in~(\ref{eq:A}) is in fact related to
scattering matrices as follows. Suppose we take a system containing
$r$ different species of particles. Consider the scattering of two
particles of types $i$ and $j$, and rapidities $\theta_i$ and
$\theta_j$ respectively. For the type of system of interest to us,
the particles merely pass through each other with some time delay
(i.e. there is no exchange of particle quantum number). This time
delay is described by the scattering matrix, with the scattering
being described by an energy-dependent phase
$$S_{ij}=e^{if_{ij}(\theta_j-\theta_i)}\ .$$
The scattering matrix takes particular values at $\pm\infty$. In
terms of these values $A$ is defined as
$$A_{ij}=\frac{f_{ij}(-\infty)-f_{ij}(+\infty)}{2\pi}\ .$$

\section{Quantum Groups and Yangians}
Quantum groups were first introduced by Fadeev and collaborators. They arose
from the quantum inverse scattering method \cite{F}, developed to construct and
solve quantum integrable systems. In their original form quantum groups
are associative algebras whose defining relations are expressed in terms of
a matrix of constants called a quantum R-matrix. This matrix depends on
the particular integrable system under consideration. Quantum groups facilitate
the understanding of solutions (R-matrices) of the quantum Yang-Baxter equation
associated with such integrable systems. Furthermore they provide a general
framework for finding new solutions. Of special importance are those solutions
that depend on a spectral parameter. In particular those which are rational
functions of this parameter arise from the family of quantum groups called
Yangians. More recently quantum groups have arisen in connection with 1+1
dimensional integrable quantum field theories, as the algebras satisfied by
certain non-local conserved currents. For example, Yangians appear as `quantum
symmetry algebras' in G-invariant Wess-Zumino-Witten models \cite{B}.

The term Yangian was introduced by V.G. Drinfeld \cite{D1} to
specify those quantum groups related to rational solutions of the
quantum Yang-Baxter equation. In fact Yangians are named after C.N.
Yang who found the simplest such solution \cite{Y}. It is worth
noting that L\"{u}scher \cite{ML} effectively found much of the
Yangian $Y(\s\of_n)$ well in advance of the general construction.

Although quantum groups first appeared in the physics literature and
many of the fundamental papers are written in the language of integrable
systems, their properties are still accessible through more mainstream
mathematical techniques. There are many unexpected connections between quantum
groups and other seemingly unrelated areas of mathematics (for example knot
theory and the representation theory of algebraic groups in characteristic p).
In recent years these connections have sparked considerable interest in quantum
groups.

\subsection{Representation theory of Yangians}
As mentioned above, the importance of Yangians stems from the fact
that their finite-dimensional representations can be used to
construct rational solutions of the quantum Yang-Baxter equation.
The problem of describing all finite-dimensional irreducible
representations of $Y(\g)$ was solved by Drinfeld himself. He gave a
classification of such representations, similar to that for the Lie
algebra $\g$ in terms of highest weights, but without giving an
`explicit' realisation of these representations. Such a realisation
was given for $\g=\s\lf_2$ by V. Chari and A. Pressley
in~\cite{CP2}.

An alternative approach to obtaining a better understanding of such
representations would be to find a Yangian character formula,
analogous to the Weyl character formula for Lie algebras.
In~\cite{CP3, CP4} Chari and Pressley give examples of such a
formula for $Y(\s\lf_2)$. Unfortunately their proof does not extend
to other cases, and at present no general formula is known.

\subsubsection{Irreducible Yangian representations}
Suppose $\g$ is a Lie algebra of rank $r$. Then $\g$ has $r$
fundamental weights, denoted $\omega_1,\ldots,\omega_r$, one
corresponding to each node on its Dynkin diagram. The {\bf fundamental
representations} of $\g$ are the $r$ irreducible representations of highest weights
$\omega_i$ ($i=1,2,\ldots,r)$. Similarly $Y(\g)$ has $r$ fundamental
(finite-dimensional) irreducible representations.

Since $\g\subset Y(\g)$, any representation of $Y(\g)$ is
automatically a representation of $\g$. However, a representation
which is $Y(\g)$-irreducible may become reducible when restricted to
$\g$. In fact this is typically the case for the fundamental
representations of $Y(\g)$ (whose $\g$-components are the
corresponding fundamental irreducible representations of $\g$ + some
other representations).

Nevertheless, in some cases an irreducible $Y(\g)$-representation
remains $\g$-irreducible. In the simplest situation, given an
irreducible representation $\rho$ of $\g$, in certain cases we can
construct a representation $\tilde{\rho}$ of $Y(\g)$ by
\begin{equation}\tilde{\rho}(I_a)=\rho(I_a)\, ,\quad
\tilde{\rho}(J_a)=0\ . \label{eq:irrep}\end{equation} These cases
are described as follows.

Let $a_i$ be the coefficient of the simple root $\alpha_i$ in the
expansion of the highest root $\theta$ of $\g$. Put
$k_i=(\theta,\theta)/(\alpha_i,\alpha_i)$, and let $\omega_i$ be the
corresponding fundamental weight of $\g$. Then the irreducible
representation of $\g$, of highest weight $\lambda$, can be extended
to an irreducible representation of $Y(\g)$ using~(\ref{eq:irrep})
in the following cases:
\begin{enumerate}
\item{$\lambda=\omega_i$, when $a_i=k_i$\ ,}
\item{$\lambda=t\omega_i$\  ($t\in\mathbb{N}$), when $a_i=1$\ .}
\end{enumerate}
Included in these cases are all fundamental representations of $A_n$ and $C_n$, and the
vector and (half)-spinor representations of $B_n$ and $D_n$.

The more general case, in which an irreducible $Y(\g)$-representation is $\g$-reducible is
significantly more complicated. For more details see \cite{MK}.

\section{The Lie Algebra $D_r$}
\subsection{Some properties of $D_r$}
The Lie algebra $D_r$ has $r$ simple roots
$\alpha_1,\ldots,\alpha_r$ given by
\begin{eqnarray*}
\alpha_1 &=& e_1-e_2\ ,\\
\alpha_2 &=& e_2-e_3\ ,\\
&\vdots& \\
\alpha_{r-1} &=& e_{r-1}-e_{r}\ ,\\
\alpha_r &=& e_{r-1}+e_r\ ,
\end{eqnarray*}
and $r$ fundamental weights $\omega_1,\ldots,\omega_r$ given by
\begin{eqnarray}
\omega_1 &=& e_1\ ,\nonumber\\
\omega_2 &=& e_1+e_2\ ,\nonumber\\
&\vdots& \nonumber\\
\omega_{r-2} &=& e_1+\ldots+e_{r-2}\ ,\label{eq:omega}\\
\omega_{r-1} &=& \frac{1}{2}(e_1+e_2+\ldots+e_{r-1}-e_r)\ ,\nonumber\\
\omega_{r} &=& \frac{1}{2}(e_1+e_2+\ldots+e_{r-1}+e_r)\ .\nonumber
\end{eqnarray}
The Weyl vector $\rho$ is given by
$$\rho = (r-1)e_1+(r-2)e_2+\ldots+2e_{r-2}+e_{r-1}\ .$$

The quantity $e_i$ is defined by its action on a diagonal $n\times n$ matrix as
$$e_i\left(\diag(a_1,\ldots,a_n)\right)=a_i\ .$$

The Weyl group of $D_r$ is denoted $W(D_r)$. It is an extension of $S_r$ by
$\left(\mathbb{Z}_2\right)^{r-1}$. Its elements act on $(e_1,\ldots,e_r)$ as
$$(e_1,\ldots,e_r)\rightarrow(\epsilon_1 e_{s(1)},\ldots,\epsilon_r e_{s(r)})\ ,$$
where $s\in S_r$, $\epsilon_i\in\{\pm1\}$, and
$\prod_{i=1}^r\epsilon_i=1$. The order of $W(D_r)$ is $r!\,2^{r-1}$.

\subsubsection{Representations of $Y(D_r)$}
In this thesis we follow the notation of Kirillov and Reshetikhin
\cite{KR}. Again $\g$ is a simple Lie algebra of rank $r$, with fundamental weights
$\omega_1,\ldots,\omega_r$. $Y(\g)$ is the corresponding Yangian.

Let $W_i^j$ denote the irreducible representation of $Y(\g)$, of highest
weight $j\omega_i$. Here $j\in\mathbb{N}$ and $i=1,2,\ldots,r$. As
mentioned in the previous section, $W_i^j$ may become reducible when restricted to the
Lie algebra $\g$.

For $\g=D_r$, the representations $\left.W_i^j\right|_\g$ are
irreducible in the cases
\begin{eqnarray*}
(i,j)&=& (1,j)\ ,\\
(i,j)&=& (r-1,j)\ ,\\
(i,j)&=& (r,j)\ .
\end{eqnarray*}
These correspond to the vector and half-spinor representations.

Define polynomials $Q_j^i$ by
$$Q_j^i=\ch\left(\left.W_i^j\right|_\g\right)\ ,$$ where $\ch$ denotes the
character of a representation.

It is claimed in \cite{KR} that for $\g=D_r$, the functions
$Q_j^i$ form the unique solution of the system of recurrence
relations
\begin{eqnarray}
(Q_j^i)^2-Q_{j-1}^i Q_{j+1}^i &=& Q_j^{i-1}
Q_j^{i+1}, \quad\quad 1 \leq i \leq r-3\ , \nonumber\\
(Q_j^{r-2})^2-Q_{j-1}^{r-2} Q_{j+1}^{r-2}
&=& Q_j^{r-3} Q_j^{r-1} Q_j^r\ ,\nonumber\\
(Q_j^{r-1})^2-Q_{j-1}^{r-1} Q_{j+1}^{r-1} &=& Q_j^{r-2}\ ,\label{eq:kr}\\
(Q_j^r)^2-Q_{j-1}^r Q_{j+1}^r &=& Q_j^{r-2}\ ,\nonumber
\end{eqnarray}
with initial data given by
\begin{eqnarray}
Q_j^0 &=& 1\ , \nonumber\\
Q_1^i &=& \ch\left(V(\omega_i)+V(\omega_{i-2})+\ldots\right), \quad
i=1, \ldots, r-2\ ,
\nonumber\\
Q_1^{r-1} &=& \ch\left(V(\omega_{r-1})\right)\ , \label{eq:initial}\\
Q_1^r &=& \ch\left(V(\omega_r)\right)\ . \nonumber
\end{eqnarray}
Here $V(\omega_i)$ denotes the $i^{th}$ fundamental representation of
$D_r$.

In future we refer to the equations~(\ref{eq:kr}) as the
Kirillov-Reshetikhin (KR) equations. In the following section we
prove a formula for the quantities $Q_1^j$ that agrees with the
Yangian interpretation.

\subsection{Solutions of the Kirillov-Reshetikhin equations}
The paper \cite{KNH} studies a class of 2-dimensional Toda equations
on discrete space-time. These arise as functional relations for
commuting families of transfer matrices in solvable lattice models
associated with any classical Lie algebra $X_r$. For $D_r$ ($r\geq
4$) the relevant system of Toda equations is
\begin{eqnarray} T_k^a(u-1)T_k^a(u+1)-T_{k+1}^a(u)T_{k-1}^a(u)
=\begin{cases}
T_k^{a-1}(u)T_k^{a+1}(u) &1 \leq a \leq r-3\ ,\\
T_k^{r-3}(u)T_k^{r-1}(u)T_k^r(u) &a = r-2\ , \label{eq:toda}\\
T_k^{r-2}(u) &a = r-1,r\ .
\end{cases}
\end{eqnarray}

Here $k\in\mathbb{Z}_{\geq0}$, $u\in\mathbb{C}$, and
$a\in\{1,2,\ldots,r\}$, and the $T_k^a(u)$ are complex numbers
(depending on $a$, $k$, and $u$). The system is considered with the
initial conditions $T_0^a(u)=1$ for any $1\leq a\leq r$. (Note that
we also take $T_k^0=1$). Later it will be useful to impose the
further initial conditions~(\ref{eq:initial}), however that is not
necessary at this stage.

The system of equations~(\ref{eq:toda}) can be solved iteratively to
express an arbitrary $T_k^a(u)$ ($k\geq 1$) as a determinant or a
Pfaffian of a matrix with entries $0$ and \linebreak$\pm
T_1^b(u+\mathrm{const.})$ for $0\leq b \leq r$. Such solutions are
given explicitly for the cases $B_r$, $C_r$, and $D_r$ in
\cite{KNH}. The proof is shown only for the case $C_r$, with the
claim that it extends to the other cases. We carry out the proof for
the $D_r$ case and it indeed works nicely. Notation and method
follow exactly that of \cite{KNH}.

\subsubsection{Notation}
For any $l\in\mathbb{C}$, put
\begin{equation}
x_l^a = \begin{cases} T_1^a(u+l) & \text{if } 1\leq a \leq r\ ,\\
1 & \text{if } a=0\ .
\end{cases} \label{eq:xla}
\end{equation}

Introduce the infinite-dimensional matrices
$\mathcal{T}=\left(\mathcal{T}_{ij}\right)_{i,j\in\,\mathbb{Z}}$ and
$\mathcal{E}=\left(\mathcal{E}_{ij}\right)_{i,j\in\,\mathbb{Z}}$ by
\begin{eqnarray*}
\mathcal{T}_{ij}&=&
\begin{cases}
x_{\frac{i+j}{2}-1}^{\frac{j-i}{2}+1} & \text{if $i\in
2\mathbb{Z}+1$ and $\frac{i-j}{2}\in\{1,0,\ldots,3-r\}$}\ ,\\
-x_{\frac{i+j-1}{2}}^{r-1} & \text{if $i\in 2\mathbb{Z}+1$ and
$\frac{i-j}{2}=\frac{5}{2}-r$}\ ,\\
-x_{\frac{i+j-3}{2}}^{r} & \text{if $i\in 2\mathbb{Z}+1$
and $\frac{i-j}{2}=\frac{3}{2}-r$}\ ,\\
-x_{\frac{i+j}{2}-1}^{\frac{i-j}{2}+2r-3} & \text{if $i\in
2\mathbb{Z}+1$ and $\frac{i-j}{2}\in\{1-r,-r,\ldots,3-2r\}$}\ ,\\
0 & \text{otherwise}\ .
\end{cases}\\
\mathcal{E}_{ij}&=&
\begin{cases}
\pm1 & \text{if $i=j-2\pm2$ and $i\in2\mathbb{Z}$}\ ,\\
x_i^{r-1} & \text{if $i=j-3$ and $i\in2\mathbb{Z}$}\ ,\\
x_{i-2}^r & \text{if $i=j-1$ and $i\in2\mathbb{Z}$}\ ,\\
0 & \text{otherwise}\ .
\end{cases}
\end{eqnarray*}
For any $1\leq a \leq r$ and any $l$, the element $\pm x_l^a$
appears exactly once in the matrix $\mathcal{T}_{u\rightarrow
u+\xi}$. Here $u\rightarrow u+\xi$ means the overall shift of lower
indices according to~(\ref{eq:xla}). Let $\mathcal{T}_k(i,j,\pm
x_l^a)$ denote the $k\times k$ sub-matrix of $\mathcal{T}$ whose
$(i,j)$ element is exactly $\pm x_l^a$. We use a similar notation
for $\mathcal{E}(i,j,\pm x_l^a)$. These definitions are unambiguous;
for a more detailed explanation refer to the original paper.

\begin{thm}
For $k\in\mathbb{Z}_{\geq1}$,
\begin{eqnarray}
T_k^a(u)&=&\det\left(\mathcal{T}_{2k-1}(1,1,x_{-k+1}^{a})+
\mathcal{E}_{2k-1}(2,3,x_{-k-r+a+4}^r) \right) \label{eq:Q1}\\
&&\hspace{57mm} 1\leq a\leq r-2\ , \nonumber\\
T_k^{r-1}(u)&=&\pf\left(\mathcal{T}_{2k}(2,1,-x_{-k+1}^{r-1})+
\mathcal{E}_{2k}(1,2,x_{-k+1}^{r-1}) \right)\ , \label{eq:Q2}\\
T_k^{r}(u)&=&(-1)^k \pf\left(\mathcal{T}_{2k}(1,2,-x_{-k+1}^{r})+
\mathcal{E}_{2k}(2,1,x_{-k+1}^{r}) \right)\ ,\label{eq:Q3}
\end{eqnarray}
solve the $D_r$ system of equations~(\ref{eq:toda}).
\end{thm}

{\bf Proof}\\
Using equations~(\ref{eq:Q1}),~(\ref{eq:Q2}) and ~(\ref{eq:Q3}) we
can show that
\begin{eqnarray}
&&T_k^{r-1}(u)T_k^r(u)=(-1)^k
\det\left(\mathcal{T}_{2k}(1,1,-x_{-k+1}^{r-1})+
\mathcal{E}_{2k}(2,2,x_{-k+1}^r) \right),\label{eq:q1}\\
&&T_k^{r-1}(u+1)T_k^r(u-1)=(-1)^k
\det\left(\mathcal{T}_{2k}(2,1,x_{1-k}^{r-2})+
\mathcal{E}_{2k}(1,1,x_{-k}^{r}) \right),\label{eq:q2}\\
&&T_{k+1}^{r-1}(u)T_k^r(u-1)=(-1)^{k+1}
\det\left(\mathcal{T}_{2k+1}(1,1,-x_{-k}^{r-1})+
\mathcal{E}_{2k+1}(2,2,x_{-k}^r) \right),\label{eq:q3}\\
&&T_k^{r-1}(u+1)T_{k+1}^r(u)=(-1)^k
\det\left(\mathcal{T}_{2k+1}(2,1,x_{-k+1}^{r-2})+
\mathcal{E}_{2k+1}(1,1,x_{-k}^r) \right).\hspace{10mm}\label{eq:q4}
\end{eqnarray}
This is done by taking matrices
\begin{eqnarray*}
M &=& \mathcal{T}_{2k+1}(2,1,-x_{-k+1}^{r-1})
+\mathcal{E}_{2k+1}(1,2,x_{-k+1}^{r-1})\ ,\\
M &=& \mathcal{T}_{2k+1}(1,2,-x_{-k}^r)
+\mathcal{E}_{2k+1}(2,1,x_{-k}^{r})\ ,\\
M &=& \mathcal{T}_{2k+2}(2,1,-x_{-k}^{r-1})
+\mathcal{E}_{2k+2}(1,2,x_{-k}^{r-1})\ ,\\
M &=& \mathcal{T}_{2k+2}(1,2,-x_{-k}^r)
+\mathcal{E}_{2k+2}(2,1,x_{-k}^{r})\ ,
\end{eqnarray*}
respectively, and using Jacobi's identity as in \cite{KNH}.

{\it Jacobi's Identity:}
\nobreak
$$D_M\left[\begin{array}{c} 1\\1\end{array}\right]
D_M\left[\begin{array}{c} n\\n\end{array}\right]
=D_MD_M\left[\begin{array}{ccc} 1&,&n\\1&,&n\end{array}\right]
+D_M\left[\begin{array}{c} 1\\n\end{array}\right]
D_M\left[\begin{array}{c} n\\1\end{array}\right]\ .
$$
Here $D_M$ is the determinant of any $n\times n$ matrix $M$, and
$D_M\left[\begin{array}{ccccc}
i_1&,&i_2&,&\ldots\\j_i&,&j_2&,&\ldots\end{array}\right]$ denotes
its minor removing the $i_k$'s rows and the $j_k$'s columns.

The following facts are used in the proof: \begin{enumerate}
\item{If $M$ is an odd anti-symmetric matrix then $\det(M)=0$\ .}
\item{If $M$ is an odd anti-symmetric matrix then\ $D_M\left[\begin{array}{c} 1\\n\end{array}\right]
=D_M\left[\begin{array}{c} n\\1\end{array}\right]$\ .}
\item{If $M$ is an even anti-symmetric matrix then\ $D_M\left[\begin{array}{c} 1\\n\end{array}\right]
=-D_M\left[\begin{array}{c} n\\1\end{array}\right]$\ .}\\
\end{enumerate}

Having done this,~(\ref{eq:toda}) can be proved by again using the
Jacobi identity, this time setting
\begin{eqnarray*}
D_M &=& T_{k+1}^a(u)\ ,\\
D_M &=& T_{k+1}^{r-2}(u)\ ,\\
D_M &=& T_{k+1}^{r-1}(u)T_{k}^r(u-1)\ ,\\
D_M &=& T_k^{r-1}(u+1)T_{k+1}^r(u)\ ,
\end{eqnarray*}
for $1\leq a\leq r$, $a=r-2$, $a=r-1$, and $a=r$ respectively. $M$
is taken as in the right hand sides of~(\ref{eq:q1}-\ref{eq:q4}).
This proves the theorem.

\subsubsection{A special case}
Of particular interest to us is the special case in which $T$ is a
constant, i.e. there is no $u$-dependence. In this case the
structure of the $T$-system is exactly that of the
Kirillov-Reshetikhin equations~(\ref{eq:kr}). In the case where each $T$ is
$u$-independent, the subindex $l$ in the quantity $x_l^a$ can be
omitted, as this refers to the shift in the $u$ argument, which is
now irrelevant . In fact from now on we will abandon the $x_l^a$
notation altogether as it is no longer necessary in this simpler
special case. We revert to using $T_1^a$ in place of $x^a$. We get
the following corollary of theorem 1.

\newtheorem{cor}{Corollary}
\begin{cor}
For $k\in\mathbb{Z}_{\geq1}$,
\begin{eqnarray}
&&T_k^a=\det\left(\mathcal{T}_{2k-1}(1,1,T_1^{a})+
\mathcal{E}_{2k-1}(2,3,T_1^r) \right)\ ;\  1\leq
a\leq r-2\ , \nonumber\\
&&T_k^{r-1}=\pf\left(\mathcal{T}_{2k}(2,1,-T_1^{r-1})+
\mathcal{E}_{2k}(1,2,T_1^{r-1}) \right)\ , \label{eq:sol2}\\
&&T_k^{r}=(-1)^k \pf\left(\mathcal{T}_{2k}(1,2,-T_1^{r})+
\mathcal{E}_{2k}(2,1,T_1^{r}) \right)\ , \nonumber
\end{eqnarray}
solve the $D_r$ Kirillov-Reshetikhin equations
\begin{equation*}
\left(T_k^a\right)^2 - T_{k+1}^aT_{k-1}^a =
\begin{cases}
T_k^{a-1}T_k^{a+1} & 1\leq a\leq r-3\ ,\\
T_k^{r-3}T_k^{r-1}T_k^r & a = r-2\ ,\\
T_k^{r-2} & a=r-1,\, r\ .
\end{cases}
\end{equation*}
\end{cor}

We now proceed to show that, under the initial conditions~(\ref{eq:initial})
used by Kirillov and Reshetikhin, the quantities $T_1^1,\ldots,T_1^r$ are exactly the Yangian
characters $Q_1^1,\ldots,Q_1^r$.

\subsubsection{Recursion relation for $T_n^1$}
(\ref{eq:sol2}) is a solution of the Kirillov-Reshetikhin equations
in terms of determinants and Pfaffians of certain matrices with
entries $0$ and $\pm T_1^1,\ldots,\pm T_1^r$. In particular, for
$n\geq1$, $T_n^1$ is given by the determinant
$$T_n^1=\det\left(\mathcal{T}_{2n-1}(1,1,T_1^1)+
\mathcal{E}_{2n-1}(2,3,T_1^r) \right)\ .$$ Expansion of this
determinant leads to the following recursion relation for the
quantity $T_n^1$ ($n\geq 1$) in terms of
$T_{n-1}^1,\,T_{n-2}^1,\ldots,T_1^1$ and $T_1^1,\ldots,T_1^r$:
\begin{eqnarray}
T_n^1&=& T_1^1T_{n-1}^1-T_1^2T_{n-2}^1+T_1^3T_{n-3}^1-\ldots
-T_1^{r-2}T_{n-r+2}^1 \nonumber\\ &+&T_1^{r-1}T_1^rT_{n-r+1}^1
+T_1^{r-2}T_{n-r}^1-\ldots-T_1^1T_{n-2r+3}^1+T_{n-2r+2}^1\nonumber\\
&-&((T_1^{r-1})^2+(T_1^r)^2)(T_{n-r}^1+T_{n-r-2}^1+\ldots\ )\nonumber\\
&+& 2T_1^{r-1}T_1^r(T_{n-r-1}^1+T_{n-r-3}^1+\ldots\ )\ . \label{eq:rec}
\end{eqnarray}
For simplicity we define the quantity $R_{n-k}$ by
$$R_{n-k}=T_{n-k}^1+T_{n-k-2}^1+T_{n-k-4}^1+\ldots$$
Then $R_n=T_n^1+R_{n-2}$ .

\begin{thm}
The recursion equation~(\ref{eq:rec}) can be written in matrix form
as
$$\left(\begin{array}{l}
T_n^1\\
T_{n-1}^1\\
T_{n-2}^1\\
\vdots\\
T_{n-2r+3}^1\\
R_{n-r+1}\\
R_{n-r}
\end{array}\right)=\mathcal{M}
\left(\begin{array}{l}
T_{n-1}^1\\
T_{n-2}^1\\
T_{n-3}^1\\
\vdots\\
T_{n-2r+2}^1\\
R_{n-r}\\
R_{n-r-1}
\end{array}\right)$$
where $\mathcal{M}$ is the $2r\times 2r$ matrix
$$\footnotesize{\left(\begin{array}{cccccccccccc} T_1^1 & -T_1^2  & \ldots&
-T_1^{r-2}&T_1^{r-1}T_1^r& T_1^{r-2} &\ldots& T_1^2& -T_1^1&1&
-(T_1^{r-1})^2-(T_1^r)^2 & 2T_1^{r-1}T_1^r\\
1&0&\ldots&0&0&0&\ldots&0&0&0&0&0\\
0&1&\ldots&0&0&0&\ldots&0&0&0&0&0\\
\vdots&\vdots&\ldots&\vdots&\vdots&\vdots&\ldots&\vdots&\vdots&\vdots&\vdots&\vdots\\
0&0&\ldots&1&0&0&\ldots&0&0&0&0&0\\
0&0&\ldots&0&1&0&\ldots&0&0&0&0&0\\
0&0&\ldots&0&0&1&\ldots&0&0&0&0&0\\
\vdots&\vdots&\ldots&\vdots&\vdots&\vdots&\ldots&\vdots&\vdots&\vdots&\vdots&\vdots\\
0&0&\ldots&0&0&0&\ldots&0&1&0&0&0\\
0&0&\ldots&0&1&0&\ldots&0&0&0&0&1\\
0&0&\ldots&0&0&0&\ldots&0&0&0&1&0\\
\end{array}\right)}$$

with initial values $T_0^1=T_{-2r+2}^1=1$ and $T_{-j}^1=0$ for
$j=1,2,\ldots,2r-3$.

{\it Note: This is the matrix $\mathcal{M}$ for $r$ even. For $r$
odd the matrix has the same structure with slight changes of sign.
This does not affect subsequent calculations, and results are
identical.\\}
\end{thm}

The eigenvalues of the matrix $\mathcal{M}$ are the solutions of the
equation
\begin{eqnarray}
0 &=& 1+\lambda^{2r} -T_1^1(\lambda^{2r-1}+\lambda)
+(T_1^2-1)(\lambda^{2r-2}+\lambda^2)\nonumber\\
&-& (T_1^3-T_1^1)(\lambda^{2r-3}+\lambda^3) +\ldots + (T_1^{r-2}-T_1^{r-4})(\lambda^{r+2}+\lambda^{r-2})\nonumber\\
 &-& (T_1^{r-1}T_1^r-T_1^{r-3})(\lambda^{r+1}+\lambda^{r-1})
+ ((T_1^{r-1})^2+(T_1^r)^2-2T_1^{r-2})\lambda^r \ .\quad\qquad\label{eq:lambda}
\end{eqnarray}

Suppose the equation~(\ref{eq:lambda}) has roots
$a_1,\ldots,a_{2r}$. Notice that $0,\pm1$ are not roots. By the
symmetry of equation~(\ref{eq:lambda}), and the fact that
$\prod_{i=1}^{2r}a_i=1$, the roots can be written as $a_1,\ldots
a_1^{-1}$,\ $a_r,\ldots,a_r^{-1}$, where for the purposes of
ordering we identify $a_{r+1}=a_1^{-1},\ldots,a_{2r}=a_r^{-1}$. Then
the following equations must be satisfied:
\begin{eqnarray*}
T_1^1 &=& \sum_{i=1}^{2r} a_i\ ,\\
T_1^2-1 &=& \sum_{i<j} a_ia_j\ ,\\
T_1^k-T_1^{k-2}&=&\sum_{i_1<\ldots<i_k} a_{i_1}\ldots a_{i_k}\quad{\rm for}
\ k=3,\ldots,r-2\ ,\\
T_1^{r-1}T_1^r-T_1^{r-3}&=&\sum_{i_1<\ldots<i_{r-1}} a_{i_1}\ldots a_{i_{r-1}}\ ,\\
(T_1^{r-1})^2+(T_1^r)^2-2T_1^{r-2}&=&\sum_{i_1<\ldots<i_r}
a_{i_1}\ldots a_{i_r}\ .\\
\end{eqnarray*}

We consider the orthogonal group, $SO(2r)$, of linear
transformations of $\mathbb{C}^{2r}$ which preserve the form
$z_1z_2+z_3z_4+\ldots+z_{2r-1}z_{2r}$. A maximal torus consists of
elements $g=\diag(a_1,a_1^{-1},\ldots,a_r,a_r^{-1}),\
a_i\in\mathbb{C}$.

\begin{thm}
For the choice~(\ref{eq:initial}) of initial conditions, and
$g=\diag(a_1,a_1^{-1},\ldots,a_r,a_r^{-1})$, the quantities $T_1^i$
are exactly the Yangian characters $Q_1^i(g)$ for $i=1,2,\ldots,r$.
\end{thm}

{\bf Proof}\\
It is a well known fact that for $1\leq k\leq r-2$, the characters
of the Lie algebra $D_r$ have the structure
$$\chi(\omega_k)=\sum_{i_1<\ldots<i_k}a_{i_1}\ldots a_{i_k}\ ,$$ where
$a\in\{a_1,a_1^{-1},\ldots a_r,a_r^{-1}\}$, as before. The final two
characters have a slightly different structure: $\chi(\omega_{r-1})
= \sum {a_1}^{\pm\frac{1}{2}}\ldots {a_r}^{\pm\frac{1}{2}}$, where
the sum is taken over all possible combinations with an odd number
of negative exponents, and $\chi(\omega_r)$ is defined similarly but
with an even number of negative exponents. This follows immediately
from the structure of the fundamental weights of $D_r$, and the
action of the Weyl group, see~(\ref{eq:omega}).

Now suppose we take initial conditions as in
equations~(\ref{eq:initial}). We can then conclude the following:
\begin{eqnarray*}
T_1^1=\sum_{i=1}^{2r} a_i &\Rightarrow& T_1^1=\chi(\omega_1)=Q_1^1\ ,\\
&\vspace{5mm}&\\
T_1^2-1=\sum_{i<j}a_ia_j &\Rightarrow& T_1^2-1=\chi(\omega_2)\\
&\Rightarrow& T_1^2=\chi(\omega_2)+1\\
&\Rightarrow& T_1^2=Q_1^2\ ,\\
&\vspace{5mm}&\\
T_1^3-T_1^1=\sum_{i<j<k} a_ia_ja_k &\Rightarrow& T_1^3-T_1^1=\chi(\omega_3)\\
&\Rightarrow&T_1^3=\chi(\omega_3)+T_1^1\\
&\Rightarrow&T_1^3=\chi(\omega_3)+\chi(\omega_1)\\
&\Rightarrow&T_1^3=Q_1^3\ ,\\
&\vspace{5mm}&\\
T_1^k-T_1^{k-2}=\sum_{i_1<\ldots<i_k} a_{i_1}\ldots a_{i_k}
&\Rightarrow&
T_1^k-T_1^{k-2}=\chi(\omega_k)\\
&\Rightarrow& T_1^k=T_1^{k-2}+\chi(\omega_k)\\
&\Rightarrow& T_1^k=Q_1^{k-2}+\chi(\omega_k)\quad \mbox{by induction
on k}\\
&\Rightarrow&T_1^k=Q_1^k\quad {\rm for}\ 4\leq k\leq r-2\ .
\end{eqnarray*}
The final two equations
\begin{eqnarray*}
T_1^{r-1}T_1^r-T_1^{r-3}&=&\sum_{i_1<\ldots<i_{r-1}} a_{i_1}\ldots
a_{i_{r-1}}\\
&&\\
\Rightarrow
T_1^{r-1}T_1^r&=&\sum_{i_1<\ldots<i_{r-1}} a_{i_1}\ldots a_{i_{r-1}}+Q_1^{r-3}\ ,\\
&{\rm and}&\\
(T_1^{r-1})^2+(T_1^r)^2-2T_1^{r-2}&=&\sum_{i_1<\ldots<i_r}
a_{i_1}\ldots a_{i_r}\\
&&\\ \Rightarrow
(T_1^{r-1})^2+(T_1^r)^2&=&2Q_1^{r-2}+\sum_{i_1<\ldots<i_r}
a_{i_1}\ldots a_{i_r}\ ,
\end{eqnarray*}
are satisfied by $$T_1^{r-1}=\chi(\omega_{r-1})=Q_1^{r-1}\ ,$$ and
$$T_1^r=\chi(\omega_r)=Q_1^r\ .$$

Hence we can conclude that $T_1^1,\ldots,T_1^r$ are in fact the
Yangian characters $Q_1^1,\ldots,Q_1^r$.\\

{\it Remark: Using the Weyl character formula and the recursion
relation~(\ref{eq:rec}), we can conclude that
$$T_i^1(g)=Q_i^1(g)\ ,$$ for $g=\diag\left(a_1,a_1^{-1},\ldots,a_r,a_r^{-1}\right)$.
This is done in the following theorem.}

\begin{thm}
$T_i^1(g)=Q_i^1(g)$ for all $i$, where
$g=\diag\left(a_1,a_1^{-1},\ldots,a_r,a_r^{-1}\right)$.
\end{thm}

{\bf Proof}\\
Since $Q_i^1$ is irreducible for all $i$, we can write
$$Q_i^1=\ch\left(V(i\omega_1)\right)=\frac{\sum_{w\in W(D_r)}\sgn(w)w
\left(a_1^{r-1+i}a_2^{r-2}\ldots a_{r-1}\right)}{Q_0}\ ,$$ by the
Weyl character formula. Then clearly $Q_i^1=1$ for $i=0\ \text{and}\
-2r+2$, and $Q_i^1=0$ for $i=-1,-2,\ldots,-2r+3$.

We already know that $T_{i+1}^1=\mathcal{M}T_i^1$, where
$\mathcal{M}$ is a $2r\times 2r$ matrix. Hence, if
$T_i^1=\ch\left(V(i\omega_1)\right)$ is true for any $2r$ values of
$i$, then $T_i^1=\ch\left(V(i\omega_1)\right)$ must be true for all
$i$.

Notice that the recursion relation~(\ref{eq:rec}) is true for
$n=1,0,-1,\ldots,-2r+2$ if one puts $T_0^1=T_{-2r+2}^1=1$,
$R_0=R_{-1}=0$, and $T_{-j}^1=0$ for $j=1,\ldots,2r-3$.

Then clearly for the $2r-1$ values $i=0,\ldots,-2r+2$ we have
$Q_i^1(g)=T_i^1(g)$. We also know from the previous theorem that
$T_1^1=Q_1^1$. Hence $Q_i^1(g)=T_i^1(g)$ for all $i$. This proves
the theorem.

\section{Solving the Equations of the $(D_m,A_n)$ Model}
In this section we study the integrable model described by the pair
of Dynkin diagrams $(D_m,A_n)$. Using the representation theory of
Yangians we solve the equations of the model.

\subsection{Equations of the model}
The equations of the model $(D_m,A_n)$ are $AU=V$, where
$A=C(D_m)^{-1}\otimes\,C(A_n)$, $U=\log(x)$, $V=\log(1-x)$, and
$x=(x_{11},\ldots,x_{mn})$. By exponentiation and a change of
variables they can be rewritten as
\begin{equation}
z^{2-C(D_m)}+z^{2-C(A_n)}=z^2\ . \label{eq:zm}
\end{equation}
We seek a matrix $g\in SO(2m)$ whose Yangian characters satisfy
$Q_{n+1}^i(g)=1$ for $i=1,2,\ldots,m$. This leads to the relation
$$z_{ij}=Q_j^i(g)\ ,$$
when $z_{1j}=Q_j^1(g)$ as above. Here $z_{ij}$ are the components of
the solutions of~(\ref{eq:zm}). Write
$$g=\diag(a_1,a_1^{-1},\ldots,a_m,a_m^{-1}) \in SO(2m)\ .$$

\begin{thm}
Suppose $g\in SO(2m)$ is a matrix that satisfies
\begin{equation}
Q_{n+1}^i(g)=1\ , \label{eq:solve}
\end{equation}
for $i=1,2,\ldots,m$, and moreover $Q_j^i(g)\neq0$ for
$j=1,\ldots,n$. Then $g$ also satisfies the set of equations
\begin{eqnarray}
Q_{n+2}^i(g) &=& 0\ ,\nonumber\\
Q_{n+3}^i(g) &=& 0\ ,\nonumber\\
&\vdots& \label{eq:zeros}\\
Q_{n+2m-2}^i(g) &=& 0\ ,\nonumber\\
Q_{n+2m-1}^i(g) &=& \pm1\ , \nonumber
\end{eqnarray}
for $i=1,2,\ldots,m$. In particular
\begin{equation*}
Q_{n+2m-1}^i = \begin{cases} +1 & {\rm for}\ i=1,2,\ldots,m-2\ ,\\
+1 & {\rm for}\ i=m-1,m\ ,\ {\rm if}\ m\equiv0\  {\rm or}\ 1\quad (\mathrm{mod}\ 4)\ ,\\
-1 & {\rm for}\ i=m-1,m\ ,\ {\rm if}\ m\equiv2\  {\rm or}\ 3 \quad
(\mathrm{mod}\ 4)\ .
\end{cases}
\end{equation*}
\end{thm}

{\bf Proof}\\
This has been proved by Nahm. The idea of the proof is the
following. First one considers the variety of generic solutions of
equations~(\ref{eq:kr}) with initial data given
by~(\ref{eq:initial}) and its closure. This excludes solutions with
an unwanted pattern of vanishing $Q^i_j$. Interchanging $i$ and $j$
in equations~(\ref{eq:kr}) allows to write down explicit algebraic
relations between $Q^1_j$ and $Q^i_{n+2}$. These yield $Q^1_{n+j}=0$
for $j=2,\ldots,m-1$. Then one considers an analogous algebraic
variety of solutions of equations~(\ref{eq:kr}) without imposing
initial data at $j=0$. The character formula for $Q^1_j$ is easily
generalised to this larger variety. The algebraic equations remain
valid and yield $dQ^1_{n+m}=0$. When one uses the character formula
for $Q^1_{n+m}$ this implies $Q^1_{n+m+j}=Q^1_{n+m-j}$.

\begin{lemma}
Any matrix $g\in SO(2m)$ that satisfies the equations
\begin{eqnarray}
Q_{n+1}^1(g)&=&1\ ,\nonumber\\
Q_{n+2}^1(g)&=&0\ ,\nonumber\\
Q_{n+3}^1(g)&=&0\ ,\nonumber\\
&\vdots& \label{eq:solve2}\\
Q_{n+2m-2}^1(g)&=&0\ ,\nonumber\\
Q_{n+2m-1}^1(g)&=&1\ , \nonumber
\end{eqnarray}
and the two equations
\begin{eqnarray}
Q_{n+1}^m&=&1\ , \label{eq:solve3}\\
Q_{n+2m-1}^m&=&\pm1\ ,\label{eq:solve4}
\end{eqnarray}
also satisfies the set of equations~(\ref{eq:zeros}). \end{lemma}

{\bf Proof}\\
This follows immediately from substitution of~(\ref{eq:solve2}, \ref{eq:solve3}, \ref{eq:solve4})
in the Kirillov-Reshetikhin equations~(\ref{eq:kr}) for $D_m$.

One can conclude that the problem of finding a matrix $g$ whose
characters satisfy~(\ref{eq:solve}) is equivalent to the problem of
finding a matrix $g$ whose characters satisfy~(\ref{eq:solve2},
\ref{eq:solve3}, \ref{eq:solve4}). We choose to work with the second
set of conditions as these involve only those irreducible Yangian
representations that remain irreducible as representations of the
corresponding Lie algebra. This simplifies matters greatly. To find
a matrix $g\in SO(2m)$ whose Yangian characters satisfy the
equations~(\ref{eq:solve2}, \ref{eq:solve3}, \ref{eq:solve4}), we
first find a solution of the set of equations~(\ref{eq:solve2}), and
show that it is unique. We then show that the same solution
satisfies~(\ref{eq:solve3}) and~(\ref{eq:solve4}) when square roots
are chosen appropriately.

\subsection{Some useful facts}
{\bf Notation}\\
Let $W(D_{m-1})[k]$ denote the Weyl group of $D_{m-1}$ acting on the
elements
$$\pm1,\pm2,\ldots,\pm(k-1),\pm(k+1),\ldots,\pm m\ .$$
Any element of $W(D_m)$ can be factorised uniquely as
$\sigma_k\omega,\ k=1,\ldots,m$, where $\sigma_1=1$, $\sigma_k=(1k)$
for $k\neq 1$, and $\omega\in W(D_{m-1})[k]$. Define the quantity
$A_k$ by
$$A_k=\sum_{w\in \,W(D_{m-1})[k]}\sgn(1k).\sgn(w).
w\left((1k)(a_2^{m-2}a_3^{m-3}\ldots a_{m-2}^2a_{m-1})\right)\ .$$

\vspace*{5mm}

\begin{lemma}
$$A_k=(-1)^{k-1}\prod_{t=1}^m a_t^{2-m}
\prod_{i<j}(a_ia_j-1)(a_i-a_j)\ ,$$ where the first sum is taken
over $t\neq k$, and the second is taken over $i,j\in\{1,\ldots,m\}$
with $i,j\neq k$.
\end{lemma}

{\bf Proof}\\
For $k\neq 1$ we have
\begin{eqnarray*}
A_k &=& -\sum_{w\in \,W(D_{m-1})[k]}\sgn(w).
w\left((1k)(a_2^{m-2}a_3^{m-3}\ldots a_{m-2}^2a_{m-1})\right)\\
&=& -\sum_{w\in \,W(D_{m-1})[k]}\sgn(w).
w\left((a_2^{m-2}a_3^{m-3}\ldots a_{k-1}^{m-k+1}a_1^{m-k}a_{k+1}^{m-k-1}\ldots a_{m-2}^2a_{m-1})\right)\\
&=& -(-1)^{k-2}\sum_{w\in \,W(D_{m-1})[k]}\sgn(w).
w\left((a_1^{m-2}a_2^{m-3}\ldots a_{k-1}^{m-k}a_{k+1}^{m-k-1}\ldots
a_{m-2}^2a_{m-1})\right)\ .
\end{eqnarray*}
This final expression in fact holds for all $k$ (including $k=1$). Up to a
factor of $(-1)^{k-1}$ this is the Weyl denominator for
$D_{m-1}[k]$. Therefore, by the multiplicative formula for the Weyl
denominator we can write
\begin{eqnarray*}
A_k &=& (-1)^{k-1}(a_1^{2-m}a_2^{3-m}\ldots
a_{k-1}^{k-m}a_{k+1}^{1+k-m}\ldots a_m^{-1})
\prod_{i<j}(a_ia_j-1)(a_ia_j^{-1}-1)\\
&=& (-1)^{k-1}\prod_{t=1}^m a_t^{2-m} \prod_{i<j}(a_ia_j-1)(a_i-a_j)\ ,
\end{eqnarray*}
where the first sum is taken over $t\neq k$, and the second is taken
over $i,j\in\{1,2,\ldots,m\}$ with $i,j\neq k$.

\begin{lemma}
$$A_1A_2\ldots A_m =
\pm\frac{\prod_{i>j}(a_i-a_j)^{m-2}(a_ia_j-1)^{m-2}}{\prod_{i=1}^ma_i^{(m-1)(m-2)}}\
.$$
\end{lemma}

{\bf Proof}
\begin{eqnarray*}
A_1A_2\ldots A_m &=& \left(\prod_{k=1}^m(-1)^{k-1}\right)
\left(\prod_{i=1}^m
a_i^{2-m}\right)^{m-1}\prod_{i<j}(a_ia_j-1)^{m-2}(a_i-a_j)^{m-2}\\
&&\\ &=& \left(\prod_{k=1}^m(-1)^{k-1}\right)\left(
\frac{\prod_{i<j}(a_ia_j-1)^{m-2}(a_i-a_j)^{m-2}}{\prod_{i=1}^m
a_i^{(m-1)(m-2)}}\right)\\
&&\\
&=& (-1)^{\frac{(m-1)m}{2}}\left(
\frac{\prod_{i<j}(a_ia_j-1)^{m-2}(a_i-a_j)^{m-2}}{\prod_{i=1}^m
a_i^{(m-1)(m-2)}}\right)\\
&&\\
&=&(-1)^{\frac{(m-1)m}{2}}\left(
\frac{\prod_{i>j}(a_ia_j-1)^{m-2}(a_i-a_j)^{m-2}(-1)^{m-2}}{\prod_{i=1}^m
a_i^{(m-1)(m-2)}}\right)\\
&&\\
&=&(-1)^{\frac{(m-1)m}{2}}\left((-1)^{m-2}\right)^{\frac{m(m-1)}{2}}\left(
\frac{\prod_{i>j}(a_ia_j-1)^{m-2}(a_i-a_j)^{m-2}}{\prod_{i=1}^m
a_i^{(m-1)(m-2)}}\right)\\
&&\\
&=&(-1)^{\frac{m(m-1)^2}{2}}\left(\frac{\prod_{i>j}(a_ia_j-1)^{m-2}(a_i-a_j)^{m-2}}{\prod_{i=1}^m
a_i^{(m-1)(m-2)}}\right)\\
&&\\
&=&
\pm\frac{\prod_{i>j}(a_ia_j-1)^{m-2}(a_i-a_j)^{m-2}}{\prod_{i=1}^m
a_i^{(m-1)(m-2)}}\ .\\
\end{eqnarray*}

\begin{lemma}
$$\prod_{i=1}^m
a_i^{m-1}\left|\begin{array}{cccc}
1 & \ldots & \ldots & 1\\
(a_1+a_1^{-1}) & \ldots & \ldots & (a_m+a_m^{-1})\\
\vdots & \vdots & \vdots & \vdots\\
(a_1^{m-1}+a_1^{1-m}) & \ldots & \ldots &
(a_m^{m-1}+a_m^{1-m})\end{array}\right|=\prod_{i>j}(a_i-a_j)(a_ia_j-1)\ .$$
\end{lemma}

{\bf Proof}\\
Clearly both sides of this equation are polynomials. To prove that
they are equivalent, we must show that their zeros coincide.
This implies equality up to a constant, which we show to be 1.
The proof is in three steps.
\begin{enumerate}
\item{Show that every zero of the RHS is also a zero of the LHS,}
\item{Show that both sides of the equation have the same order,}
\item{Show that the constant is equal to 1.}
\end{enumerate}

The RHS has a zero when $a_i=a_j$ or when $a_i=a_j^{-1}$ for $i\neq
j$. Both of these cases lead to rows $i$ and $j$ being equal in the
determinant on the LHS, which implies the whole LHS is zero. Hence
every zero of the RHS is also a zero of the LHS.

The left hand side can be rewritten as
$$LHS=\left|\begin{array}{cccc}
a_1^{m-1} & \ldots & \ldots & a_m^{m-1}\\
a_1^m+a_1^{m-2} & \ldots & \ldots & a_m^m+a_m^{m-2}\\
\vdots & \vdots & \vdots & \vdots\\
a_1^{2m-2}+1 & \ldots & \ldots & a_m^{2m-2}+1\end{array}\right|\ .$$
The term of highest order on the LHS is $a_1^{m-1}a_2^m\ldots
a_m^{2m-2}$. Hence the LHS has order
$$\sum_{k=m-1}^{2m-2}k=\sum_{k=1}^{2m-2}k-\sum_{k=1}^{m-2}k=\frac{3m(m-1)}{2}=3\binom{m}{2}\ .$$

Now for the order of the RHS. There are $\binom{m}{2}$ terms of the
form $(a_i-a_j)(a_ia_j-1)$, and each has order three. Hence the RHS
has order $3\binom{m}{2}$. Thus the LHS and the RHS have the same
order. This proves that the two sides of the equation are equivalent
up to a constant. To calculate this constant compare the coefficient
of the term $a_1^{m-1}a_2^m a_3^{m+1}\ldots
a_{m-1}^{2m-3}a_m^{2m-2}$ on both sides of the equation. On the LHS
this term occurs exactly once, coming from a product along the main
diagonal of the matrix. On the RHS it arises from the product
$$\prod_{i>j}(a_ia_j)(a_i)=\left(\prod_{i=1}^m a_i^{m-1}\right)\left(a_2a_3^2\ldots
a_{m-1}^{m-2}a_m^{m-1}\right)=a_1^{m-1}a_2^m\ldots a_m^{2m-2}\ .$$
Hence it occurs exactly once on both sides. If the coefficients on
both sides agree for one term they must agree for every term,
hence the constant is 1. This proves the theorem.

\begin{lemma}
Let $\chi$ denote some character of a Lie algebra $\g$, and let $g$ be
an element of some maximal torus. Suppose $w_1\in W(\g)$ satisfies $sgn(w_1)=-1$ such that
$\chi(w_1(g))=\chi(g)$. Then $\sum_{w\in W} sgn(w)\chi(w(g))=0$.
\end{lemma}

{\bf Proof}
\begin{eqnarray*}
\sum_{w\in W} \sgn(w)\chi(w(g)) &=& \sum_{w\in W}
\sgn(ww_1)\chi(ww_1(g))\\
&=& -\sum_{w\in W}\sgn(w)\chi(w(g))\\
&=&0\ .
\end{eqnarray*}

\begin{lemma}
$$\sum_{k=1}^m\left(a_k^{m-j}+a_k^{j-m}\right)A_k=0$$ for
$j=2,\ldots,m$.
\end{lemma}

{\bf Proof}\\
For notational simplicity set $\xi_j^{m-i}=(a_j^{m-i}+a_j^{-m+i})$.
Then $A_k$ can be written as
$$A_k=\sgn(1k)\sum_{S_{m-1}}\sgn(w)\,w\left((1k)\xi_2^{m-2}
\xi_3^{m-3}\ldots\xi_{m-1}^1\right)\ .$$ Then for $j=2,\ldots,m$ we
get the following:
\begin{eqnarray*}
&&\sum_{k=1}^m \left(a_k^{m-j}+a_k^{-m+j}\right)A_k\\
&=& \sum_{k=1}^m \xi_k^{m-j}A_k\\
&=& \xi_1^{m-j}A_1+\sum_{k=2}^m\xi_k^{m-j}A_k\\
&=&\sum_{S_{m-1}}\sgn(w)\xi_1^{m-j}w\left(\xi_2^{m-2}\xi_3^{m-3}
\ldots \xi_{m-1}^1\right)\\
&-&\sum_{k=2}^m
\left(\sum_{S_{m-1}}\sgn(w)\xi_k^{m-j}w\left(\xi_2^{m-2} \ldots
\xi_{k-1}^{m-k+1}\xi_1^{m-k}\xi_{k+1}^{m-k-1} \ldots
\xi_{m-1}^1\right)\right)\\
&=&\sum_{S_m}\sgn(w)w\left(\xi_1^{m-j}\xi_2^{m-2}\ldots\xi_j^{m-j}\ldots\xi_k^{m-k}
\ldots\right)\\
&=&0\ .
\end{eqnarray*}

\subsection{Solution of the equations}
By theorem 5 we need a matrix
$g=\diag(a_1,a_1^{-1},\ldots,a_m,a_m^{-1})\in \nolinebreak SO(2m)$
whose characters satisfy
$$\begin{array}{lcl}
Q_{n+1}^1(g) &=&1\ ,\\
Q_{n+j}^1(g) &=&0\ ,\ \text{for}\ j=2,\ldots,m-1\ ,\\
Q_{n+m}^1(g) &=&0\ ,\\
Q_{n+2m-j}^1(g) &=&0\ ,\ \text{for}\ j=2,\ldots,m-1\ ,\\
Q_{n+2m-1}^1(g) &=&1\ .\\
\end{array}$$

Using the Weyl character formula these can be rewritten as follows
\begin{eqnarray*}
&&Q_{n+1}^1(g)=1\\
&\Leftrightarrow& \sum_{i=1}^m(a_i^{n+m}+a_i^{-n-m})A_i =
\sum_{i=1}^m(a_i^{m-1}+a_i^{1-m})A_i\\
&\Leftrightarrow&
\sum_{i=1}^m\left((a_i^{n+2m-1}-1)a_i^{1-m}+(a_i^{-n-2m+1}-1)a_i^{m-1}\right)A_i
=0\ ,\\
&&\\
&&Q_{n+j}^1(g)=0\qquad j=2,\ldots,m-1\\
&\Leftrightarrow& \sum_{i=1}^m(a_i^{n+m+j-1}+a_i^{-n-m-j+1})A_i=0\\
&\Leftrightarrow&
\sum_{i=1}^m\left((a_i^{n+2m-1}-1)a_i^{j-m}+(a_i^{-n-2m+1}-1)a_i^{m-j}\right)A_i
=0\ ,\\
&&\\
&&Q_{n+m}^1(g)=0\\
&\Leftrightarrow& \sum_{i=1}^m(a_i^{n+2m-1}+a_i^{-n-2m+1})A_i=0\\
&\Leftrightarrow& \sum_{i=1}^m(a_i^{n+2m-1}-2+a_i^{-n-2m+1})A_i=0\ ,\\
&&\\
&&Q_{n+2m-j}^1(g)=0\qquad j=2,\ldots,m-1\\
&\Leftrightarrow& \sum_{i=1}^m(a_i^{n+3m-j-1}+a_i^{-n-3m+j+1})A_i=0\\
&\Leftrightarrow&
\sum_{i=1}^m\left((a_i^{n+2m-1}-1)a_i^{m-j}+(a_i^{-n-2m+1}-1)a_i^{j-m}\right)A_i
=0\ ,\\
&&\\
&&Q_{n+2m-1}^1(g)=1\\
&\Leftrightarrow& \sum_{i=1}^m(a_i^{n+3m-2}+a_i^{-n-3m+2})A_i =
\sum_{i=1}^m(a_i^{m-1}+a_i^{1-m}) A_i\\
&\Leftrightarrow&
\sum_{i=1}^m\left((a_i^{n+2m-1}-1)a_i^{m-1}+(a_i^{-n-2m+1}-1)a_i^{1-m}\right)A_i
=0\ .
\end{eqnarray*}

{\it Note: In obtaining the above equations we used Lemma 6.}

We now have a system of $2m-1$ equations in $2m$ variables. The
variables are $a_i$ and $a_i^{-1}$ with $i=1,\ldots,m$. By summing
the equations for $Q_{n+j}^1$ and $Q_{n+2m-j}^1$ for
$j=1,\ldots,m-1$, we get a system of $m$ equations in $m$ variables.
This time the variables are $a_i^{n+2m-1}+a_i^{-n-2m+1}$ for
$i=1,\ldots,m$. These equations can be solved exactly. They can be
written in matrix form as
$$\left(\begin{array}{cccc}
A_1 & \ldots & \ldots & A_m\\
(a_1+a_1^{-1})A_1 & \ldots & \ldots & (a_m+a_m^{-1})A_m\\
\vdots & \vdots & \vdots & \vdots\\
(a_1^{m-1}+a_1^{1-m})A_1 & \ldots & \ldots &
(a_m^{m-1}+a_m^{1-m})A_m\\
\end{array}\right)
\left(\begin{array}{c}
a_1^{n+2m-1}-2+a_1^{-n-2m+1}\\
a_2^{n+2m-1}-2+a_2^{-n-2m+1}\\
\vdots\\
a_m^{n+2m-1}-2+a_m^{-n-2m+1}\\
\end{array}
 \right)$$

\begin{equation}
=\left(\begin{array}{c} 0\\
\vdots\\
0\\
0\\
\end{array}\right)\label{eq:matrix}
\end{equation}
For $i=1,\ldots,m$ this implies
\begin{eqnarray*}
a_i^{n+2m-1}-2+a_i^{-n-2m+1}&=&0\\
a_i^{-n-2m+1}\left(a_i^{n+2m-1}-1\right)^2&=&0\\
a_i^{n+2m-1}&=&1
\end{eqnarray*}
since $a_i\neq0$.\\

\begin{thm}
Any matrix $g=\diag(a_1,a_1^{-1},\ldots,a_m,a_m^{-1})\in SO(2m)$,
whose entries satisfy
\begin{equation}
a_i^{n+2m-1}=1\ , \label{eq:soln}
\end{equation} for $i=1,\ldots,m$, is a solution
of the equations~(\ref{eq:solve2}).
\end{thm}

{\bf Proof}
$$\begin{array}{lcl}
a_i^{n+2m-1}=1 &\Rightarrow& a_i^{n+m}=a_i^{1-m}\\
&\Rightarrow& \sum_{w\in W(D_m)}\sgn(w).w(a_1^{n+m}a_2^{m-2}\ldots a_{m-1})\\
&=& \sum_{w\in W(D_m)}\sgn(w).w(a_1^{1-m}a_2^{m-2}\ldots a_{m-1})\\
&=& \sum_{w\in W(D_m)}\sgn(w).w(w_1(a_1^{1-m}a_2^{m-2}\ldots
a_{m-1}))\\
&\quad& \mbox{where $w_1$ is the identity with signs
$(-+\ldots+-)$}\\
&=& \sum_{w\in W(D_m)}\sgn(w).w(a_1^{m-1}a_2^{m-2}\ldots a_{m-1})\\
&\Rightarrow& Q_{n+1}^1(g)=1\ .\\
\end{array}$$

$$\begin{array}{lcl}
a_i^{n+2m-1}=1 &\Rightarrow& a_i^{n+3m-2}=a_i^{m-1}\\
&\Rightarrow& \sum_{w\in W(D_m)}\sgn(w).w(a_1^{n+3m-2}a_2^{m-2}\ldots a_{m-1})\\
&=& \sum_{w\in W(D_m)}\sgn(w).w(a_1^{m-1}a_2^{m-2}\ldots a_{m-1})\\
&\Rightarrow& Q_{n+2m-1}^1(g)=1\ .\\
\end{array}$$

$$\begin{array}{lcl}
a_i^{n+2m-1}=1 &\Rightarrow& a_i^{n+m+j-1}=a_i^{j-m}\\
&\Rightarrow& \sum_{w\in W(D_m)}\sgn(w).w(a_1^{n+m+j-1}a_2^{m-2}\ldots a_{m-1})\\
&=& \sum_{w\in W(D_m)}\sgn(w).w(a_1^{-m+j}a_2^{m-2}\ldots a_{m-1})\\
&=&0\ ,\ \text{by Lemma 5 using $w_1=(1j)$}\\
&\Rightarrow& Q_{n+j}^1(g)=0\quad\mbox{for}\quad j=2,\ldots,m-1\ .\\
\end{array}$$

$$\begin{array}{lcl}
a_i^{n+2m-1}=1 &\Rightarrow& a_i^{n+3m-j-1}=a_i^{m-j}\\
&\Rightarrow& \sum_{w\in W(D_m)}\sgn(w).w(a_1^{n+3m-j-1}a_2^{m-2}\ldots a_{m-1})\\
&=& \sum_{w\in W(D_m)}\sgn(w).w(a_1^{m-j}a_2^{m-2}\ldots a_{m-1})\\
&=&0\ ,\ \text{by Lemma 5 using $w_1=(1j)$}\\
&\Rightarrow& Q_{n+2m-j}^1(g)=0\quad\mbox{for}\quad j=2,\ldots,m-1\ .\\
\end{array}$$

$$\begin{array}{lcl}
a_i^{n+2m-1}=1 &\Rightarrow& \sum_{w\in W(D_m)}\sgn(w).w(a_1^{n+2m-1}a_2^{m-2}\ldots a_{m-1})\\
&=& \sum_{w\in W(D_m)}\sgn(w).w(a_2^{m-2}\ldots a_{m-1})\\
&=&0\ ,\ \text{by Lemma 5 using $w_1=(1m)$}\\
&\Rightarrow& Q_{n+m}^1(g)=0\ .\\
\end{array}$$

\begin{thm}
Condition~(\ref{eq:soln}) is necessary for solutions of the
$(D_m,A_n)$ equations, provided $a_i\neq a_j^{\pm 1}$ for
$i,j=1,2,\ldots,m$ and $i\neq j$.
\end{thm}

{\bf Proof}\\
The solution is unique if and only if the determinant of the matrix
in equation~(\ref{eq:matrix}) is non-zero.

\begin{eqnarray*}
\det&=&\left|\begin{array}{cccc}
A_1 & \ldots & \ldots & A_m\\
(a_1+a_1^{-1})A_1 & \ldots & \ldots & (a_m+a_m^{-1})A_m\\
\vdots & \vdots & \vdots & \vdots\\
(a_1^{m-1}+a_1^{1-m})A_1 & \ldots & \ldots &
(a_m^{m-1}+a_m^{1-m})A_m\\
\end{array}\right|\\
&&\\ &=& \prod_{i=1}^m A_i \left|\begin{array}{cccc}
1 & \ldots & \ldots & 1\\
(a_1+a_1^{-1}) & \ldots & \ldots & (a_m+a_m^{-1})\\
\vdots & \vdots & \vdots & \vdots\\
(a_1^{m-1}+a_1^{1-m}) & \ldots & \ldots &
(a_m^{m-1}+a_m^{1-m})\end{array}\right|\\
&&\\ &=& \prod_{i=1}^m A_i \left(
\frac{\prod_{i>j}(a_i-a_j)(a_ia_j-1)}{\prod_{i=1}^m a_i^{m-1}}
\right)\\
&&\\ &=&
\left(\pm\frac{\prod_{i>j}(a_i-a_j)^{m-2}(a_ia_j-1)^{m-2}}{\prod_{i=1}^m
a_i^{(m-1)(m-2)}}\right)\left(
\frac{\prod_{i>j}(a_i-a_j)(a_ia_j-1)}{\prod_{i=1}^m
a_i^{m-1}}\right)
\end{eqnarray*}
This is non-zero provided $a_i\neq a_j$ and $a_i\neq a_j^{-1}$ for
$i\neq j$.

{\it Remark: Nahm's proof of theorem 5 also implies that $a_i\neq
a_j^{\pm1}$ is necessary for $i\neq j$.}

We now show that this particular choice of the matrix $g$ also satisfies the
equations~(\ref{eq:solve3}, \ref{eq:solve4}).

\begin{thm}
A matrix $g=\diag(a_1,a_1^{-1},\ldots,a_m,a_m^{-1})\in SO(2m)$,
whose entries satisfy $a_i^{n+2m-1}=1$ for $i=1,2,\ldots,m$, is a
solution of the equation $Q_{n+1}^m(g)=1$, provided square roots of
the $a_i$ are chosen to satisfy
\begin{equation}
\prod_{i=1}^m a_i^{\frac{n+2m-1}{2}}=
\begin{cases}
+1& \text{if\ \ $m\equiv 0\, ({\rm mod}\,4)$\ \ or\ \  $m\equiv 1\, ({\rm mod}\,4)$}\ ,\\
-1& \text{if\ \ $m\equiv 2\, ({\rm mod}\,4)$\ \ or\ \  $m\equiv 3\,
({\rm mod}\,4)$}\ .
\end{cases}
\end{equation}
\end{thm}

{\bf Proof}
\begin{eqnarray*}
&&Q_{n+1}^m(g)=1\\&&\\ &\Leftrightarrow& \sum_{w\in W(D_m)}\sgn(w).
w\left(a_1^{\frac{n+2m-1}{2}}a_2^{\frac{n+2m-1}{2}-1}\ldots
a_{m-1}^{\frac{n+2m-1}{2}-(m-2)}a_m^{\frac{n+2m-1}{2}-(m-1)}\right)\\
&=& \sum_{w\in W(D_m)} \sgn(w). w\left(a_1^{m-1} a_2^{m-2}\ldots
a_{m-2}^2a_{m-1}\right)\\
&&\\ &\Leftrightarrow& \sum_{w\in W(D_m)}\sgn(w).
w\left(a_1^{\frac{n+2m-1}{2}}\ldots
a_m^{\frac{n+2m-1}{2}}\right).w\left(a_2^{-1}a_3^{-2}\ldots
a_{m-1}^{-m+2}a_m^{-m+1}\right)\\
&=&\sum_{w\in W(D_m)} \sgn(w). w\left(a_1^{m-1} a_2^{m-2}\ldots
a_{m-2}^2a_{m-1}\right)\\
&&\\ &\Leftrightarrow& a_1^{\frac{n+2m-1}{2}}\ldots
a_m^{\frac{n+2m-1}{2}} \sum_{w\in W(D_m)} \sgn(w). w\left(a_2^{-1}
a_3^{-2}\ldots a_{m-1}^{-m+2} a_m^{-m+1}\right)\\
&=& \sum_{w\in W(D_m)} \sgn(w). w\left(a_1^{m-1} a_2^{m-2}\ldots
a_{m-2}^2a_{m-1}\right)\\
&&\\ &\Leftrightarrow& a_1^{\frac{n+2m-1}{2}}\ldots
a_m^{\frac{n+2m-1}{2}} \sum_{w\in W(D_m)} \sgn(w). w\left(a_2^{-1}
a_3^{-2}\ldots a_{m-1}^{-m+2} a_m^{-m+1}\right)\\
&=& \sum_{w\in W(D_m)} \sgn(w). \sgn(\hat{w}). w\left(a_m^{m-1}
a_{m-1}^{m-2}\ldots a_3^2a_2\right)\ ,
\end{eqnarray*}
where
\begin{equation}
\hat{w}=
\begin{cases}
(1m)(2,m-1)\ldots(\frac{m}{2},\frac{m}{2}+1)& \text{for $m$ even}\ ,\\
(1m)(2,m-1)\ldots(\frac{m-1}{2},\frac{m+1}{2})& \text{for $m$ odd}\ .
\end{cases}
\end{equation}

Clearly for the equation $Q_{n+1}^m(g)=1$ to be satisfied, we must
choose the signs $\pm\sqrt{a_i}$ such that the equation
$$\prod_{i=1}^m a_i^{\frac{n+2m-1}{2}}=sgn(\hat{w})$$
is satisfied. This amounts to choosing square root signs that
satisfy
\begin{equation}
\prod_{i=1}^m a_i^{\frac{n+2m-1}{2}}=
\begin{cases}
+1& \text{if $m\equiv\,0 \, ({\rm mod} \,4)$ or $m\equiv \,1\, ({\rm mod} \,4)$}\ ,\\
-1& \text{if $m\equiv\,2\, ({\rm mod}\, 4)$ or $m\equiv \,3\, ({\rm
mod}\, 4)$}\ . \label{eq:mod}\\
\end{cases}
\end{equation}

\begin{thm}
Let $g=\diag(a_1,a_1^{-1},\ldots,a_m,a_m^{-1})$ be a matrix whose
entries satisfy the conditions~(\ref{eq:soln}) and~(\ref{eq:mod}).
Then $g$ satisfies the equation $Q_{n+2m-1}^m=\pm1$.
\end{thm}

{\bf Proof}\\
The character $Q_{n+2m-1}^m(g)$ has numerator
\begin{eqnarray*}
&&\sum_{w\in W(D_m)}\sgn(w). w\left(\prod_{i=1}^m
a_i^{\frac{n+2m-1}{2}}\right).
w\left(a_1^{m-1}a_2^{m-2}\ldots a_{m-2}^2 a_{m-1}\right)\\
&=& \pm \sum_{w\in W(D_m)}\sgn(w). w\left(a_1^{m-1}a_2^{m-2}\ldots
a_{m-2}^2 a_{m-1}\right)\\
&=&\pm\ Q_0\ ,
\end{eqnarray*}
where $Q_0$ denotes the Weyl denominator. Hence by~(\ref{eq:mod})
\begin{equation*}
Q_{n+2m-1}^m(g)=
\begin{cases}
+1& \text{if $m\equiv\,0 \, ({\rm mod} \,4)$ or $m\equiv \,1\, ({\rm mod} \,4)$}\ ,\\
-1& \text{if $m\equiv\,2\, ({\rm mod}\, 4)$ or $m\equiv \,3\, ({\rm
mod}\, 4)$}\ .
\end{cases}
\end{equation*}

\section{Effective Central Charge Calculations}
Given a model described by a pair of Dynkin diagrams, we can
calculate the effective central charge of the corresponding
conformal field theory using the dilogarithm formulae. In this
section we carry out a detailed example of this calculation for the
$(D_3,A_2)$ case, and we summarise the results of such calculations
for numerous other cases.

\subsection{Detailed example}
To show how these calculations work we look at one particular
example in more detail. Consider the model described by the pair of
Dynkin diagrams $(D_3,A_2)$. We solve the equations $AU=V$ and use
the solutions to calculate the effective central charge (and other
values of $c-24h$) of the corresponding conformal field theory.

The equations of this model are $AU=V$ where
$A=C(D_3)^{-1}\otimes\,C(A_2)$, \linebreak $U=\log(x)$,
$V=\log(1-x)$, and $x=(x_{11},x_{12},x_{21},x_{22},x_{31},x_{32})$.
Exponentiation leads to a set of algebraic equations
\begin{equation}
x^A=1-x\ . \label{eq:model}
\end{equation}
By the change of variable
\begin{equation}
x=z^{-C(D_3)\otimes I_2}\ , \label{eq:cov}
\end{equation}
the equations~(\ref{eq:model}) can be written as
\begin{equation*}
z^{(2-C(D_3))\otimes\,I_2}+z^{I_3 \otimes\,(2-C(A_2))}=z^2\ ,
\end{equation*}
or more explicitly as
\begin{equation}
(z_{ij})^2 = z_{ij^{\star}}+
\begin{cases}
z_{2j}z_{3j} & \mathrm{if}\  i=1\ ,\\
z_{1j} & \mathrm{if}\ i=2,3\ ,
\end{cases}
\label{eq:modelz}
\end{equation}
where $z=(z_{11},z_{12},z_{21},z_{22},z_{31},z_{32})$. Here $\star$
denotes interchanging the indices 1 and 2 (i.e. $j^{\star}=3-j$).
The first step is to solve these equations. The $z_{ij}$ are the
characters $Q_j^i(g)$ of representations of the Yangian $Y(D_3)$,
for some specially chosen matrix $g\in SO(6)$.

\subsubsection{Solution} Choose the matrix
$g=\diag(a_1,a_1^{-1},a_2,a_2^{-1},a_3,a_3^{-1})\in SO(6)$ so that
the entries satisfy
$$a_i^7=1, \mbox{ with } a_i \neq a_j^{\pm1} \mbox{ for } i\neq j\ .$$
As discussed earlier the square root signs $\pm\sqrt{a_i}$ must be
carefully chosen in order to satisfy the equations
$a_1^{\frac{7}{2}}a_2^{\frac{7}{2}}a_3^{\frac{7}{2}}=-1$. We rule
out any choice of $g$ that does not satisfy $Q_j^i(g)\neq 0$ for
$j=1,\ldots,n$ (these are the non-admissable solutions mentioned
earlier). This results in four choices for the matrix $g$ (up to
permutation).
$$(a_1,a_2,a_3)=
\left\{\begin{array}{l}
\left(1,e^{\frac{2\pi i}{7}},e^{\frac{4\pi i}{7}}\right)\ ,\\
\left(1,e^{\frac{4\pi i}{7}},e^{\frac{6\pi i}{7}}\right)\ ,\\
\left(1,e^{\frac{2\pi i}{7}},e^{\frac{6\pi i}{7}}\right)\ ,\\
\left(e^{\frac{2\pi i}{7}},e^{\frac{4\pi i}{7}}, e^{\frac{6\pi
i}{7}} \right)\ .\end{array}\right.
$$

\subsubsection{Characters} The three fundamental Yangian characters
are written in terms of entries of $g$ as
$$\begin{array}{rcl}
Q_1^1(g) &=& a_1+a_2+a_3+a_1^{-1}+a_2^{-1}+a_3^{-1}\ ,\\
Q_1^2(g) &=& a_1^{-1/2}a_2^{-1/2}a_3^{-1/2} +
a_1^{-1/2}a_2^{1/2}a_3^{1/2}+a_1^{1/2}a_2^{-1/2}a_3^{1/2}+a_1^{1/2}a_2^{1/2}a_3^
{-1/2}\ ,\\
Q_1^3(g)&=&a_1^{1/2}a_2^{1/2}a_3^{1/2}+a_1^{-1/2}a_2^{-1/2}a_3^{1/2}
+a_1^{1/2}a_2^{-1/2}a_3^{-1/2}+a_1^{-1/2}a_2^{1/2}a_3^{-1/2}\ .
\end{array}$$
We use the Weyl character formula to compute these characters. The
corresponding values of $Q_2^1(g)$, $Q_2^2(g)$ and $Q_2^3(g)$ can
then be calculated using the Kirillov-Reshetikhin equations. The KR
equations for $D_3$ are
\begin{eqnarray*}
(Q_j^1)^2-Q_{j-1}^1Q_{j+1}^1&=&Q_j^2Q_j^3\ ,\\
(Q_j^2)^2-Q_{j-1}^2Q_{j+1}^2&=&Q_j^1\ ,\\
(Q_j^3)^2-Q_{j-1}^3Q_{j+1}^3&=&Q_j^1\ .
\end{eqnarray*}
The identification $z_{ij}=Q_j^i(g)$ means that we have now found the solutions
$z=(z_{11},z_{12},z_{21},z_{22},z_{31},z_{32})$ of the equations
$z^{(2-C(D_3))\otimes\,I_2}+z^{I_3 \otimes\,(2-C(A_2))}=z^2$.

We compute the values of $x_{ij}$ using the relation
$$x=z^{-C(D_3)\otimes I_2} \quad
\Rightarrow\quad\left\{\begin{array}{rcl}
x_{11} & = & z_{11}^{-2}z_{21}z_{31}\ ,\\
x_{12} & = & z_{12}^{-2}z_{22}z_{32}\ ,\\
x_{21} & = & z_{11}z_{21}^{-2}\ ,\\
x_{22} & = & z_{12}z_{22}^{-2}\ ,\\
x_{31} & = & z_{11}z_{31}^{-2}\ ,\\
x_{32} & = & z_{12}z_{32}^{-2}\ .
\end{array}\right.$$
The logarithms of these solutions must be chosen so as to satisfy
$$\log(x)=(-C(D_3)\otimes I_2)\log(z)\ ,$$
which is equivalent to the equations
$$\left\{\begin{array}{rcl}
u_{11}=\log(x_{11}) & = & -2\log(z_{11})+\log(z_{21})+\log(z_{31})\ ,\\
u_{12}=\log(x_{12}) & = & -2\log(z_{12})+\log(z_{22})+\log(z_{32})\ ,\\
u_{21}=\log(x_{21}) & = & \log(z_{11})-2\log(z_{21})\ ,\\
u_{22}=\log(x_{22}) & = & \log(z_{12})-2\log(z_{22})\ ,\\
u_{31}=\log(x_{31}) & = & \log(z_{11})-2\log(z_{31})\ ,\\
u_{32}=\log(x_{32}) & = & \log(z_{12})-2\log(z_{32})\ ,
\end{array}\right.$$
and
$$\log(1-x)=(-I_3\otimes C(A_2))\log(z)\ ,$$
which is equivalent to $$\left\{\begin{array}{rcl}
v_{11}=\log(1-x_{11}) & = & -2\log(z_{11})+\log(z_{12})\ ,\\
v_{12}=\log(1-x_{12}) & = & \log(z_{11})-2\log(z_{12})\ ,\\
v_{21}=\log(1-x_{21}) & = & -2\log(z_{21})+\log(z_{22})\ ,\\
v_{22}=\log(1-x_{22}) & = & \log(z_{21})-2\log(z_{22})\ ,\\
v_{31}=\log(1-x_{31}) & = & -2\log(z_{31})+\log(z_{32})\ ,\\
v_{32}=\log(1-x_{32}) & = & \log(z_{31})-2\log(z_{32})\ .
\end{array}\right.$$

Then the pairs
$(u_{jk},v_{jk})\equiv\left(\log(x_{jk}),\log(1-x_{jk})\right)$ are
the solutions of the equations $AU=V$.

In this particular example there are four solutions of the equations
$x^A=1-x$. We label these $x^i=(x_{11}^i,\ldots,x_{32}^i)$ for
$i=0,1,2,3$. We calculate the values of $c-24h_i$ using the formula
$$c-24h_i=\frac{6}{\pi^2}\sum_{jk=11,\ldots,32}
L(u_{jk}^i,v_{jk}^i)\ .$$ For any given model, the effective central
charge, $c_{\eff}$, is the value of $c-24h_i$ arising from the
unique solution $x^0$, whose components $x_{jk}^0\in\mathbb{R}$ all
satisfy $0<x_{jk}^0<1$.

For the case $(D_3,A_2)$ the four choices of the matrix $g$ give rise to four
different values of $c-24h_i$. These are
$$
\begin{array}{lcl}
g=\left(1,e^{\frac{2\pi i}{7}},e^{\frac{4\pi i}{7}}\right)
&\Rightarrow&c_{\eff}=\frac{24}{7}\ ,\\
g=\left(1,e^{\frac{4\pi i}{7}},e^{\frac{6\pi i}{7}}\right)
&\Rightarrow&c-24h_1=-\frac{72}{7}\mod24\mathbb{Z}\ ,\\
g=\left(1,e^{\frac{2\pi i}{7}},e^{\frac{6\pi i}{7}}\right)
&\Rightarrow&c-24h_2=-\frac{120}{7}\mod24\mathbb{Z}\ ,\\
g=\left(e^{\frac{2\pi i}{7}},e^{\frac{4\pi i}{7}},
e^{\frac{6\pi i}{7}} \right) &\Rightarrow& c-24h_3=0\mod24\mathbb{Z}\ .\\
\end{array}
$$

\subsection{Effective central charge for other models}
These calculations have been carried out for many different models.
The results are summarised in the following table:
\begin{center}
\begin{tabular}{|c|c|c|c|c|}\hline
Pair  & $c_{\eff}$ & $c-24h_1$ & $c-24h_2$ & $c-24h_3$\\\hline
$(A_1,A_1)$ & $1/2$ &-&-&-\\
$(A_1,A_2)$ & $4/5$ & $-44/5$ &-&-\\
$(A_1,A_3)$ & $1$ &-&-&-\\
$(A_1,A_4)$ & $8/7$ & $32/7$ & $-40/7$&-\\
$(A_2,A_1)$ & $6/5$ & $-54/5$ &-&-\\
$(A_2,A_2)$ & $2$ & - &-&-\\
$(A_2,A_3)$ & $18/7$ & $6$ & $162/7$&-\\
$(A_2,A_4)$ & $3$ & $9$ & $27$ & $-9$\\
$(A_3,A_1)$ &$2$ &-&-&-\\
$(A_3,A_2)$ &$24/7$ & $0$ & $-72/7$ & $-120/7$\\
$(A_3,A_3)$ &$9/2$ &-&-&-\\
$(D_3,A_1)$ &$2$ &-&-&-\\
$(D_3,A_3)$ &$9/2$ &-&-&-\\
$(D_4,A_1)$ &$3$ &-&-&-\\
$(D_4,A_2)$ &$16/3$ & $-32/3$ & $-16$ & $-307/3$\\
$(D_4,A_3)$ & $36/5$ & $24/5$ &-&-\\\hline
\end{tabular}
\end{center}

\section{$(D_m,A_n)$ as coset models}
Consider the model described by the pair of Dynkin diagrams $(X,Y)$.
Its effective central charge is known or conjectured to be
$$c_{\eff}(X,Y)=\frac{r(X)r(Y)h(X)}{h(X)+h(Y)}\ ,$$
where $r$ denotes the rank, and $h$ the dual Coxeter number of a Lie
algebra. Then for the case $(D_m,A_n)$ we expect the effective central
charge to be
$$c_{\eff}(D_m,A_n)=\frac{(2m-2)mn}{2m+n-1}\ .$$

Now consider the coset model
\begin{equation}
\frac{(D_m)_{n+1}}{u(1)^m}\ .\label{eq:coset}
\end{equation}
This model has central charge
$$\frac{(n+1)(2m^2-m)}{n+1+2m-2}-m=\frac{(2m-2)mn}{2m+n-1}\ .$$
Clearly this coincides with the value of $c_{\eff}(D_m,A_n)$.

This is extremely good evidence to suggest that the model described
by the Dynkin diagrams $(D_m,A_n)$ is a unitary model
described by the coset
$$\frac{(D_m)_{n+1}}{u(1)^m}\ .$$
In all cases checked so far, the h-values calculated from the dilog formulae~(\ref{eq:chi},~\ref{eq:ch0})
for the model $(D_m,A_n)$ also arise as h-values of the coset model~(\ref{eq:coset}).

\chapter{Nahm's Conjecture} In the previous chapter we studied a family of matrices related to
pairs of Dynkin diagrams. Each matrix $A$ had the special property
that solutions of $A\log x=\log(1-x)$ gave torsion elements of the
Bloch group. This lead to a nice relationship with conformal field
theory. In this chapter we consider a number of $2\times2$ matrices
that don't fall into this Cartan matrix pattern but nevertheless
yield torsion elements of the Bloch group, at least for the special
`minimal' solution. We show that these matrices are again related to
conformal field theory.

\section{Overview}
A {\bf q-hypergeometric series} is a series of the form
$\sum_{n=0}^{\infty}A_n(q)$, where $A_0(q)$ is a rational function,
and $A_n(q)=R(q,q^n)A_{n-1}(q)$ for all $n\geq 1$ for some rational
function $R(x,y)$ with $\lim_{x\rightarrow 0}\lim_{y\rightarrow
0}R(x,y)=0$.

The question of when a q-hypergeometric series is also modular is an
interesting open question in mathematics. There are a handful of
known examples of such series, the most famous ones being given by
the Rogers-Ramanujan identities
\begin{eqnarray*}
\sum_{n=0}^{\infty}\frac{q^{n^2}}{(q)_n} &=& \prod_{n\,\equiv\,\pm
1\,{\rm mod}\,5}\frac{1}{1-q^n} \qquad (|q|<1)\ ,\\ &&\\
\sum_{n=0}^{\infty}\frac{q^{n^2+n}}{(q)_n} &=& \prod_{n\,\equiv\,\pm
2\,{\rm mod}\,5}\frac{1}{1-q^n} \qquad (|q|<1)\ .
\end{eqnarray*}
These are modular functions up to factors $q^{-1/60}$ and
$q^{11/60}$ respectively.

In general, the problem of understanding the overlap between
q-hypergeometric series and modular functions is still completely
unsolved. Nahm's conjecture takes a first step towards tackling this
problem by considering a special case involving r-fold
hypergeometric series. (These are defined as above but with $n$
running over $(\mathbb{Z}_{\geq0})^r$ rather than
$\mathbb{Z}_{\geq0}$).

Let $A$ be a positive definite symmetric $r\times r$ matrix, $B$ a
vector of length r, and $C$ a scalar, all three with rational
coefficients. Define a function $f_{A,B,C}$ by the r-fold
q-hypergeometric series
$$f_{A,B,C}(z)=\sum_{n=(n_1,\ldots,n_r)\in(\mathbb{Z}_{\geq0})^r}
\frac{q^{\frac{1}{2}nAn^t+Bn+C}}{(q)_{n_1}\ldots (q)_{n_r}}\ ,$$
where $(q)_n=(1-q)(1-q^2)\ldots(1-q^n)$. We can ask the question,
when is $f_{A,B,C}$ a modular function? Nahm's conjucture doesn't
fully answer this question, but predicts which matrices $A$ can
occur.

Given any positive definite symmetric $r\times r$ matrix
$A=(A_{ij})$ with real coefficients, we can consider the system of
equations
\begin{equation}
x_i=\prod_{j=1}^{r}(1-x_j)^{A_{ij}}\ . \label{eq:xeqn}
\end{equation}

This system has a finite number of solutions $x=(x_1,\ldots,x_r)$.
Again the unique solution whose components are all real and between
0 and 1 is denoted by $x^0=(x_1^0,\ldots,x_r^0)$.\\

{\it Note: The related set of equations
$1-x_i=\prod_{j=1}^{r}{x_j}^{A_{ij}}$ corresponds to the set of
equations~(\ref{eq:xeqn}) with $A$ replaced by $A^{-1}$. There is a
duality between these two cases. In particular the effective central
charges are related by $$c_{\eff}(A)+c_{\eff}(A^{-1})=r\ .$$}

Let $F$ denote the number field $\mathbb{Q}(x_1,\ldots,x_r)$. Given
any solution $x=(x_1,\ldots,x_r)$ of~(\ref{eq:xeqn}), define the
element $\xi_x\in \mathbb{Z}(F)$ by $\xi_x=[x_1]+\ldots+[x_r]$.
$\xi_x$ in an element of the Bloch group $\mathcal{B}(F)$.

\newtheorem{conj}{Conjecture}
\begin{conj}{\bf (Nahm's Conjecture)}
Let $A$ be a positive definite symmetric $r\times r$ matrix with
rational coefficients. Then the following are equivalent:
\begin{enumerate}
\item{The element $\xi_x$ is a torsion element of $\mathcal{B}(F)$
for all solutions $x=(x_1,\ldots,x_r)$.}
\item{There exist $B\in\mathbb{Q}^r$ and $C\in\mathbb{Q}$ such that
$f_{A,B,C}(z)$ is a modular function.}
\end{enumerate}
\end{conj}
For the case $r=1$ this conjecture is proved in \cite{DZ}. It is
expected that modular functions $f_{A,B,C}$ that arise in this way
are characters of certain rational conformal field theories. We show
later that this is certainly true in at least one case.

\section{Terhoeven's Matrices}
In his PhD thesis \cite{MT}, Michael Terhoeven looked for rational
$2\times2$ matrices for which $\xi_{x^0}$ was a torsion element of
the Bloch group. To do this he carried out a systematic search of
all $2\times2$ matrices of the form
$$A=\frac{1}{m}\left(\begin{array}{cc}a&b\\b&c\\ \end{array}\right) \in
M_2(\mathbb{Q})\ ,$$ with $a,b,c,m\leq 11$. Due to time constraints
we study only nine of Terhoeven's matrices:
$$\begin{array}{ccc}
A = \left(\begin{array}{cc} 11&9\\9&8 \end{array}\right)& A =
\left(\begin{array}{cc} 8&5\\5&4 \end{array}\right)&
A = \left(\begin{array}{cc} 4&3\\3&3 \end{array}\right)\\
&&\\ A = \left(\begin{array}{cc} 8&3\\3&2 \end{array}\right)& A =
\frac{1}{2}\left(\begin{array}{cc} 5&4\\4&4 \end{array}\right)&
A = \frac{1}{3}\left(\begin{array}{cc} 8&1\\1&2 \end{array}\right)\\
&&\\ A = \frac{1}{9}\left(\begin{array}{cc} 8&3\\3&0
\end{array}\right)& A = \left(\begin{array}{cc} 4&1\\1&1
\end{array}\right)&
A = \frac{1}{2}\left(\begin{array}{cc} 1&1\\1&0 \end{array}\right)\\
\end{array}$$
We chose to look at this particular subset of Terhoeven's matrices
since less is known about them than about the remaining matrices. In
particular, for his other matrices it was already known either that
all solutions gave rise to torsion elements or that all solutions
gave rise to non-torsion elements. These nine matrices are
particularly interesting in that they give rise to a mixture of
torsion and non-torsion elements.

A similar search carried out by Don Zagier, with $a,b,c,m\leq100$,
resulted in one further matrix, namely
$$A=\left(\begin{array}{cc} 24&19\\19&16 \end{array}\right)\ .$$
We don't repeat the calculations in this case as they are more
lengthy and it is already known~\cite{DZ} that this matrix doesn't
satisfy the stronger condition of having all solutions
of~(\ref{eq:xeqn}) being torsion.

Notice that the matrices
$$A = \frac{1}{9}\left(\begin{array}{cc} 8&3\\3&0
\end{array}\right) \quad \mathrm{and} \quad
A = \frac{1}{2}\left(\begin{array}{cc} 1&1\\1&0
\end{array}\right)\ ,$$ are not positive definite. We have included
the calculations carried out in these cases although it is not yet
clear exactly how they fit with Nahm's conjecture.

For each of these matrices we solve the equations $x=(1-x)^A$, and
in cases where $\xi_x$ is a torsion element of the Bloch group for
all solutions $x$, we calculate the values of B and C for which
$f_{A,B,C}$ is modular.

{\it Note: For simplicity we solve the equations $x^A=1-x$ instead
of $x=(1-x)^A$ in some cases.  To recover the desired solutions we
need only replace $x$ by $1-x$.}

\subsection{Equations}
\fbox{$A=\left(\begin{array}{cc} 11&9\\9&8 \end{array}\right)$}
\vspace*{5mm}\\
We get the following set of equations in two variables
\begin{eqnarray*}
1-x_1 &=& x_1^{11}x_2^9\ ,\\
1-x_2 &=& x_1^9x_2^8\ .
\end{eqnarray*}
These reduce to the one-variable equation
$$x^{77}(x_1^8-x_1^7+8x_1^6-9x_1^5-x_1^4+2x_1^3+3x_1^2-3x_1+1)(x_1^4-8x_1^3+3x_1^2+2x_1+1)=0\ .$$
$\xi_x$ is not a torsion element of $\mathcal{B}(F)$ for all
solutions $x$. Let $x_1$ be any solution of the equation
$$x_1^8-x_1^7+8x_1^6-9x_1^5-x_1^4+2x_1^3+3x_1^2-3x_1+1=0\ .$$ Then for
all the corresponding pair $(x_1,x_2)$ we have checked that
$D(x_1)+D(x_2)\neq 0$ in each case. Hence $\xi_x$ is not torsion for
these particular solutions. The solution $x_1=0$ is again torsion,
although somewhat trivial since it just leads to the pair
$(x_1,x_2)=(0,1)$ which gives $L(\xi_x)=L(1)$. Although we don't
always refer to it specifically, this trivial solution appears in
many of the cases below.

In the case where $x_1$ is a solutions of
$x_1^4-8x_1^3+3x_1^2+2x_1+1=0$, we have checked numerically that the
corresponding pair $(x_1,x_2)$ satisfy $D(x_1)+D(x_2)=0$. This
suggests that $\xi_x$ is torsion for these four solutions.
This second factor has Galois group $D_4$.\\

\fbox{$A=\left(\begin{array}{cc} 8&5\\5&4 \end{array}\right)$}
\vspace*{5mm}\\
We get the following set of equations in two variables
\begin{eqnarray*}
1-x_1 &=& x_1^8x_2^5\ ,\\
1-x_2 &=& x_1^5x_2^4\ .
\end{eqnarray*}
These reduce to the one-variable equation
$$x^{24}(x_1^4-x_1^3+3x_1^2-3x_1+1)(x_1^4+x_1^3+3x_1^2-3x_1-1)=0\ .$$
$\xi_x$ is not a torsion element of $\mathcal{B}(F)$ for all of the
solutions. In particular we showed that $\xi_x$ is not torsion for
solutions that annihilate the first factor (i.e. $D(\xi_x)\neq 0$).
We checked numerically that solutions annihilating the second factor
give rise to $\xi_x$ that are torsion ($D(\xi_x)=0$ numerically).
This second factor has Galois
group $D_4$.\\

\fbox{$A=\left(\begin{array}{cc} 4&3\\3&3
\end{array}\right)$}
\vspace*{5mm}\\
We get the following set of equations in two variables
\begin{eqnarray*}
1-x_1 &=& x_1^4x_2^3\ ,\\
1-x_2 &=& x_1^3x_2^3\ .
\end{eqnarray*}
These reduce to the one-variable equation
$$x^8(4x_1^2-2x_1-1)(2x_1^2-2x_1+1)=0\ .$$
We checked numerically that $\xi_x$ is a torsion element of
$\mathcal{B}(F)$ for the two solutions that annihilate the first
factor, and proved that $\xi_x$ is not torsion for the remaining
solutions. The first factor has Galois group $S_2$.\\

\fbox{$A=\left(\begin{array}{cc} 8&3\\3&2
\end{array}\right)$}
\vspace*{5mm}\\
We get the following set of equations in two variables
\begin{eqnarray*}
1-x_1 &=& x_1^8x_2^3\ ,\\
1-x_2 &=& x_1^3x_2^2\ .
\end{eqnarray*}
These reduce to the one-variable equation
$$x^8(x_1^2-x_1+1)^2(x_1^4+2x_1^3+x_1^2-2x_1-1)=0\ .$$
We checked numerically that $\xi_x$ is a torsion element of
$\mathcal{B}(F)$ for the four solutions that annihilate the second
factor. For the first factor, if $x_1$ is a root of $x_1^2-x_1+1$,
then $(x_1,x_2)=(x,x)$ and $(x,1/x)$ solve the set of two-variable
equations. The element $(x,x)$ is not torsion since $2D(x)\neq0$,
but the element $(x,1/x)$ seems to be torsion. The second
factor has Galois group $D_4$.\\

\fbox{$A=1/2\left(\begin{array}{cc} 5&4\\4&4
\end{array}\right)$}
\vspace*{5mm}\\
We get the following set of equations in two variables
\begin{eqnarray*}
1-x_1 &=& x_1^{5/2}x_2^2\ ,\\
1-x_2 &=& x_1^2x_2^2\ .
\end{eqnarray*}
These reduce to the one-variable equation
$$x^{5/2}(1-x_1+2x_1^2-2x_1^3-x_1^{3/2}+x_1^{5/2}-x_1^{7/2})=0\ .$$
Since solving the equations in this case requires a careful choice
of square roots, it is perhaps more appropriate to consider this
equation as a polynomial of degree 7 in the variable $y=x_1^{1/2}$.
The resulting equation is $$y^5(y^3-y+1)(y^4+2y^3-y-1)=0\ .$$ Again
$\xi_x$ is not torsion for all solutions. In particular $\xi_x$ is
not torsion for those solutions that annihilate the first factor,
but (numerically) seems to be torsion for solutions that annihilate
the second factor. The Galois group of the second factor is $D_4$.\\

\fbox{$A=1/3\left(\begin{array}{cc} 8&1\\1&2
\end{array}\right)$}
\vspace*{5mm}\\
We get the following set of equations in two variables
\begin{eqnarray*}
1-x_1 &=& x_1^{8/3}x_2^{1/3}\ ,\\
1-x_2 &=& x_1^{1/3}x_2^{2/3}\ .
\end{eqnarray*}
These reduce to the one-variable equation
$$(x_1^2-x_1+1)(x_1^6+x_1^5-2x_1^3+2x_1-1)=0\ .$$
$\xi_x$ is (numerically) a torsion element for six of the eight
solutions, namely the six that annihilate the second factor. The
Galois group of the second factor is $A_4\times C_2$.\\

\fbox{$A=1/9\left(\begin{array}{cc} 8&3\\3&0
\end{array}\right)$}
\vspace*{5mm}\\
We get the following set of equations in two variables
\begin{eqnarray*}
1-x_1 &=& x_1^{8/9}x_2^{1/3}\ ,\\
1-x_2 &=& x_1^{1/3}\ .
\end{eqnarray*}
These reduce to the one-variable equation
$$x_2(x_2^3-2x_2^2+x_2+1)(x_2^4-6x_2^3+12x_2^2-9x_2+1)=0\ .$$
$\xi_x$ is (numerically) a torsion element for four of the seven
non-zero solutions, the four annihilating the second factor. This
factor has Galois group $D_4$.

\fbox{$A=\left(\begin{array}{cc} 4&1\\1&1
\end{array}\right)$}
\vspace*{5mm}\\
We get the following set of equations in two variables
\begin{eqnarray*}
1-x_1 &=& x_1^4x_2\ ,\\
1-x_2 &=& x_1x_2\ .
\end{eqnarray*}
These reduce to the one-variable equation
$$x_1^4+x_1^2-1=0\ ,$$
whose Galois group is again $D_4$. We checked numerically that
$\xi_x$ is a torsion element of $\mathcal{B}(F)$ for all solutions.\\

\fbox{$A=1/2\left(\begin{array}{cc} 1&1\\1&0
\end{array}\right)$}
\vspace*{5mm}\\
We get the following set of equations in two variables
\begin{eqnarray*}
1-x_1 &=& x_1^{1/2}x_2^{1/2}\ ,\\
1-x_2 &=& x_1^{1/2}\ .
\end{eqnarray*}
These reduce to the one-variable equation
$$x_2(x_2^3-5x_2^2+6x_2-1)=0\ .$$
$\xi_x$ is a (numerically) torsion element of $\mathcal{B}(F)$ for
all solutions. Ignoring the trivial factor (whose Galois group is
$S_1$), this
equation has Galois group $A_3$.\\

{\it Note: All Galois group calculations were carried out using the
programme Pari. This can be downloaded free from
http://pari.math.u-bordeaux.fr/.}

\subsection{Torsion elements of the Bloch group}
In the context of Nahm's conjecture, the two cases of greatest
interest are $A=\left(\begin{array}{cc} 4&1\\1&1
\end{array}\right)$ and $A=1/2\left(\begin{array}{cc} 1&1\\1&0
\end{array}\right)$. In these cases {\bf all}
solutions of the equations $x^A=1-x$ yield torsion elements of the
Bloch group, whereas for the other matrices $A$ considered above,
only some of the solutions yield torsion elements. Hence the
conjecture suggests that, for these two values of $A$, there exist
values $B\in\mathbb{Q}^2$ and $C\in\mathbb{Q}$ such that the
function $f_{A,B,C}(z)$ is modular. We now proceed to calculate
these $B$ and $C$ values.

Notice that the matrix $A=1/2\left(\begin{array}{cc} 1&1\\1&0
\end{array}\right)$ is not in fact positive definite. For this
reason we are not entirely sure how it fits in with Nahm's
conjecture. However, since all solutions yield torsion elements of
the Bloch group, this case is clearly worth examining in more
detail.

\subsection{$B$ and $C$ values}
Suppose $A$ is an $r\times r$ matrix. Then in the individual terms
defining $f_{A,B,C}$, the denominators involve finite sub-products
of $\eta(z)^r$ where, (for $q=e^{2\pi iz}$),
$$\eta(z) = q^{\frac{1}{24}}\prod_{n=1}^{\infty}\left(1-q^n\right)\ .$$
Hence, for a $2\times 2$ matrix $A$, we
can expect $\eta(z)^2 f_{A,B,0}(z)$ to be a holomorphic modular
form. For values of $B$ giving rise to modular $f_{A,B,C}$, the
series $\eta(z)^2 f_{A,B,0}(z)$ should have small coefficients,
whereas for other $B$ the coefficients will be much bigger. By
considering the coefficients in this series for different values of
$B$, it is possible to identify those $B$-values that give rise to
modular $f_{A,B,C}$.

For example, consider the matrix $A=\left(\begin{array}{cc} 4&1\\1&1
\end{array}\right)$. $B=(0,1/2)$ leads to modular $f_{A,B,C}$, while
$B=(0,0)$ does not lead to modular $f_{A,B,C}$. In each case let us
look at the coefficients of $q^{25}$, $q^{50}$, $q^{75}$, $q^{100}$
in the series $\eta(z)^2 f_{A,B,0}(z)$ (divided by $q^{1/12}$ for
simplicity). $B=(0,1/2)$ gives $1q^{25}$, $1q^{50}$, $3q^{75}$, and
$5q^{100}$, while $B=(0,0)$ gives $-3q^{25}$, $-20q^{50}$,
$171q^{75}$, and $110q^{100}$. Clearly the coefficients in the
non-modular case grow significantly more quickly than those in the
modular case.

Furthermore, as a useful check, we expect that any B-value that
leads to modular $f_{A,B,0}$, should give rise to a rational value
of C.

Given the vector $B=(B_1,B_2)$, $C$ can then be calculated \cite{MT}
using the equation:
\begin{eqnarray*}
C&=&\frac{\phi_2(B_i)}{2}\left(\frac{1-x_i}{x_i}\right)-\frac{1}{2}\phi_1(B_i)
\left(\frac{1-x_i}{x_i}\right)T_{ij}^{-1}\phi_1(B_j)\left(\frac{1-x_j}{x_j}
\right)\\
&&+\frac{1}{2}\phi_1(B_i)\left(\frac{1-x_i}{(x_i)^2}\right)T_{ij}^{-1}-
\frac{1}{2}\left(\frac{1-x_i}{(x_i)^2}\right)T_{ii}^{-1}T_{ij}^{-1}\phi_1(B_j)
\left(\frac{1-x_j}{x_j}\right)\\
&&+\frac{1}{8}\left(\frac{(1-x_i)(2-x_i)}{(x_i)^3}
\right)(T_{ii})^{-1}\\
&&-\frac{1}{12}\left(\frac{1-x_i}{(x_i)^2}\right)(T_{ij}^{-1})^3
\left(\frac{1-x_j}{(x_j)^2}\right)\\
&&-\frac{1}{8}\left(\frac{1-x_i}{(x_i)^2}
\right)T_{ii}^{-1}T_{ij}^{-1}T_{jj}^{-1}\left(\frac{1-x_j}{(x_j)^2}\right)\
.
\end{eqnarray*}
Here $\phi_i$ denotes the $i^{th}$ Bernoulli polynomial, and the
matrix $T=(T_{ij})$ is given by
$$T_{ij}=(A)_{ij}^{-1}+\delta_{ij}\left(\frac{1-x_i}{x_i}\right)\ .$$

For the matrix $A=(4,1,1,1)$, we find that the values of $B$ and $C$
for which $f_{A,B,C}(z)$ is modular are
$$\begin{array}{ll}
B=(0,1/2)\ ,& C=1/120\ ,\\
B=(2,1/2)\ ,& C=49/120\ .
\end{array}$$
These values agree with the calculations of Don Zagier \cite{DZ}.

For the matrix $A=1/2(1,1,1,0)$, the corresponding $B$ and $C$
values are
$$\begin{array}{ll}
B=(0,1/2)\ ,& C=-1/84\ ,\\
B=(1/2,1/2)\ ,& C=5/84\ ,\\
B=(0,1)\ ,& C=-1/21\ .
\end{array}$$

\subsection{$f_{A,B,C}$ as characters of rational CFTs}
For the case $A=\left(\begin{array}{cc} 4&1\\1&1
\end{array}\right)$ we calculate the functions
$f_{A,B,C}$ explicitly. Up to order 6 these are
\begin{eqnarray*}
f_{A,\,\left(2,\frac{1}{2}\right),\frac{49}{120}} &=&
q^{\frac{49}{120}}\left(1+q+q^2+2q^3+3q^4+4q^5+6q^6+\ldots\right)\ ,\\
f_{A,\,\left(0,\frac{1}{2}\right),\frac{1}{120}} &=&
q^{\frac{1}{120}}\left(1+q+2q^2+3q^3+4q^4+6q^5+8q^6+\ldots\right)\ .
\end{eqnarray*}
These expansions seem to be the same as the characters of the
$(5,4)$-minimal model given on page 243 of~\cite{diF}. This leads us
to expect that the matrix $A=\left(\begin{array}{cc} 4&1\\1&1
\end{array}\right)$ describes some integrable perturbation of the
$(5,4)$-minimal model.

Recall that a perturbation of a CFT gives rise to a theory that, in
general, is no longer conformally invariant. However, there are some
cases in which the new theory turns out to be integrable, meaning
that it has an infinite number of conservation laws, and hence a
much greater chance of being solved exactly. So, what we assume
above is that the matrix $A=\left(\begin{array}{cc} 4&1\\1&1
\end{array}\right)$ describes some theory that arises as a perturbation
of the $(5,4)$-minimal model. As mentioned earlier, the scattering
matrix of any quantum field theory provides important insight into
the theory itself. We expect that, using the matrix
$A=\left(\begin{array}{cc} 4&1\\1&1 \end{array}\right)$, it should
be possible to construct the scattering matrix of the corresponding
integrable theory. Having done this, we could then hope to be able
to identify the particular integrable perturbation. It would be
interesting to explore this connection in more detail at a later
stage.

For the case $A=1/2(1,1,1,0)$ the functions $f_{A,B,C}$ are more
complicated. They are

\begin{eqnarray*}
f_{A,\,\left(0,\frac{1}{2}\right),-\frac{1}{84}} &=&
q^{-\frac{1}{84}}\left((1+2q+3q^2+6q^3+10q^4+16q^5+\ldots)\right.\\
&&+ q^{\frac{1}{4}}(1+2q+5q^2+8q^3+15q^4+24q^5+\ldots)\\
&&+\left.q^{\frac{1}{2}}(1+2q+4q^2+7q^3+12q^4+19q^5+\ldots)\right)\ ,\\
f_{A,\left(\frac{1}{2},\frac{1}{2}\right),\frac{5}{84}} &=&
q^{\frac{5}{84}}\left((1+q+3q^2+5q^3+8q^4+13q^5+\ldots)\right.\\
&&+ q^{\frac{1}{2}}(1+2q+3q^2+6q^3+10q^4+16q^5+\ldots)\\
&&+\left.q^{\frac{3}{4}}(1+2q+4q^2+8q^3+13q^4+22q^5+\ldots)\right)\ ,\\
f_{A,\left(0,1\right),-\frac{1}{21}} &=&
q^{-\frac{1}{21}}\left((1+2q+3q^2+6q^3+10q^4+16q^5+\ldots)\right.\\
&&+ q^{\frac{1}{4}}(1+q+2q^2+3q^3+5q^4+8q^5+\ldots)\\
&&+\left.q^{\frac{3}{4}}(q+2q^2+3q^3+6q^4+9q^5+\ldots)\right)\ .
\end{eqnarray*}
These expansions seem to have some modular properties, and indeed
the fact that the B-values gave rise to rational C-values is good
evidence in favour of this theory. As yet we have not managed to
find a conformal field theory with these characters, nor even
managed to identify the expansions themselves as modular forms.
However, we are optimistic that with more work these expansions can
be identified as the characters of some conformal field theory.

\subsection{Effective central charge calculations}
We have already seen that, given a solution $x=(x_1,\ldots,x_r)$
of~(\ref{eq:xeqn}) for which $\xi_x$ is a torsion element of the
Bloch group, we can apply the mapping
$$\frac{6}{\pi^2}\sum_{i=1}^r L(x_i)=c-24h\mod24\ .$$
This gives some information about the corresponding conformal field
theory.

Since $\xi_{x^0}$ is a torsion element of the Bloch group for each
of Terhoeven's matrices (and their inverses), we can calculate the
corresponding value of the effective central charge in each case.
Furthermore we can calculate $c-24h$ (mod $24$) for any solution $x$
of~(\ref{eq:xeqn}) for which $\xi_x$ is a torsion element. The
results of some such calculations are summarised below.
\begin{center}
\begin{tabular}{|c|c|c|c|}\hline $A^{-1}$ & $c_{\eff}$& $c-24h_1$ & $c-24h_2$\\
\hline $(4,1,1,1)^{-1}$ & $7/10$ & $127/10$ & $343/10$\\
$1/2(1,1,1,0)$ & $6/7$ & $150/7$ & $54/7$\\
$(11,9,9,8)^{-1}$ & $3/10$ & $27/10$ &-\\
$(8,5,5,4)$ & $8/5$ & $32/5$ &-\\
$(4,3,3,3)$ & $3/2$ &- & -\\
$(8,3,3,2)$ & $3/2$ & $27/2$ &-\\
$1/2(5,4,4,4)$ & $7/5$& $103/5$ &-\\
$1/3(8,1,1,2)$ & $8/7$ & $128/7$ & $116/7$\\
$1/9(8,3,3,0)$ & $4/5$ & $76/5$&-\\ \hline
\end{tabular}
\end{center}

\vspace*{5mm}

\subsection{Significance of the Galois group}
The {\bf Kronecker-Weber theorem} states that every finite abelian
extension $F$ of $\mathbb{Q}$ (i.e. every algebraic number field $F$
whose Galois group over $\mathbb{Q}$ is abelian) is a subfield of a
cyclotomic field.

For a given abelian extension $F$ of $\mathbb{Q}$ there is in fact a
minimal cyclotomic field that contains $F$. The {\bf conductor} of
$F$ is defined to be the smallest integer $n$ such that $F$ lies
inside the field generated by the $n^{th}$ roots of unity. Hence we
can say that $\gal(F)$ is abelian if and only if $F$ is generated by
$n^{th}$ roots of unity. We can conclude that if an equation has
abelian Galois group, its solutions can be expressed in terms of
roots of unity.

Now let $G$ be a group and $\com(G)$ be its commutator subgroup.
$G/\com(G)$ is abelian, so $G$ is an extension of an abelian group
by $\com(G)$. Of the cases considered so far, the pairs of Cartan
matrices give rise to equations with abelian Galois groups, while
the equations arising from Terhoeven's matrices have Galois groups
that are abelian or have $\com(G)=\mathbb{Z}_2\ {\rm or}\
\mathbb{Z}_2\times\mathbb{Z}_2$. The pattern is not yet understood
completely.

\chapter{Integrable Models Described by Exceptional Lie Algebras} In this chapter we consider the integrable models described by pairs
of Dynkin diagrams $(E_m,T_1)$. There are three distinct models of
this type, since $m$ can take values 6, 7, or 8. For each model the
equations are of the form $U=AV$, where the matrix $A$ is given by
$A=C(E_m)\otimes C(T_1)^{-1}$, $U=\log(x)$, $V=\log(1-x)$, and
$x=(x_1,\ldots,x_m)$. In each case we solve the equations and use
these solutions to compute the effective central charge of the
corresponding conformal field theory. Since the matrices we deal
with in this chapter are bigger than the $2\times2$ matrics
considered in the previous chapter, the calculations are more
difficult. We do not go as far as to calculate the explicit modular
forms associated to each $(E_m,T_1)$ theory, neither do we try to
identify the CFTs themselves. Both of these would be interesting
future projects.

\section{Exceptional Lie Algebras}
The $E$ family of exceptional Lie algebras has three members, namely
$E_6$, $E_7$, and $E_8$. Their Dynkin diagrams are
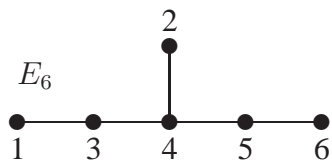
\begin{center}
\setlength{\unitlength}{1mm}
\begin{picture}(40,10)
\put(0,5){$E_6$}
\put(0,0){\circle*{2}}
\put(-1,-5){1}
\put(10,0){\circle*{2}}
\put(9,-5){3}
\put(20,0){\circle*{2}}
\put(19,-5){4}
\put(30,0){\circle*{2}}
\put(29,-5){5}
\put(40,0){\circle*{2}}
\put(39,-5){6}
\put(20,10){\circle*{2}}
\put(19,12){2}
\put(0,0){\line(1,0){10}}
\put(10,0){\line(1,0){10}}
\put(20,0){\line(1,0){10}}
\put(30,0){\line(1,0){10}}
\put(20,0){\line(0,1){10}}
\end{picture}
\end{center}

\vspace{5mm}

\begin{center}
\setlength{\unitlength}{1mm}
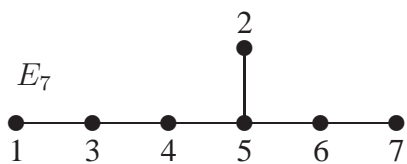
\begin{picture}(50,10)
\put(0,5){$E_7$}
\put(0,0){\circle*{2}}
\put(-1,-5){1}
\put(10,0){\circle*{2}}
\put(9,-5){3}
\put(20,0){\circle*{2}}
\put(19,-5){4}
\put(30,0){\circle*{2}}
\put(29,-5){5}
\put(40,0){\circle*{2}}
\put(39,-5){6}
\put(50,0){\circle*{2}}
\put(49,-5){7}
\put(30,10){\circle*{2}}
\put(29,12){2}
\put(0,0){\line(1,0){10}}
\put(10,0){\line(1,0){10}}
\put(20,0){\line(1,0){10}}
\put(30,0){\line(1,0){10}}
\put(40,0){\line(1,0){10}}
\put(30,0){\line(0,1){10}}
\end{picture}
\end{center}

\vspace{5mm}

\begin{center}
\setlength{\unitlength}{1mm}
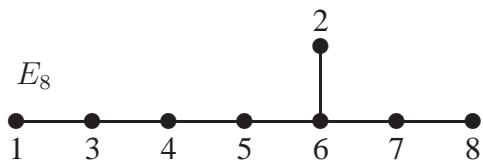
\begin{picture}(60,10)
\put(0,5){$E_8$}
\put(0,0){\circle*{2}}
\put(-1,-5){1}
\put(10,0){\circle*{2}}
\put(9,-5){3}
\put(20,0){\circle*{2}}
\put(19,-5){4}
\put(30,0){\circle*{2}}
\put(29,-5){5}
\put(40,0){\circle*{2}}
\put(39,-5){6}
\put(50,0){\circle*{2}}
\put(49,-5){7}
\put(60,0){\circle*{2}}
\put(59,-5){8}
\put(40,10){\circle*{2}}
\put(39,12){2}
\put(0,0){\line(1,0){10}}
\put(10,0){\line(1,0){10}}
\put(20,0){\line(1,0){10}}
\put(30,0){\line(1,0){10}}
\put(40,0){\line(1,0){10}}
\put(50,0){\line(1,0){10}}
\put(40,0){\line(0,1){10}}
\end{picture}
\end{center}

\vspace*{10mm}

The corresponding Cartan matrices are
$$
C(E_6) = \left(\begin{array}{rrrrrr}
2 & 0 & -1 & 0 & 0 & 0 \\
0 & 2 & 0 & -1 & 0 & 0 \\
-1 & 0 & 2 & -1 & 0 & 0 \\
0 & -1 & -1 & 2 & -1 & 0 \\
0 & 0 & 0 & -1 & 2 & -1 \\
0 & 0 & 0 & 0 & -1 & 2 \\
\end{array}\right)$$

$$
C(E_7) = \left(\begin{array}{rrrrrrr}
2 & 0 & -1 & 0 & 0 & 0 & 0\\
0 & 2 & 0 & -1 & 0 & 0 & 0\\
-1 & 0 & 2 & -1 & 0 & 0 & 0\\
0 & -1 & -1 & 2 & -1 & 0 & 0\\
0 & 0 & 0 & -1 & 2 & -1 & 0\\
0 & 0 & 0 & 0 & -1 & 2 & -1\\
0 & 0 & 0 & 0 & 0 & -1 & 2\\
\end{array}\right)$$

$$C(E_8) = \left(\begin{array}{rrrrrrrr}
2 & 0 & -1 & 0 & 0 & 0 & 0 & 0\\
0 & 2 & 0 & -1 & 0 & 0 & 0 & 0\\
-1 & 0 & 2 & -1 & 0 & 0 & 0 & 0\\
0 & -1 & -1 & 2 & -1 & 0 & 0 & 0\\
0 & 0 & 0 & -1 & 2 & -1 & 0 & 0\\
0 & 0 & 0 & 0 & -1 & 2 & -1 & 0\\
0 & 0 & 0 & 0 & 0 & -1 & 2 & -1\\
0 & 0 & 0 & 0 & 0 & 0 & -1 & 2\\
\end{array}\right)$$

\vspace*{10mm}

Their Coxeter numbers are
\begin{eqnarray*}
h(E_6) &=& 12\ ,\\
h(E_7) &=& 18\ ,\\
h(E_8) &=& 30\ .
\end{eqnarray*}

\section{Dynkin Diagrams $T_r$}
The `Tadpole' diagram, $T_r$, is got by folding the diagram $A_{2r}$ in the
middle, to get a pairwise identification of the vertices
$$T_r=A_{2r}/\mathbb{Z}_2\ .$$

This gives the Dynkin diagram
\begin{center}
\setlength{\unitlength}{1mm}
\begin{picture}(40,5)(-15,-18)
\put(0,0){\circle*{2}} \put(-1,-5){1} \put(10,0){\circle*{2}}
\put(9,-5){2} \put(30,0){\circle*{2}} \put(40,0){\circle*{2}}
\put(37,-4){r} \put(44,0){\circle{8}} \put(0,0){\line(1,0){10}}
\put(30,0){\line(1,0){10}} \thicklines \dottedline{2}(10,0)(20,0)
\dottedline{2}(20,0)(30,0)
\end{picture}
\end{center}
\vspace{-15mm} The Cartan matrix of $T_r$ is identical to that of
$A_r$ except for the entry $C(T_r)_{rr}=1$. In particular for $T_1$
we get the $1\times 1$ Cartan matrix
$$C(T_1)=1\ .$$

The Coxeter number of $T_r$ is $h(T_r)=2r+1$.

\section{Pairs of Dynkin Diagrams $(E_m,T_1)$}
\subsection{Solving the equations of the model}
\subsubsection{Case 1: $\mathbf{A=A(E_6,T_1)}$}
Here we consider the pair of Dynkin diagrams $\,X=E_6\,$ and
$\,Y=T_1\,$. The matrix $A=A(X,Y)$ is given by
$$A(E_6,T_1)=C(E_6)\otimes C(T_1)^{-1}=C(E_6)\ .$$
Exponentiating the equations $U=AV$ leads to the following set of
algebraic equations:
\begin{eqnarray*}
x_1 &=& \frac{(1-x_1)^2}{1-x_3}\ ,\\
x_2 &=& \frac{(1-x_2)^2}{1-x_4}\ ,\\
x_3 &=& \frac{(1-x_3)^2}{(1-x_1)(1-x_4)}\ ,\\
x_4 &=& \frac{(1-x_4)^2}{(1-x_2)(1-x_3)(1-x_5)}\ ,\\
x_5 &=& \frac{(1-x_5)^2}{(1-x_4)(1-x_6)}\ ,\\
x_6 &=& \frac{(1-x_6)^2}{1-x_5}\ .
\end{eqnarray*}
To solve these equations we write them in terms of the two variables
$x_6$ and $x_1$. This leads to two possible solutions; either
$x_6=x_1$, and $x_6$ satisfies the equation
\begin{equation}(5x_6^2-5x_6+1)(x_6-1)^3=0\
,\label{eq:e61}\end{equation} or $x_6\neq x_1$, and $x_6$ satisfies
the following polynomial of degree 7
\begin{equation} x_6(x_6^2-3x_6+1)(x_6-1)^4=0\ .\label{eq:e62}\end{equation}
We exclude any `solutions' that imply $x_i=\infty$ for any
$i=1,\ldots,6$ (this excludes all solutions arising
from~(\ref{eq:e62}) and some of those arising from~(\ref{eq:e61})).
This leaves two distinct solutions given by
\begin{equation}
x_1=\frac{1}{2}\pm\frac{\sqrt{5}}{10}\ .\label{eq:E6soln}
\end{equation}
The remaining $x_i$ are given by
\begin{eqnarray*}
x_2 &=& \frac{5}{2}x_1-1\ ,\\
x_3 &=& 4x_1-2\ ,\\
x_4 &=& 10-\frac{25}{2}x_1\ ,\\
x_5 &=& 4x_1-2\ ,\\
x_6 &=& x_1\ .
\end{eqnarray*}
Clearly these solutions reflect the symmetry of the $E_6$ Dynkin
diagram.

More explicitly we have
\begin{eqnarray*}
x_1=x_6=\frac{1}{2}+\frac{\sqrt{5}}{10} &\Rightarrow& x_2=\frac{1+\sqrt{5}}{4}\\
&\Rightarrow& x_3=\frac{2}{\sqrt{5}}=x_5\\
&\Rightarrow& x_4=\frac{15-5\sqrt{5}}{4}\ ,
\end{eqnarray*}
where $\sqrt{5}$ denotes either square root of $5$.

\subsubsection{Case 2: $\mathbf{A=A(E_7,T_1)}$}
In this case $(X,Y)=(E_7,T_1)$. The matrix $A(X,Y)$ is given by
$$A(E_7,T_1)=C(E_7)\otimes C(T_1)^{-1}=C(E_7)\ .$$
We want to solve to following set of algebraic equations arising
from the model
\begin{eqnarray*}
x_1 &=& \frac{(1-x_1)^2}{1-x_3}\ ,\\
x_2 &=& \frac{(1-x_2)^2}{1-x_4}\ ,\\
x_3 &=& \frac{(1-x_3)^2}{(1-x_1)(1-x_4)}\ ,\\
x_4 &=& \frac{(1-x_4)^2}{(1-x_2)(1-x_3)(1-x_5)}\ ,\\
x_5 &=& \frac{(1-x_5)^2}{(1-x_4)(1-x_6)}\ ,\\
x_6 &=& \frac{(1-x_6)^2}{(1-x_5)(1-x_7)}\ ,\\
x_7 &=& \frac{(1-x_7)^2}{(1-x_6)}\ .
\end{eqnarray*}

As before, these equations can be solved analytically. We reduce the
equations to a polynomial of degree 14 in $x_7$. This can be
factorised using MAPLE. We then solve the equation
$$(2x_7-1)(3x_7^2-9x_7+5)(x_7^3-x_7^2-2x_7+1)(x_7^2-3x_7+1)(x_7-1)^6=0\ .$$
The solutions are given by
\begin{eqnarray*}
x_7 &=& -\frac{3}{2}+\frac{\sqrt{21}}{2}\ ,\\
x_7 &=& -6\cos\left(\frac{2\pi}{7}\right)
+2\cos\left(\frac{3\pi}{7}\right) +4\cos\left(\frac{\pi}{7}\right)\
,
\end{eqnarray*}
and their Galois conjugates (given explicitly in later
calculations). Again we have excluded any solutions with
$x_i=\infty$ for any $i=1,\ldots,7$.

\subsubsection{Case 3: $\mathbf{A=A(E_8,T_1)}$}
In this case $(X,Y)=(E_8,T_1)$. The matrix $A$ is given by
$$A(X,Y)=C(E_8)\otimes C(T_1)^{-1}=C(E_8)\ .$$
The algebraic equations obtained from the model are
\begin{eqnarray*}
x_1 &=& \frac{(1-x_1)^2}{1-x_3}\ ,\\
x_2 &=& \frac{(1-x_2)^2}{1-x_4}\ ,\\
x_3 &=& \frac{(1-x_3)^2}{(1-x_1)(1-x_4)}\ ,\\
x_4 &=& \frac{(1-x_4)^2}{(1-x_2)(1-x_3)(1-x_5)}\ ,\\
x_5 &=& \frac{(1-x_5)^2}{(1-x_4)(1-x_6)}\ ,\\
x_6 &=& \frac{(1-x_6)^2}{(1-x_5)(1-x_7)}\ ,\\
x_7 &=& \frac{(1-x_7)^2}{(1-x_6)(1-x_8)}\ ,\\
x_8 &=& \frac{(1-x_8)^2}{(1-x_7)}\ .
\end{eqnarray*}

\noindent As in the previous case we get a polynomial of degree 17
in $x_8$. After factorisation this is
$$(2x_8-1)(x_8^3-2x_8^2-x_8+1)(x_8^5-3x_8^4-3x_8^3+4x_8^2+x_8-1)(x_8-1)^8=0\ .$$
There relevant solutions are
\begin{equation}
x_8=4\cos\left(\frac{2\pi}{11}\right)-2\cos\left(\frac{4\pi}{11}\right)-2\cos\left(
\frac{5\pi}{11}\right)-2\cos\left(\frac{\pi}{11}\right)\ ,\\
\end{equation}
and its four algebraic conjugates, got by replacing $\cos(2n\pi/11)$
by $\cos(2an\pi/11)$ with $a$ ranging over the quadratic residues
modulo 11.

\subsection{Effective central charge calculations}
Each $x=(x_1,x_2,\ldots,x_m)$ satisfies the set of algebraic
equations $x=(1-x)^A$. Logarithms of these solutions must be chosen
so that they satisfy the original equations of the model, namely
$\log(x)=A\log(1-x)$. To do this we define $\log(x)$ in terms of
$\log(z)$, where $z=(z_1,\ldots,z_m)$ is the variable introduced
earlier
\begin{eqnarray*}
1-x=z^{-C(Y)} &\Rightarrow& v_i=\log(1-x)=-C(Y)\log(z)\ ,\\
x=z^{-C(X)} &\Rightarrow& u_i=\log(x)=-C(X)\log(z)\ .
\end{eqnarray*}
The effective central charge is calculated using the formula
\begin{equation}
c_{\eff}=\frac{6}{\pi^2}\sum_{i=1}^mL(u_i^0,v_i^0)\ .
\label{eq:ceff}
\end{equation}
Here $(u_i^0,v_i^0)$ corresponds to the special solution whose
components satisfy $x_i\in\mathbb{R}$ and $0<x_i<1$ for all $i$.

Other values of $c-24h$ can be calculated (modulo $24\mathbb{Z}$)
from the remaining solutions by
\begin{equation}
c-24h=\frac{6}{\pi^2}\sum_{i=1}^mL(u_i,v_i)\ . \label{eq:ch}
\end{equation}

\subsubsection{Case 1: $A(E_6,T_1)$}
The relations
$$1-x=z^{-C(T_1)\otimes I_6}\qquad \mbox{and} \qquad x=z^{-C(E_6)}\ ,$$
give rise to the following set of equations for $\log(x_i)$ and
$\log(1-x_i)$.
\begin{eqnarray*}
v_i=\log(1-x_i)&=&-\log(z_i)\ ,\ \text{for}\ i=1,\ldots,6\ ,\\
&\text{and}&\\
u_1=\log(x_1)&=&-2\log(z_1)+\log(z_3)\ ,\\
u_2=\log(x_2)&=&-2\log(z_2)+\log(z_4)\ ,\\
u_3=\log(x_3)&=&\log(z_1)-2\log(z_3)+\log(z_4)\ ,\\
u_4=\log(x_4)&=&\log(z_2)+\log(z_3)-2\log(z_4)+\log(z_5)\ ,\\
u_5=\log(x_5)&=&\log(z_4)-2\log(z_5)+\log(z_6)\ ,\\
u_6=\log(x_6)&=&\log(z_5)-2\log(z_6)\ .
\end{eqnarray*}

This choice of logs must (and does) satisfy the following equations
($U=AV$).
\begin{eqnarray*}
\log(x_1)&=&2\log(1-x_1)-\log(1-x_3)\ ,\\
\log(x_2)&=&2\log(1-x_2)-\log(1-x_4)\ ,\\
\log(x_3)&=&-\log(1-x_1)+2\log(1-x_3)-\log(1-x_4)\ ,\\
\log(x_4)&=&-\log(1-x_2)-\log(1-x_3)+2\log(1-x_4)-\log(1-x_5)\ ,\\
\log(x_5)&=&-\log(1-x_4)+2\log(1-x_5)-\log(1-x_6)\ ,\\
\log(x_6)&=&-\log(1-x_5)+2\log(1-x_6)\ .
\end{eqnarray*}

Substituting the two solutions~(\ref{eq:E6soln}) into the
equations~(\ref{eq:ceff}) and~(\ref{eq:ch}) gives the following
results:

\begin{eqnarray*}
x_1=\frac{1}{2}-\frac{\sqrt{5}}{10} &\Rightarrow& c-24h=-\frac{24}{5}\mod24\mathbb{Z}\ ,\\
x_1=\frac{1}{2}+\frac{\sqrt{5}}{10} &\Rightarrow& c_{\eff}=\frac{24}{5}\ .\\
\end{eqnarray*}

\subsubsection{Case 2: $A(E_7,T_1)$}
Using the same method in this case gives the following results:
\begin{eqnarray*}
x_1=-\frac{3}{2}+\frac{\sqrt{21}}{2} &\Rightarrow& c_{\eff}=6\ ,\\
x_1=-\frac{3}{2}-\frac{\sqrt{21}}{2} &\Rightarrow& c-24h=-18\mod24\mathbb{Z}\ ,\\
x_1=-6\cos\left(\frac{2\pi}{7}\right)+2\cos\left(\frac{3\pi}{7}\right)+4\cos\left(
\frac{\pi}{7}\right) &\Rightarrow&c-24h=-\frac{30}{7}\mod24\mathbb{Z}\ ,\\
x_1=-4\cos\left(\frac{2\pi}{7}\right)+6\cos\left(\frac{3\pi}{7}\right)+2\cos\left(
\frac{\pi}{7}\right) &\Rightarrow&c-24h=-\frac{78}{7}\mod24\mathbb{Z}\ ,\\
x_1=-2\cos\left(\frac{2\pi}{7}\right)+4\cos\left(\frac{3\pi}{7}\right)+6\cos\left(
\frac{\pi}{7}\right)&\Rightarrow&c-24h=-\frac{102}{7}\mod24\mathbb{Z}\
.
\end{eqnarray*}

\subsubsection{Case 3: $A(E_8,T_1)$}
Again the same calculation gives
\begin{eqnarray*}
&&x_1=4\cos\left(\frac{2\pi}{11}\right)-2\cos\left(\frac{4\pi}{11}\right)-2\cos\left(
\frac{5\pi}{11}\right)-2\cos\left(\frac{\pi}{11}\right)\\
&\Rightarrow&
c-24h=-\frac{40}{11}\mod24\mathbb{Z}\ ,\\
&&\\
&&x_1=2\cos\left(\frac{2\pi}{11}\right)+4\cos\left(\frac{4\pi}{11}\right)
+2\cos\left(\frac{3\pi}{11}\right)-2\cos\left(\frac{\pi}{11}\right)\\
&\Rightarrow&c-24h=-\frac{160}{11}\mod24\mathbb{Z}\ ,\\
&&\\
&&x_1=-4\cos\left(\frac{5\pi}{11}\right)+2\cos\left(\frac{4\pi}{11}\right)-
2\cos\left(\frac{3\pi}{11}\right)+2\cos\left(\frac{\pi}{11}\right)\\
&\Rightarrow&c_{\eff}=\frac{80}{11}\ ,\\
&&\\
&&x_1=-4\cos\left(\frac{3\pi}{11}\right)+2\cos\left(\frac{4\pi}{11}\right)
+2\cos\left(\frac{2\pi}{11}\right)+2\cos\left(\frac{5\pi}{11}\right)\\
&\Rightarrow&c-24h=-\frac{112}{11}\mod24\mathbb{Z}\ ,\\
&&\\
&&x_1=-4\cos\left(\frac{\pi}{11}\right)-2\cos\left(\frac{2\pi}{11}\right)
-2\cos\left(\frac{5\pi}{11}\right)-2\cos\left(\frac{3\pi}{11}\right)\\
&\Rightarrow&c-24h=-\frac{208}{11}\mod24\mathbb{Z}\ .
\end{eqnarray*}

{\it Note: Similar calculations have been carried out by Klassen and
Melzer for the cases $(A_1,E_6)$, $(A_1,E_7)$, and $(A_1,E_8)$. For
more details see~\cite{KM1}.}


\chapter*{Summary} Certain integrable models are described by pairs $(X,Y)$ of ADET
Dynkin diagrams. At high energy these models are expected to have a
conformally invariant limit. The S-matrix of the model determines
algebraic equations, whose solutions are mapped to the central
charge and scaling dimensions of the corresponding conformal field
theory. We study the equations of the $(D_m,A_n)$ model and find all
solutions explicitly using the representation theory of Lie algebras
and related Yangians. These mathematically rigorous results are in
agreement with the expectations arising from physics. We also
investigate the overlap between certain q-hypergeometric series and
modular functions. We study a particular class of 2-fold
q-hypergeometric series, denoted $f_{A,B,C}$. Here $A$ is a positive
definite, symmetric, $2\times 2$ matrix, $B$ is a vector of length
$2$, and $C$ is a scalar, all three with rational entries. It turns
out that for certain choices of the matrix $A$, the function
$f_{A,B,C}$ can be made modular. We calculate the corresponding
values of $B$ and $C$. It is expected that functions $f_{A,B,C}$
arising in this way are characters of some rational conformal field
theory. We show that this is true in at least one case, namely
$A=\left(\begin{array}{cc} 4&1\\1&1
\end{array}\right)$.

\bibliographystyle{unsrt}

\bibliography{bibliography}

\nocite{*}

\appendix

\chapter{Useful Method of Solving Equations} This appendix describes a useful method of solving equations. In
chapter 5 it is used to simplify certain solutions arising from
equations of the $(E_m,T_1)$ model.
\section{General Method}
Suppose we wish to find the roots of an irreducible polynomial
\begin{equation}
x^3+a_1x^2+a_2x+a_3=0\ , \label{eq:poly}
\end{equation}
with rational $a_i$. This equation has three roots, say $x_1$,
$x_2$, and $x_3$. Suppose that
$$x_1=\sum_{i=1}^3b_i(\omega^i+\omega^{-i})=b_1(\omega+\omega^{-1})+b_2(\omega^2
+\omega^{-2})+b_3(\omega^3+\omega^{-3})\ ,$$ with rational $b_i$,
and $\omega=exp\left(\frac{2\pi i}{7}\right)$.\\

{\bf Galois group}\\
The Galois group of $\omega$ is generated by
$\,\gamma:\omega\mapsto\omega^3$. This action extends to other
elements as
$$\gamma:\omega\mapsto\omega^3\mapsto\omega^2\mapsto\omega^6\mapsto\omega^4
\mapsto\omega^5\mapsto\omega\ .$$ Hence the Galois group is
$\mathbb{Z}_6$
$$\gamma:1\mapsto3\mapsto2\mapsto6\mapsto4\mapsto5\mapsto1\ .$$

{\bf Solutions}\\
Using the action of the Galois group we can find the other two roots
by
$$x=\sum_{k=1}^3b_k\left(\omega^{\gamma(k)}+\omega^{-\gamma(k)}\right)\ .$$

It follows that the three roots of the equation are
\begin{eqnarray}
x_1 &=&
b_1(\omega+\omega^{-1})+b_2(\omega^2+\omega^{-2})+b_3(\omega^3+\omega^{-3})\ , \nonumber\\
x_2 &=&
b_1(\omega^3+\omega^{-3})+b_2(\omega+\omega^{-1})+b_3(\omega^2+\omega^{-2})\ , \label{eq:root}\\
x_3 &=&
b_1(\omega^2+\omega^{-2})+b_2(\omega^3+\omega^{-3})+b_3(\omega+\omega^{-1})\
. \nonumber
\end{eqnarray}

\pagebreak
{\bf Coefficients}\\
In terms of the three roots~(\ref{eq:root}) the
equation~(\ref{eq:poly}) becomes
\begin{eqnarray*}
&& x^3+a_1x^2+a_2x+a_3\\
&=& x^3+(b_1+b_2+b_3)x^2+\left(-2\sum{b_i^2}+(-2+5)\sum_{i<j}b_ib_j\right)x
+\ldots
\end{eqnarray*}

Solving for $\{a_1,a_2,a_3\}$ in terms of $\{b_1,b_2,b_3\}$ gives
\begin{eqnarray*}
a_1 &=& b_1+b_2+b_3\ ,\\
a_2 &=&
-2\sum{b_i^2}+3\sum_{i<j}b_ib_j=-\frac{7}{2}\sum{b_i^2}+\frac{3}{2}
\left(\sum{b_i}\right)^2\ .
\end{eqnarray*}

This can be rearranged to get
\begin{eqnarray*}
\sum_{i=1}^3{b_i} &=& a_1\ ,\\
\sum_{i=1}^3{b_i^2} &=&
-\frac{2}{7}\left(a_2-\frac{3}{2}a_1^2\right)\ .\\
\end{eqnarray*}

\section{Example}
We use the method described above to find the roots of the equation
$$x^3-6x^2+5x-1=0\ .$$

{\bf Solution}\\
We can solve for $\{b_1,b_2,b_3\}$ to get
\begin{eqnarray*}
(a_1,a_2)=(-6,5) &\Rightarrow& b_1+b_2+b_3=-6 \quad\mbox{and}\quad
b_1^2+b_2^2+b_3^2=14\\
&\Rightarrow& (b_1,b_2,b_3)=(-3,-2,-1)\ .
\end{eqnarray*}
The 3 roots of the equation turn out to be
\begin{eqnarray*}
x &=& -4\cos\left(\frac{2\pi}{7}\right)
+6\cos\left(\frac{3\pi}{7}\right)
+2\cos\left(\frac{\pi}{7}\right)\ ,\\
x &=& -2\cos\left(\frac{2\pi}{7}\right)
+4\cos\left(\frac{3\pi}{7}\right)
+6\cos\left(\frac{\pi}{7}\right)\ ,\\
x &=& -6\cos\left(\frac{2\pi}{7}\right)
+2\cos\left(\frac{3\pi}{7}\right) +4\cos\left(\frac{\pi}{7}\right)\
.
\end{eqnarray*}

\end{document}